\documentclass[journal,comsoc]{IEEEtran}

\usepackage{setspace}
\usepackage{cite}
\ifCLASSINFOpdf
\else
\fi

\usepackage{setspace}
\usepackage{cite}
\usepackage{caption}
\usepackage{subcaption}
\usepackage{nomencl}
\usepackage{parskip}
\usepackage{subfig}
\usepackage{soul}

\usepackage[T1]{fontenc}
\usepackage{lmodern} 

\usepackage[section]{placeins}

\usepackage{url}
\usepackage[dvips]{epsfig}  
\usepackage{verbatim}
\usepackage{dcolumn}
\usepackage{amsmath}
\usepackage{array}
\usepackage{color}
\usepackage{lscape}
\usepackage{latexsym}
\usepackage{gensymb}
\usepackage{amssymb}
\usepackage[section]{placeins}
\usepackage{graphicx, epstopdf, epsfig}
\usepackage{xcolor}
\usepackage{textcomp}
\DeclareGraphicsExtensions{.eps, .png}
\usepackage{subfig}
\usepackage{multirow}
\usepackage{float}
\newcolumntype{M}{>{\centering\arraybackslash}m{\dimexpr.5\linewidth-2\tabcolsep}}

\begin{document}
\pagestyle{empty}

\thispagestyle{empty}
%
\title{
Toward Next-Generation AI Data Centers: Power Delivery Architecture Shifts, Emerging Technologies, and Challenges \\[0.6em]

\centering
\small\itshape
This work has been submitted to the IEEE for possible publication.\\[-0.1em]
Copyright may be transferred without notice, after which this version may no longer be accessible.%
}


\author{
Sangwhee~Lee,~\IEEEmembership{Member,~IEEE};
~Rafal P. Wojda,~\IEEEmembership{Senior Member,~IEEE};
~Cheol-Hee Jo,~\IEEEmembership{Member,~IEEE};
~Shuntaro~Inoue,~\IEEEmembership{Senior Member,~IEEE};
~Pedro Ribeiro,~\IEEEmembership{Member,~IEEE};
~Gui-Jia~Su,~\IEEEmembership{Senior Member,~IEEE};
~Mostak~Mohammad,~\IEEEmembership{Senior Member,~IEEE}; 
~Himel~Barua,~\IEEEmembership{Member,~IEEE};
~Nishanth~Gadiyar,~\IEEEmembership{Senior Member,~IEEE};
~Praveen~Kumar,~\IEEEmembership{Senior Member,~IEEE};
~Spencer~Cochran,~\IEEEmembership{Member,~IEEE};
~Subho~Mukherjee,~\IEEEmembership{Senior Member,~IEEE};
~Whit~Vinson,~\IEEEmembership{Member,~IEEE};
~Vandana~Rallabandi,~\IEEEmembership{Senior Member,~IEEE};
~Shajjad~Chowdhury,~\IEEEmembership{Senior Member,~IEEE};
~Burak~Ozpineci,~\IEEEmembership{Fellow,~IEEE}

\thanks{S. Lee, R. Wojda, C. Jo, S. Inoue, P. Ribeiro, G. Su, M. Mohammad, H. Barua, N. Gadiyar, P. Kumar, S. Cochran, S. Mukhuerjee, W. Vinson, V. Rallbandi, S. Chowdhury, and B. Ozpineci are with Oak Ridge National Laboratory, Knoxville, TN 37932 USA (email: lees9$@$ornl.gov; chowdhuryms$@$ornl.gov; baruah$@$ornl.gov; burak$@$ornl.gov). \textit{(Corresponding author: Burak Ozpineci.)}}
\thanks{This manuscript has been authored by UT-Battelle, LLC, under contract DE-AC05-00OR22725 with the US Department of Energy (DOE). The US government retains and the publisher, by accepting the article for publication, acknowledges that the US government retains a nonexclusive, paid-up, irrevocable, worldwide license to publish or reproduce the published form of this manuscript, or allow others to do so, for US government purposes. DOE will provide public access to these results of federally sponsored research in accordance with the DOE Public Access Plan (http://energy.gov/downloads/doe-public-access-plan).}}

\markboth{}%
{Shell \MakeLowercase{\textit{et al.}}: Bare Demo of IEEEtran.cls for IEEE Journals}

\maketitle


\begin{abstract}
The rapid growth of AI workloads is driving unprecedented increases in data center power demand, current transients, and thermal stress, exposing fundamental limitations in traditional 48 V rack architectures, low-voltage AC distribution, and line-frequency transformer interfaces. This paper reviews the three stages of architectural shifts required to support next-generation AI data centers and identifies three enabling technological building blocks: high-voltage conversion-ratio DC–-DC converters, facility-level low-voltage DC distribution, and medium-voltage solid-state transformers. The advantages, technical challenges, and potential solutions associated with each building block are reviewed. Finally, future research directions and open challenges are discussed.

\end{abstract}


%

\IEEEpeerreviewmaketitle
\section{Introduction}

Traditional data centers have long supported IT services and cloud computing \cite{PES_2025,MS_History}. Their power requirements were relatively moderate, and power delivery and cooling were often secondary design considerations \cite{PES_2025,MS_History}. However, the rise of AI workloads—coupled with the end of Dennard scaling—has drastically changed power demand dynamics in data centers \cite{Dennard_scaling,NVIDIA800VDC}. Lawrence Berkeley National Laboratory (LBNL) projected that the total data center electricity demand in the United States could rise by up to about three times, representing about 12\% of the total US electricity consumption \cite{LBNL2024}. Furthermore, AI workloads changed the operational profile of data centers as well. Unlike the traditional cloud application workloads \cite{11025802,10606267}, AI training workloads generate high-magnitude power fluctuations as GPU clusters shift between compute-intensive and communication-intensive phases \cite{02886-x}. These transitions produce steep power fluctuations and extreme $di/dt$ in the power delivery system in the data centers. As a result, power infrastructure—once an afterthought—has become the primary determinant of AI data center operation and performance \cite{NVIDIA800VDC}. 

Given these unprecedented demands, incremental efficiency improvements are no longer sufficient for future AI data centers. Instead, a fundamental paradigm shift in data center power delivery architecture---ranging from compute rack bus and data center facility busways to the data center-to-grid interface---is required. Traditional data centers are equipped with in-rack 48 V DC buses and in-facility 480 V AC power distribution architecture \cite{NVIDIA800VDC}. In such architecture, the conventional in-rack 48 V DC bus should conduct over 2 kA to supply 100 kW, limiting the power scalability of the conventional racks \cite{Google2025_1MW_IT_Racks,ocp_diablo400,2025_OCP_GOOGLE}. To support the growing demand for higher compute power per rack, compute racks with 1 MW+ scale power are required for next-generation data centers \cite{NVIDIA800VDC}. The traditional in-rack 48 V DC buses can no longer support racks with such high power demand \cite{Google2025_1MW_IT_Racks,ocp_diablo400,2025_OCP_GOOGLE}. As a result, major hyperscalers have begun raising the in-rack bus voltage from the conventional 48 V DC to the low-voltage (LV) DC level such as ±400 V DC or 800 V DC to support 1 MW+ scale compute racks \cite{ocp_diablo400, NVIDIA800VDC}. 

The conventional in-facility power distribution architecture based on LV AC (e.g., 480 V AC) is also being challenged. As individual rack power is increasing toward 1 MW+ scale, the power demand for AI data centers is increasing up to multigigawatt scale. As data center power demand continues to increase, losses throughout the power distribution busway can no longer be treated as negligible. Because all IT loads require DC power, the conventional in-facility LV AC distribution architecture requires multiple power conversion stages, adding to the power delivery losses. To reduce the total number of power conversion stages in a data center's power delivery architecture, the facility-level power delivery architecture is also shifting from the conventional LV AC (e.g., 480 V AC) to the LV DC (e.g., ±400 V DC, ±750 V DC, or 800 V DC) distribution busway \cite{NVIDIA800VDC,ocp_diablo400,2025_OCP_GOOGLE}. Furthermore, on-site energy storage systems (ESS) and generators can be directly connected to the LV DC distribution architecture, further improving the efficiency of the power delivery architecture \cite{NVIDIA800VDC,ocp_diablo400,2025_OCP_GOOGLE,Cuzner_DC_ground}.

Beyond increasing the in-rack bus voltage and shifting to the LV DC distribution in the data center facility, there is a growing interest in how future AI data centers should interface with the medium-voltage (MV) AC grid. Typically, bulky line-frequency transformers (LFTs) are used to step down the MV AC voltage to LV AC followed by an AC--DC converter to form a stiff DC voltage in data centers \cite{NVIDIA800VDC,ocp_diablo400,2025_OCP_GOOGLE}. Instead, directly interfacing the MV AC grid to a data center's LV DC distribution system is getting more attention \cite{NVIDIA800VDC,2025_OCP_GOOGLE}. The enabling technology is the MV-solid-state transformer (MV-SST) \cite{NVIDIA800VDC,2025_OCP_GOOGLE}. This power electronics technology replaces a bulky LFT with more compact medium-frequency transformers, dramatically reducing the footprint of the data center's power delivery architecture. 

This article first provides an overview of the state-of-the-art architectural shift that is actively pursued in the AI data center industry and identifies three technology building blocks related to power electronics: (i) isolated DC--DC converter with high-voltage conversion ratio as LV intermediate bus converter (LV IBC), (ii) in-facility DC distribution system, and (iii) MV-SST. Afterward, the current technology status and associated challenges for each technological building block are examined. The remainder of the article is structured as follows. Section~II discusses the three major shifts in AI data center power delivery architecture and identifies the corresponding technological building blocks. Section~\ref{section:LV-IBC} examines the first building block—the LV IBC. The in-facility DC distribution system is reviewed as the second technological building block in Section~\ref{section:DC_Distribution}. The final building block—the medium-voltage solid-state transformer (MV-SST)—is discussed in Section~\ref{section:MV-SST}. Section~\ref{section:Emerging} outlines research gaps and future opportunities, and Section~\ref{Section:Conclusion} concludes the article.

\section{Evolution of Data Center Power Delivery Architecture}
\label{section:3_stages}

To understand the technological requirements of future AI data centers, it is first necessary to establish a clear framework for voltage levels used in power delivery systems. Throughout this article, voltage levels are classified according to IEC 60364 and summarized as follows:

\begin{itemize}
    \item Extra-low voltage (ELV): $\leq$ 50 V AC or 120 V DC
    \item Low voltage (LV): $\leq$ 1,000 V AC or 1,500 V DC
    \item Medium voltage (MV): $>$ 1,000 V AC or 1,500 V DC
\end{itemize}

These voltage classifications are used consistently throughout the article unless stated otherwise.

With this voltage framework in place, Section~\ref{section:3_stages} outlines three architectural shifts currently taking place for future AI data centers. This article also identifies three core technology building blocks associated with these three architectural shifts that anchor the remainder of this article.

\begin{figure*}[!t]
    \centering
    \begin{subfigure}{0.56\textwidth}
        \centering
        \includegraphics[width=0.9\linewidth]{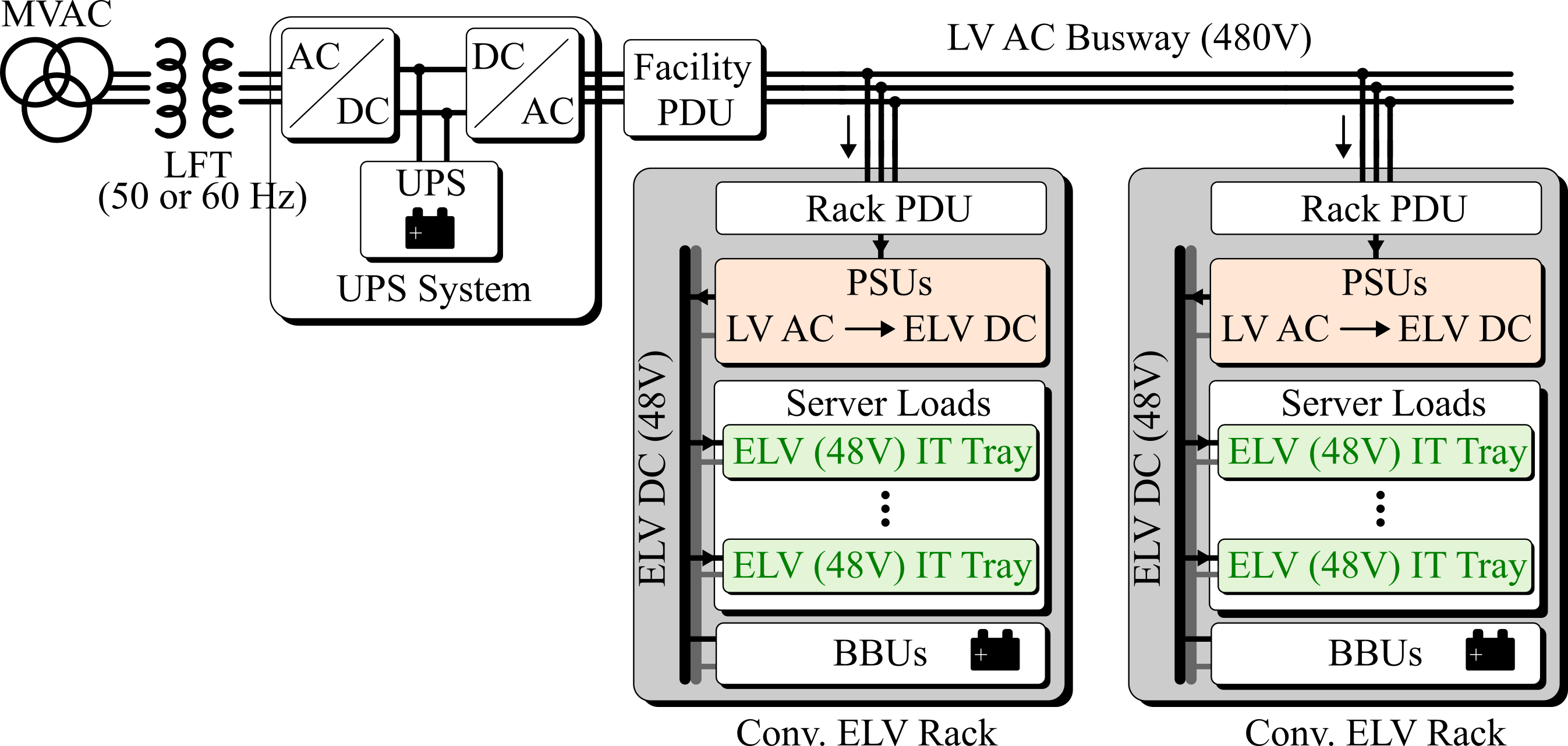}
        \caption{}
        \label{conv_a}
    \end{subfigure}
    \hfill
    \begin{subfigure}{0.42\textwidth}
        \centering
        \includegraphics[width=0.9\linewidth]{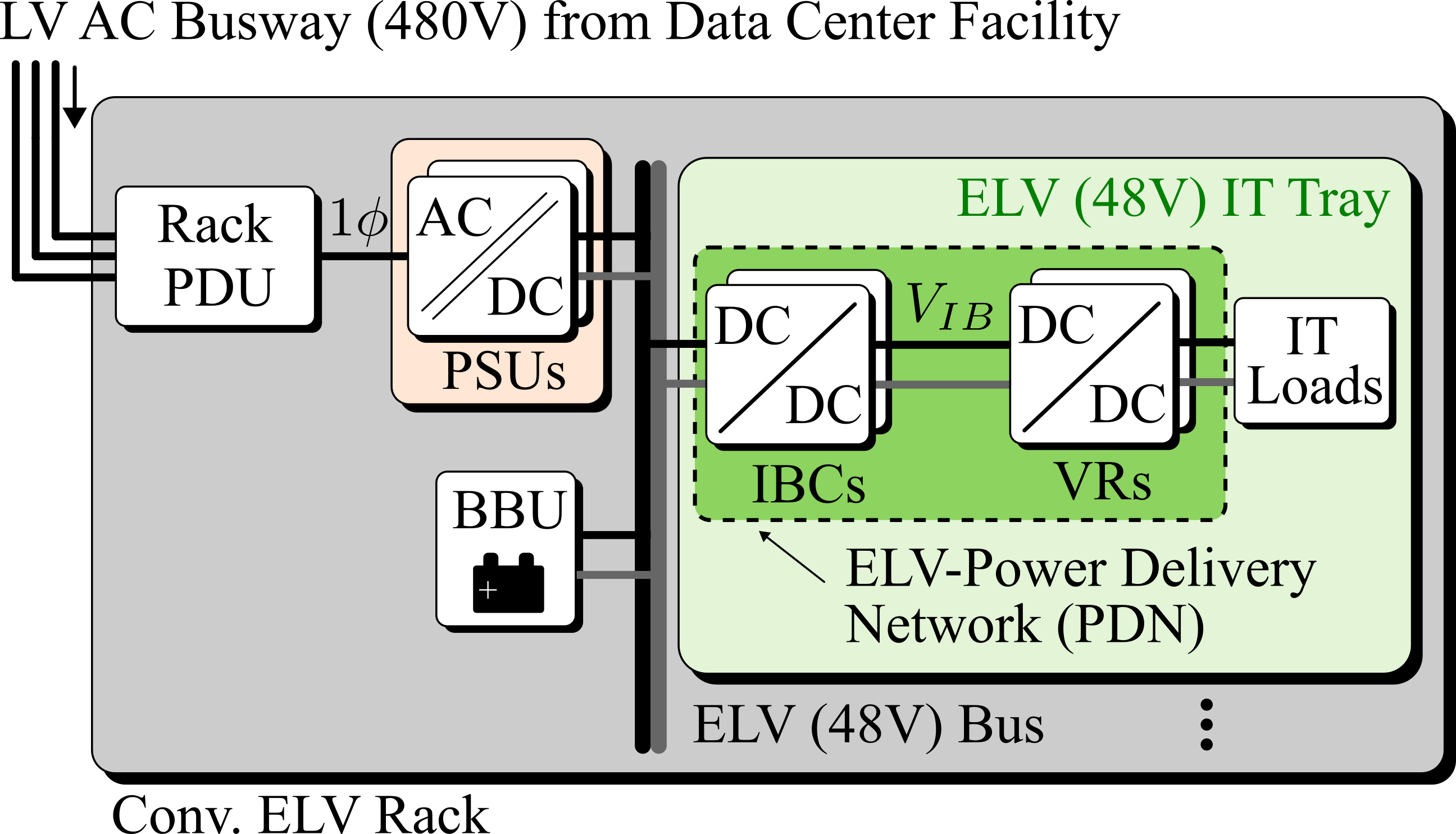}
        \caption{}
        \label{conv_b}
    \end{subfigure}      
    \caption{Conventional modern data center power delivery architecture with in-facility LV AC busway and conventional ELV racks with ELV DC bus (e.g., in-rack 48 V DC bus). (a) Overall power delivery architecture of conventional data center and (b) expanded view of conventional ELV rack with ELV-IBC on the ELV IT tray. Abbreviations: IBC = intermediate bus converter; PDB = power distribution board on IT tray.}
    \label{fig:conv}
\end{figure*}

A typical schematic of a data center and its key subsystems is shown in Fig.~\ref{fig:conv}(a) \cite{SL2}. A data center consists of the following major components:

\begin{itemize}
    \item Line-frequency transformer (LFT)
    \item Energy storage system (ESS)
    \item Power distribution units (PDUs)
    \item Power supply units (PSUs)
    \item IT trays
\end{itemize}

In conventional data center power delivery architectures, MV AC power is supplied from the distribution grid to the data center substation. Within the substation, the MV AC voltage is stepped down to a LV AC level—typically 480 V AC in the United States and 400 V AC in Europe and Asia—using an LFT. This stepped-down LV AC voltage is then delivered to the ESS (i.e., the uninterruptible power supply [UPS] system), as illustrated in Fig.~\ref{fig:conv}(a).

The LV AC from the UPS system's output is delivered to the data center facility through the PDU, forming a LV AC busway inside the data center facility, as shown in Fig.~\ref{fig:conv}(a).

From the LV AC busway, power is delivered to individual racks. In the rack, the LV AC is first delivered to the rack PDU, which is equipped with protection equipment such as fuses and circuit breakers \cite{PDU_EATON, PDU_Peral_Hu}. The rack PDU in a conventional ELV rack converts the three-phase (3$\phi$) line-to-line LV AC into three single-phase (1$\phi$) line-to-neutral voltages, as shown in Fig. \ref{fig:conv}(b) \cite{ORV2}.

Multiple PSUs (or single-phase AC--DC converters) are connected in parallel, forming power shelves (see the orange area in Fig. \ref{fig:conv}[a]) in a conventional ELV rack. These PSUs collectively convert the LV AC from the rack PDU to ELV DC, forming the ELV DC (e.g., 48 V DC) bus within the conventional ELV rack (see the ELV DC bus in Figs. \ref{fig:conv}[a] and [b]) \cite{ORV2}. The output of PSUs (i.e., ELV DC bus) delivers power to the rest of the in-rack system such as the server load's in-rack ESS (e.g., backup battery units [BBUs] in Fig. \ref{fig:conv}). This conventional in-rack bus system is often referred to as a 48 V DC bus system \cite{NVIDIA800VDC}. 

The in-rack ESS, such as the BBUs and supercapacitor banks, are connected to the ELV DC rack bus (see Fig. \ref{fig:conv}), supporting the in-rack bus under dynamic load conditions \cite{Dynamic_AI_work_load}. Note that BBUs in individual racks eliminate the centralized UPS system, potentially saving about 10\% of total energy consumption in the power delivery architecture \cite{SL3}. This distributed DC UPS system can also improve the reliability of the overall system \cite{SL3}.

The main purpose of the rack bus is to stably deliver power to the server loads, which are composed of multiple IT trays, as shown in Fig. \ref{fig:conv}(a). On a typical ELV IT tray, multiple DC--DC converters are placed to supply power to different IT loads. This power delivery network on the ELV IT tray is referred to as the ELV-power delivery network (ELV-PDN), as shown in Fig. \ref{fig:conv}(b), inside the area surrounded by the dotted line \cite{IBC_VT_NVIDIA}. Inside the ELV-PDN, the intermediate bus converters (IBCs) first convert the rack bus voltage (e.g., 48 V) down to an intermediate bus voltage $V_{IB}$, as shown in Fig. \ref{fig:conv}(b). Voltage regulators (VRs) convert $V_{IB}$ into lower voltages that are compatible with the IT load, as shown in Fig. \ref{fig:conv}(b). Details on the ELV-PDN and its DC--DC converters are discussed in Section \ref{section:LV-IBC}.

Because of the ever-increasing demand for more computational power, the conventional in-rack ELV DC bus (see Fig. \ref{fig:conv}) can no longer support the IT load. As a result, rather than incremental improvements, a fundamental architectural shift is required. The evolution of AI data center architecture can be characterized by three major stages: (i) a shift toward higher in‑rack DC bus voltage, (ii) a transition to in‑facility DC distribution, and (iii) the adoption of direct MV AC interfacing using MV‑SST technology and thereby eliminating bulky LFTs. The following subsections discuss these architectural shifts that are actively pursued by hyperscalers and OEMs for next-generation AI data centers \cite{NVIDIA800VDC, ocp_diablo400}. The current status and challenges of each technology building block are discussed in the subsequent sections (i.e., Sections \ref{section:LV-IBC}, \ref{section:DC_Distribution}, and \ref{section:MV-SST}).
\\
\subsubsection{Shift 1: Toward Higher In-Rack DC Distribution Voltage}

One of the most daunting challenges associated with increasing rack power (e.g., 1 MW or higher) in the conventional ELV rack (see Fig. \ref{fig:conv}[a]) is the enormous current that the in-rack bus must handle. Simply put, to support a 1 MW compute rack, the in-rack bus needs to conduct approximately 20 kA of current. Designing an in-rack bus capable of handling such high current levels is extremely challenging. Consequently, key hyperscalers and OEMs in the data center sector are actively pursuing in-rack bus systems with higher voltage levels \cite{NVIDIA800VDC, ocp_diablo400}. 

\begin{figure*}[!t]
    \centering
    \begin{subfigure}{0.56\textwidth}
        \centering
        \includegraphics[width=1\linewidth]{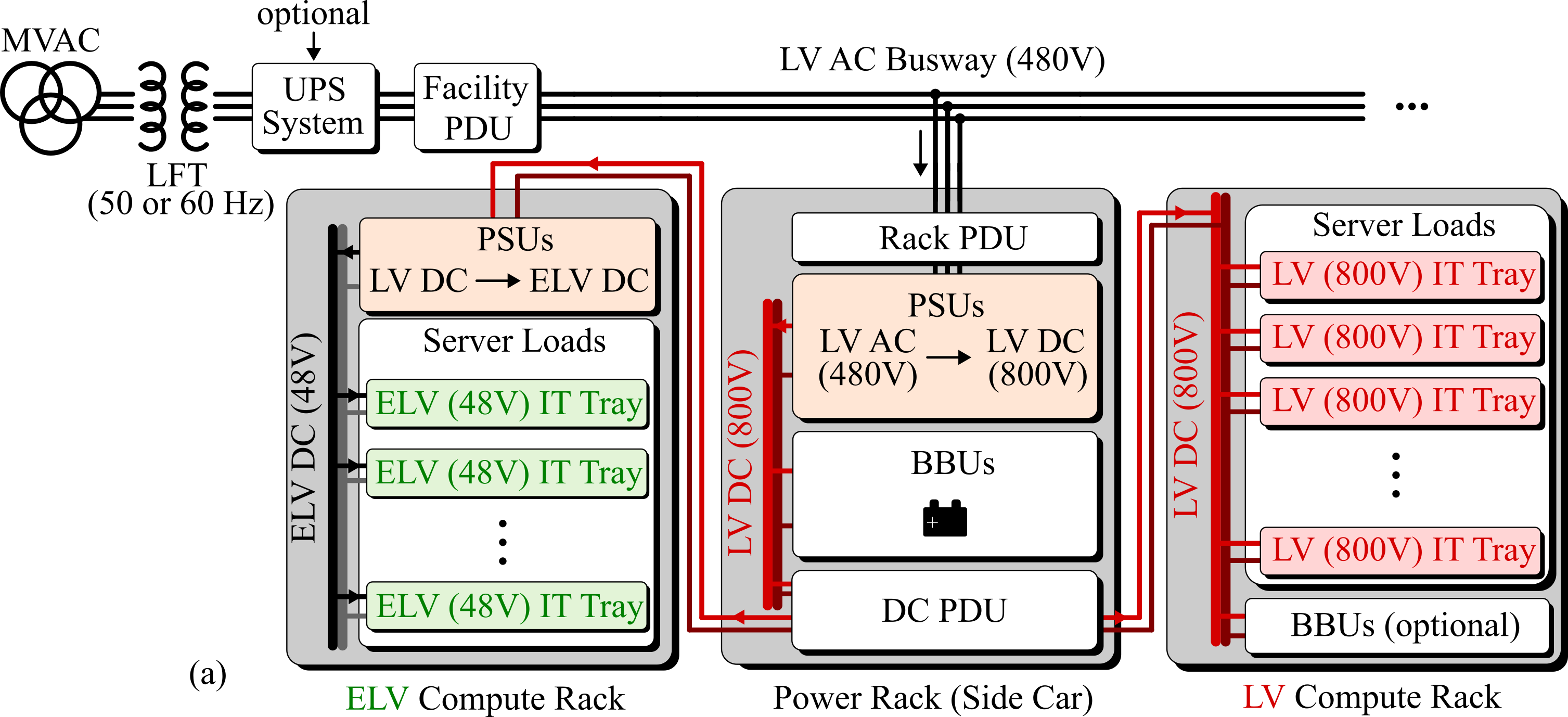}
        \label{side_car_a}
    \end{subfigure}
    \hfill
    \begin{subfigure}{0.42\textwidth}
        \centering
        \includegraphics[width=1\linewidth]{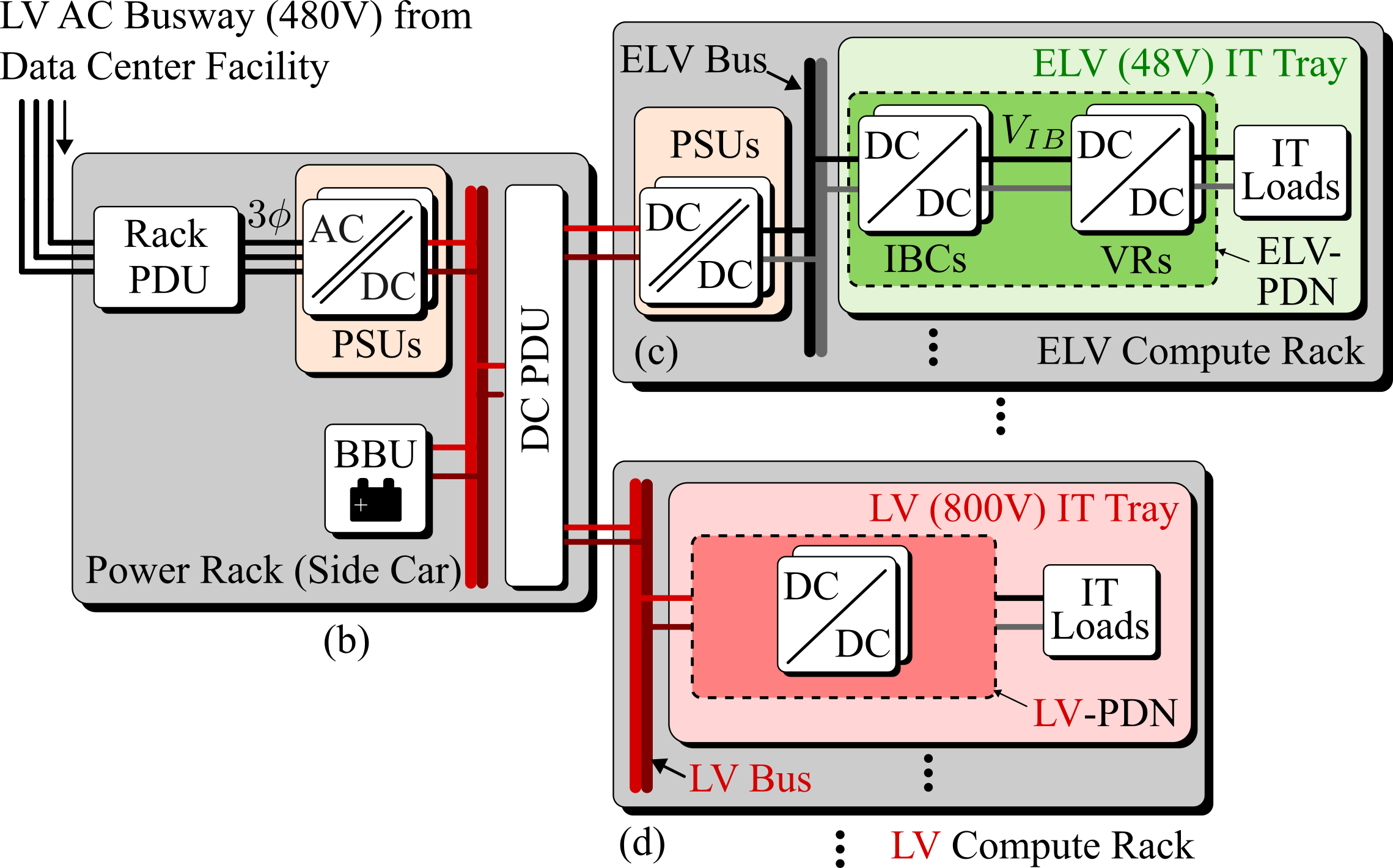}
        \label{side_car_bcd}
    \end{subfigure}      
    \caption{Data center power delivery architecture employing an in-rack LV bus (i.e., LV compute rack), a disaggregated power rack (i.e., sidecar), and an ELV compute rack. (a) Overall architecture of Shift~1 employing a conventional LV AC (e.g., 480~V AC) busway and an LV compute rack. (b) Expanded view of the power rack (sidecar). (c) Expanded view of the ELV compute rack. (d) Expanded view of the LV compute rack with an LV power distribution network (LV-PDN).}
    \label{fig:side_car}
\end{figure*}

A typical embodiment of this stage is illustrated in Fig. \ref{fig:side_car} \cite{ocp_diablo400, NVIDIA800VDC}. In this configuration, a power rack converts 480 V AC from the busway to the in-rack LV DC bus (e.g., 800 V DC) \cite{NVIDIA800VDC}. The LV DC is then delivered to compute racks, including both the LV compute rack equipped with an in-rack LV bus system and the ELV compute rack, as shown in Fig. \ref{fig:side_car}(a). 

The disaggregation of power supply and energy storage units (e.g., PSUs and BBUs) enables optimization of individual racks either for power (power rack) or computation (LV compute rack), thereby providing more space for IT loads within a single rack \cite{ocp_diablo400, NVIDIA800VDC}, as shown in Fig. \ref{fig:side_car}(a). This disaggregated rack design offers several advantages: (i) the power rack can be reused across multiple compute rack generations and configurations, reducing the need to replace power hardware; (ii) compute racks can be upgraded or replaced independently of the power rack, lowering infrastructure costs; and (iii) the disaggregated system offers scalability in rated power (from approximately 800~kW to beyond 1~MW), as well as improved redundancy and backup flexibility. 

The expanded view of the power rack is shown in Fig. \ref{fig:side_car}(b). The three-phase 480 V AC busway voltage is delivered to isolated AC–DC converters within the PSUs. The outputs of the PSUs form an in-rack LV bus inside the power rack. The LV DC is then distributed to other compute racks, such as the ELV compute rack (see Fig. \ref{fig:side_car}[c]) and the LV compute rack (see Fig. \ref{fig:side_car}[d]).

The ELV compute rack (see the leftmost rack in Fig. \ref{fig:side_car}[a]) represents an intermediate solution that can be used during the transition toward a full in-rack LV bus system. It retains the conventional in-rack ELV bus system. However, instead of the PSUs converting the 480 V AC busway voltage to ELV DC (see the conventional ELV rack in Fig. \ref{fig:conv}), the LV DC supplied from the power rack is converted to ELV DC to form the in-rack bus system. Because conventional server loads are compatible with the ELV bus voltage, the existing IT trays can be directly used in the ELV compute rack, as shown in the expanded view of Fig. \ref{fig:side_car}(c). Although the ELV compute rack benefits from additional space for server loads because of the disaggregation of power components \cite{ocp_diablo400}, it still employs the ELV bus system and is therefore insufficient to support megawatt-scale compute loads. 

The LV compute rack (see the rightmost rack in Fig. \ref{fig:side_car}[a]), which natively operates with an in-rack LV bus system, represents the ultimate goal of increasing the in-rack bus voltage. The LV DC supplied from the power rack directly forms the in-rack LV bus system, delivering power to the IT trays, as shown in Fig. \ref{fig:side_car}(a).

A key technical challenge in the LV compute rack is developing an IT tray compatible with the in-rack LV bus system, as illustrated in the LV compute rack shown in Fig. \ref{fig:side_car}(a) and its expanded view in Fig. \ref{fig:side_car}(d). Compared with the conventional ELV IT tray that interfaces with the ELV bus (see the ELV IT tray in Fig. \ref{fig:side_car}[c]), the PDN on the IT tray must be upgraded to an LV-PDN (see Fig. \ref{fig:side_car}[d]) capable of interfacing with the in-rack LV bus system. Three different architectures are suitable for the next-generation LV-PDN (see the area in dotted box in Fig.\ref{fig:side_car}[d]), and the state-of-the-art DC--DC converter technologies with voltage step-down capability are reviewed in Section \ref{section:LV-IBC}. 
\\
\subsubsection{Shift 2: Toward In-Facility LV DC Distribution}

\begin{figure*}[!t]
    \centering
    \begin{subfigure}{0.49\textwidth}
        \centering
        \includegraphics[width=1\linewidth]{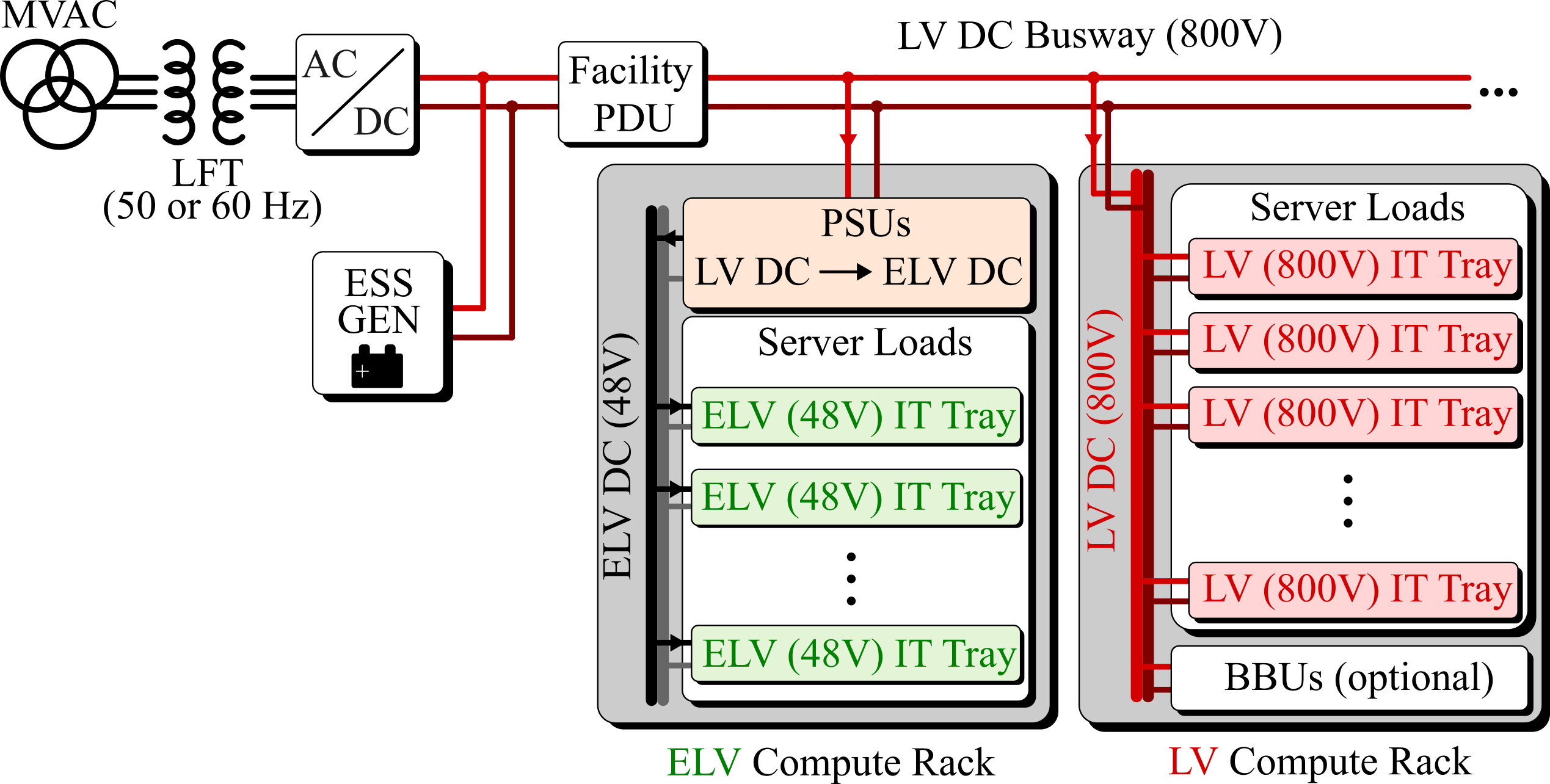}
        \caption{}
    \end{subfigure}
    \hfill
    \begin{subfigure}{0.49\textwidth}
        \centering
        \includegraphics[width=1\linewidth]{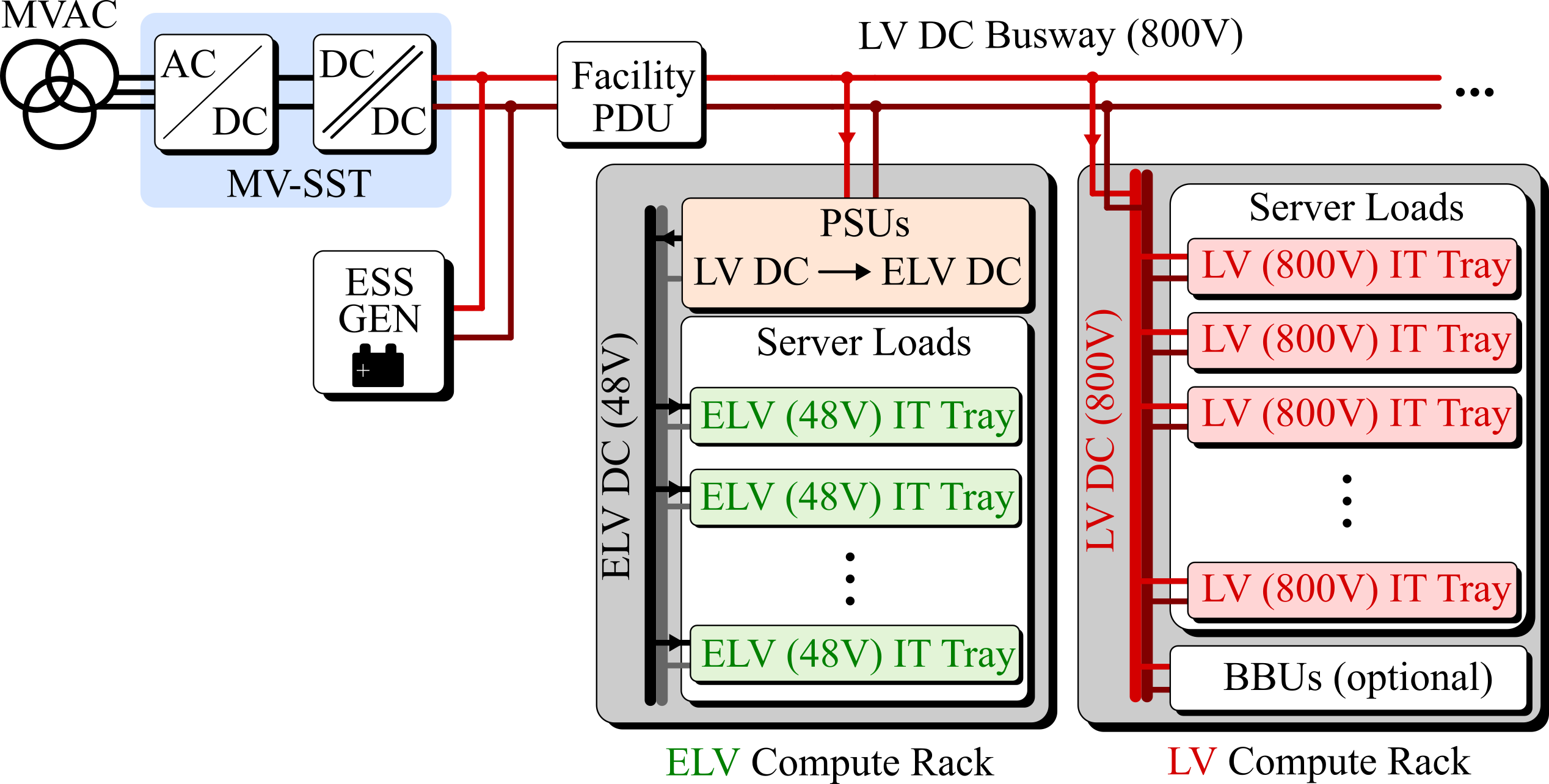}
        \caption{}
    \end{subfigure}      
    \caption{Shift 2 and Shift 3: Future LV DC in-facility distribution architectures for AI data centers and emerging direct connections to the MV AC grid through MV-SSTs. (a) Power delivery architecture employing an LV DC busway (e.g., 800~V). (b) Power delivery architecture directly connected to the MV-SST.}
    \label{fig:stage2_3}
\end{figure*}

The second shift is the movement from conventional LV AC distribution to the LV DC distribution busway system (e.g., 800 V DC, ±400 V DC, or ±750 V DC) that is directly compatible with the in-rack LV bus system in the LV compute racks \cite{NVIDIA800VDC, ocp_diablo400,2025_OCP_GOOGLE}, as shown in Fig. \ref{fig:stage2_3}(a). 

Compared with the conventional in-facility LV AC busway (see Fig.~\ref{fig:conv}), the in-facility LV \emph{DC} busway can be directly connected to the LV compute rack (see Fig.~\ref{fig:stage2_3}), eliminating the need for an additional power rack that is otherwise required to convert the in-facility LV AC voltage (e.g., 480~V AC) to LV DC. Instead of multiple distributed power racks, the in-facility LV DC distribution system employs a centralized AC--DC active front-end converter that converts the AC distribution voltage (e.g., 480~V AC) to LV DC, thereby forming the in-facility LV DC busway, as shown in Fig.~\ref{fig:stage2_3}. Furthermore, the AC--DC and DC--AC conversion stages in the conventional UPS system can be replaced by a battery energy storage system (BESS), as illustrated in Fig.~\ref{fig:stage2_3}. The DC distribution system is identified as the second technology building block that supports future AI data centers. Accordingly, Section~\ref{section:DC_Distribution} is devoted to this topic.\\

\subsubsection{Shift 3: Toward MV-Solid-State Transformers}

To further improve data center power density, increasing attention is being devoted to technologies that eliminate the LFT and directly interface the MV AC grid with the data center \cite{SL8}. The key enabling technologies include the recent development of 10--15~kV SiC MOSFETs with superior switching characteristics and solid-state transformers (SSTs). Eliminating the LFT can significantly reduce both the size and losses of the data center power distribution system. However, the reliability of SSTs must be improved before they can completely replace conventional LFTs. Therefore, the MV-SST is identified as the final technological building block necessary to power future AI data centers. In Section~\ref{section:MV-SST}, the topologies and design challenges related to MV-rated SSTs (MV-SSTs) are discussed.

This section establishes that the evolution of AI data centers can be understood as a sequence of three tightly coupled architectural shifts: (i) increasing the in-rack DC distribution voltage to enable megawatt-scale compute racks, (ii) transitioning from facility-level AC distribution to LV DC busways to reduce conversion stages and improve efficiency, and (iii) directly interfacing data centers with the MV grid through SST technology. Each architectural shift is enabled by a corresponding power-electronics technology building block. Accordingly, the following sections (Sections~\ref{section:LV-IBC}--\ref{section:MV-SST}) review these technology building blocks that support the three architectural shifts in AI data centers.

Section~\ref{section:LV-IBC} discusses DC--DC converters with high-voltage conversion ratios for LV-IBC applications in compute racks, which enable the first architectural shift toward higher in-rack DC distribution voltages. Section~\ref{section:DC_Distribution} reviews in-facility LV DC distribution systems as the enabling technology for the second architectural shift, with emphasis on system architectures, grounding schemes, and protection challenges. Section~\ref{section:MV-SST} examines MV-SSTs as the grid-interface technology that enables the third architectural shift toward direct MV grid integration. Together, these sections reflect a hierarchical progression from rack-level power conversion to facility- and grid-level integration, consistent with the architectural framework established in Section~\ref{section:3_stages}.

\section{LV-PDN Architectures and High Voltage Conversion Ratio DC--DC Converters}
\label{section:LV-IBC}

The first architectural shift identified in Section~\ref{section:3_stages} is the transition toward higher in-rack DC bus voltages to enable megawatt-scale compute racks. This shift in the in-rack bus voltage fundamentally alters the design requirements of the on-tray power distribution network (PDN) architecture and its associated power converters. 

\begin{figure}[!t]
     \centering
      \includegraphics[width=0.55\linewidth]{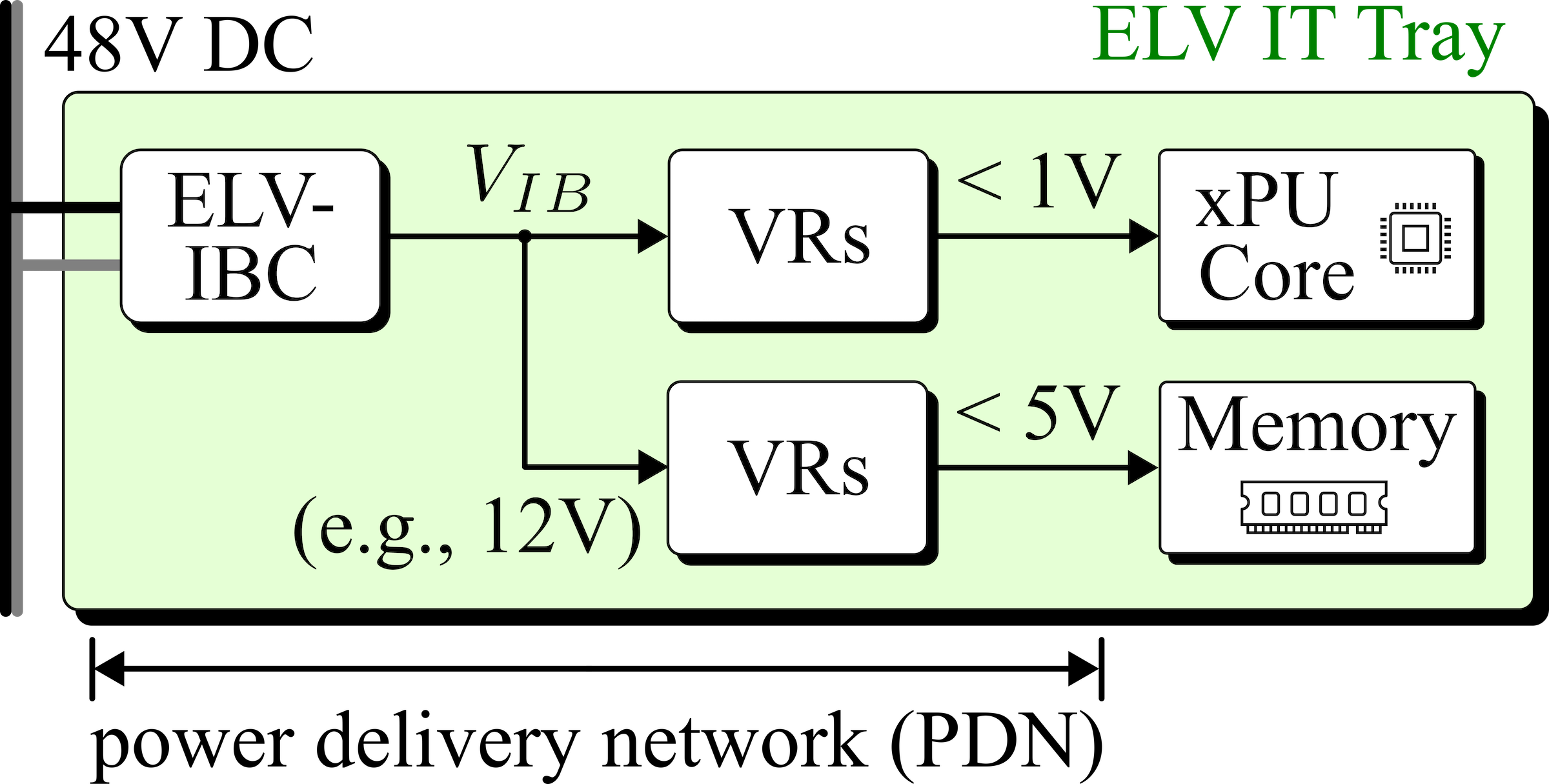}
     \caption{Typical two-stage on-tray PDN of a conventional ELV (48 V) IT tray.}
     \label{fig:ELV_PDN}
 \end{figure}

Designing power converters for the PDN has been a challenging and active research topic in the power electronics domain owing to the unique and stringent design constraints imposed on these converters. Because IT loads require hundreds to thousands of amperes at sub-1~V voltage levels, voltage conversion from the rack bus voltage to lower voltage levels must occur physically close to the IT loads. This requirement necessitates that PDN power converters be co-located in close physical proximity to the IT loads \cite{48V_IBC_power_density}. Furthermore, because IT tray loads such as xPU cores and memory devices are highly sensitive to voltage noise, voltage regulators (VRs) are placed close to the loads to enable tight voltage regulation \cite{Minjie_VR_review,IBC_VT_NVIDIA}. As a result, PDN power converters are embedded on the IT tray, as shown in Fig.~\ref{fig:conv}(b) and Fig.~\ref{fig:ELV_PDN} \cite{IBC_VT_NVIDIA}. However, this placement of PDN converters introduces the following unique design requirements \cite{Minjie_VR_review,STC_TPEL}:

\begin{itemize}
    \item High power density because of limited space on IT trays \cite{IBC_VT_NVIDIA}
    \item High efficiency \cite{Minjie_VR_review}
    \item Stringent electromagnetic compatibility (EMC) requirements owing to electromagnetic interference (EMI)-sensitive IT loads \cite{48V_IBC_power_density,CPSS_IBC}
    \item High transient dynamic performance because of highly dynamic AI training workloads \cite{Dynamic_AI_work_load,Minjie_VR_review,IBC_VT_NVIDIA}
\end{itemize}

The conventional PDN architecture is shown in Fig.~\ref{fig:ELV_PDN}. Although single-stage PDN architectures offer the potential for improved efficiency and power density, achieving the required high step-down voltage ratio demands sophisticated converter topologies and advanced control strategies. Consequently, two-stage architectures with an intermediate bus have been widely adopted owing to their greater implementation flexibility and enhanced transient performance \cite{need_buck_for_start_up,Minjie_VR_review,IBC_review_2014}. 

In the conventional PDN architecture (see Fig.~\ref{fig:ELV_PDN}), the ELV-IBC converter steps down the ELV DC bus voltage to an intermediate bus voltage $V_{IB}$, which is typically 12~V \cite{IBC_TI,IBC_VT_NVIDIA,Minjie_VR_review}. The ELV-IBC is unregulated and focuses primarily on efficient power conversion. Because the operation of the ELV-IBC typically does not involve active regulation, it is often referred to as a DC transformer \cite{Minjie_VR_review,DCX_1,DCX_2}. The second stage of the PDN is the VR, as shown in Fig.~\ref{fig:ELV_PDN}. In general, VRs are implemented as multiphase buck converters \cite{Minjie_VR_review,VRM_Multi_phase_buck,IBC_TI}. Because of the bulky inductors required in multiphase buck converters, VRs occupy the majority of the footprint on a GPU card, leaving extremely limited board area for the IBC \cite{IBC_VT_NVIDIA}. VRs step down $V_{IB}$ to lower voltage levels required by different loads within the IT tray (e.g., sub-1~V for xPU cores) \cite{IBC_TI,DCX_1,3239516}. It should be noted that multiple voltage levels are required within the IT tray, including on xPU boards \cite{IBC_TI,DCX_1,3239516,IBC_VT_NVIDIA}, which necessitates dedicated VRs for different loads. 

As data center architectures evolve toward higher in-rack bus voltages (see Fig.~\ref{fig:stage2_3}), next-generation PDN (i.e., LV-PDN) architectures and power converters that are native to the LV DC (e.g., 800~V) in-rack bus must be developed. These next-generation PDN architectures and converters must satisfy the aforementioned design requirements while achieving a significantly higher voltage conversion ratio. For example, in a conventional rack with a 48~V DC in-rack bus voltage, the PDN requires a voltage conversion ratio of 48 to achieve 1~V, which is relatively modest \cite{3422672}. In contrast, PDNs and power converters native to an 800~V in-rack bus voltage must achieve a voltage conversion ratio of 800 to step down to 1~V. This corresponds to an approximately 17-fold increase in the required voltage conversion ratio compared with that of conventional PDNs. Therefore, it is essential to review PDN architectures and power converter topologies that can address this unique challenge in the design of next-generation PDNs.

\begin{table*}[!t]
\centering
\caption{Comparison of design requirements between conventional ELV-PDN and LV-PDN}
\label{tab:PDN_comparison}
\renewcommand{\arraystretch}{1.25}
\begin{tabular}{lcc}
\hline
Type of PDN                 & Conventional ELV-PDN & LV-PDN            \\
\hline
In-rack bus voltage         & ELV DC (e.g., 48 V)  & LV DC (e.g., 800 V)                  \\
Power density               & high          & very high                     \\
EMI/EMC requirement         & moderate      & very high (higher voltage)    \\
Transient dynamics          & high          & very high (higher rack power) \\
Voltage conversion ratio    & moderate      & very high                          \\
Galvanic isolation          & optional      & preferred                      \\
\hline
\end{tabular}
\end{table*}

Table~\ref{tab:PDN_comparison} summarizes the key differences in design requirements between conventional ELV-PDNs used in ELV racks and LV-PDNs operating with LV in-rack DC buses. This comparison highlights that LV-PDNs are not a simple extension or scale-up of existing ELV-PDN designs; rather, they represent a fundamentally different architecture and power converter design. Section~\ref{section:LV-IBC}-\textit{A} reviews PDN architectures that can be employed for next-generation LV-PDNs native to the LV DC in-rack bus voltage. The subsequent sections discuss voltage step-down converter topologies that can serve as the building blocks of next-generation LV-PDNs. 

\subsection{Architectures for Next-Generation LV-PDNs}
\begin{figure}[!t]
     \centering
      \includegraphics[width=0.80\linewidth]{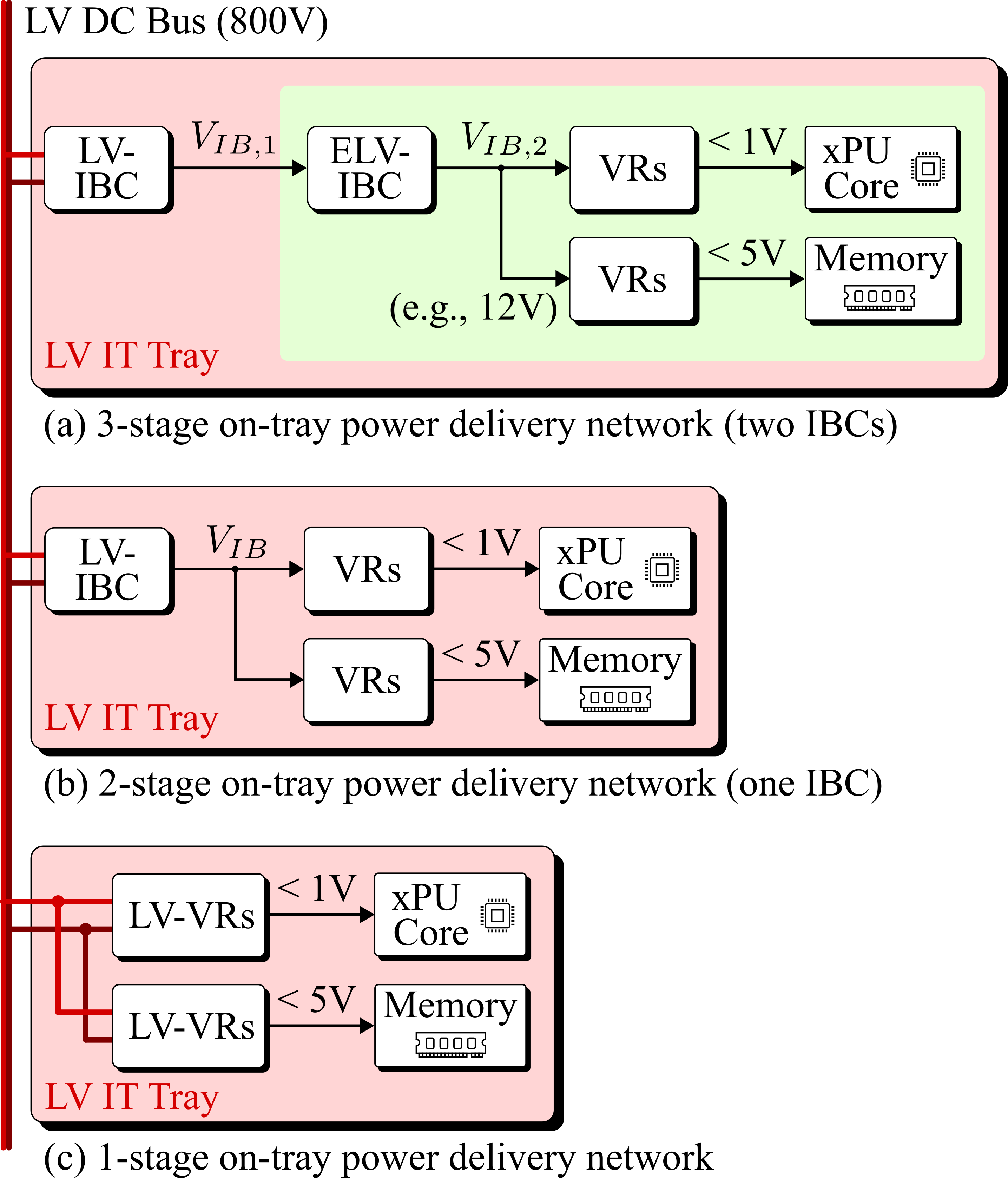}
     \caption{Different on-tray PDN architectures for LV IT trays. (a) Three-stage PDN with two IBCs. (b) Two-stage PDN with one IBC. (c) Single-stage PDN with one LV voltage regulator (LV-VR), which directly converts the LV DC bus voltage to IT load–compatible voltages (e.g., sub-1~V).}
     \label{fig:LV_PDN}
 \end{figure}

Three different architectures suitable for next-generation LV-PDNs are shown in Fig.~\ref{fig:LV_PDN}. The first option (see Fig.~\ref{fig:LV_PDN}[a]) is a three-stage LV-PDN architecture that builds upon the conventional ELV-PDN. The first stage is an LV-IBC that converts the in-rack LV DC bus voltage to an intermediate bus voltage $V_{IB,1}$, as shown in Fig.~\ref{fig:LV_PDN}(a). If $V_{IB,1}$ is set to 48~V, which is typically used in conventional ELV-PDNs, the remainder of the LV-PDN can be implemented using existing ELV-PDN designs. Specifically, the green shaded region in the LV-PDN of Fig.~\ref{fig:LV_PDN}(a) corresponds to the conventional PDN shown in Fig.~\ref{fig:ELV_PDN}. The primary advantage of this LV-PDN architecture is that it can leverage mature and widely available technologies by adding only one additional conversion stage (i.e., the LV-IBC). However, this architecture requires two IBC stages, which limits the achievable power density and efficiency of the LV-PDN—both of which are critical design metrics for PDNs.

The second LV-PDN architecture is the two-stage architecture, which topologically is the same as the conventional ELV-PDN architecture shown in Fig. \ref{fig:ELV_PDN}. The first stage, which is the LV-IBC, converts the LV DC in-rack voltage into a $V_{IB}$. VRs tightly regulate the output voltage needed for the IT loads. The advantage of the two-stage architecture is that the LV-IBC can be designed for high voltage conversion efficiency, and the VRs can be designed to fast control dynamics \cite{IBC_review_2014}. If $V_{IB}$ is chosen to be 12 V, which is the same as the $V_{IB}$ used in the conventional ELV-PDN (see Fig. \ref{fig:ELV_PDN}), the conventional VR technologies can be exploited. However, to maximize the performance of the two-stage LV-PDN architecture, the optimal $V_{IB}$ should be reconsidered. In fact, different $V_{IB}$ values have been considered in the literature, drifting away from the conventional 12 V $V_{IB}$ (24 V \cite{24V_V_IB}, 6 V \cite{6V_V_IB}, 4 V \cite{4V_V_IB}, 3.3 V \cite{3.3V_V_IB}, and 1.8 V \cite{IBC_VT_NVIDIA}). 

In general, increasing $V_{IB}$ reduces conduction losses between the IBC and VRs, making it possible to place the IBC farther away from the VRs. However, because of the higher input voltage applied to the VRs, switching losses in the VRs increase, which limits the maximum achievable switching frequency. This, in turn, degrades dynamic performance. In addition, a lower switching frequency increases the size of the filtering components, thereby limiting the power density of the VRs.

On the other hand, reducing $V_{IB}$ can significantly reduce switching losses in the VRs, enabling higher switching frequencies, smaller filter components, and higher control bandwidth \cite{IBC_VT_NVIDIA}. However, a lower $V_{IB}$ leads to higher conduction losses between the IBC and VRs, forcing the IBC to be placed closer to the VRs. Furthermore, a higher voltage conversion ratio is required for the IBC, making its design more challenging \cite{HVCR_review}. This challenge becomes particularly critical for future data centers as the in-rack bus voltage increases from ELV (e.g., 48~V) to LV (e.g., 800~V), thereby demanding an even higher voltage conversion ratio for the IBC. Consequently, determining the optimal $V_{IB}$ for two-stage LV-PDNs—considering its impact on overall LV-PDN performance—remains an open research problem.

The third LV-PDN architecture is the single-stage architecture, as shown in Fig.~\ref{fig:LV_PDN}(c). In this architecture, a single DC--DC converter, denoted as the LV voltage regulator (LV-VR) in Fig.~\ref{fig:LV_PDN}(c), must simultaneously achieve a high voltage conversion ratio and tight output voltage regulation \cite{IBC_MPEL}. Owing to the reduced number of conversion stages, this architecture can achieve the lowest component count. Moreover, because the LV DC bus voltage is stepped down to sub-1~V levels in close proximity to the IT load, the excessive conduction losses between the IBC and the load—present in multistage PDN architectures—are eliminated, thereby improving overall PDN efficiency \cite{IBC_MPEL}. One major VR supplier has demonstrated that direct 48~V-to-1~V conversion in conventional ELV-PDNs can save approximately \$500{,}000 per data center per year \cite{mouser_vr12_whitepaper,single_stage_PDN_1}.

However, employing a single-stage LV-PDN architecture introduces several challenges. Because of the extremely high voltage conversion ratio (e.g., from 800~V to sub-1~V), the LV-VRs must operate at extremely low duty ratios, which can lead to imbalanced loss distribution within the converter and reduced reliability \cite{single_stage_PDN_1}. In addition, switching devices with higher blocking voltages $V_{B}$ (i.e., rated voltages) tend to exhibit higher conduction losses \cite{single_stage_PDN_2} and higher switching losses \cite{Optimal_cell_number}. For example, \cite{Optimal_cell_number} shows that switching loss scales approximately as $P_{\text{sw}} \propto V_{B}^{3}$. This inherent trend in switching device performance limits the achievable switching frequency of LV-VRs, potentially degrading their dynamic performance. From the perspective of the overall IT tray, the absence of an intermediate bus in the PDN necessitates multiple LV-VRs to support individual loads, as shown in Fig.~\ref{fig:LV_PDN}. Given that converters employing higher-$V_{B}$ devices generally exhibit reduced performance \cite{single_stage_PDN_2,Optimal_cell_number}, integrating multiple LV-VRs on a space-constrained IT tray poses significant challenges.

Beyond the choice of PDN architecture, another critical consideration in selecting an optimal PDN for next-generation LV IT trays is the need for galvanic isolation and its practical implementation. IT trays must support hot-swapping, meaning that trays can be inserted or removed without shutting down rack-level power (i.e., the rack bus remains \emph{hot}). In conventional ELV-PDNs, the typical 48~V bus voltage is considered ELV (i.e., touch-safe), and therefore galvanic isolation between the rack bus and the IT tray is not required. This allows the use of nonisolated converter solutions in ELV-PDNs \cite{3239516,STC_TPEL,STC_2,STC_cascaded_voltage_divider_JESTPE,sigma_converter_LLC}. However, when the bus voltage increases from ELV (e.g., 48~V) to LV (e.g., 800~V), which is not touch-safe, the requirement for galvanic isolation in the PDN and its converters must be considered when selecting an LV-PDN architecture. Currently, no dedicated standards explicitly address galvanic isolation requirements for LV IT trays and power converters in LV-PDNs. Moreover, if galvanic isolation is required, its placement within multistage PDN architectures (e.g., Fig.~\ref{fig:LV_PDN}) must also be carefully considered.

This subsection reviewed three PDN architectures suitable for LV IT trays, along with their respective advantages and challenges. In addition, the need for galvanic isolation and its practical implementation in next-generation LV-PDNs was identified as an important future research direction. The following three subsections review converter topologies reported in the state-of-the-art literature that feature high voltage step-down capability.

\subsection{Class of Buck-Derived Converter Topologies}
The buck converter is the simplest topology that can be used to step down the input voltage. Its high switching frequency capability and pulse-width modulation (PWM) operation enable high control bandwidth, making it suitable as the final stage of the on-tray PDN, which requires high dynamic performance \cite{Minjie_VR_review}. However, when a high voltage conversion ratio is required, a conventional buck converter suffers from an extremely low duty ratio, leading to high output current ripple and low efficiency \cite{Coupled_Ind_2024}. As a result, numerous approaches have been proposed in the literature to develop buck converters with high voltage conversion ratios.\\ 

\subsubsection{Buck Converter with Tapped or Coupled Inductor} 
The first approach to achieving a high voltage conversion ratio in buck converters is to replace the output inductor with a tapped or coupled inductor \cite{Tapped_Ind_2005,Tapped_Ind_2014,Coupled_Ind_2012,Coupled_Ind_2024}. 

\begin{figure}[!t]
    \centering
    \begin{subfigure}{0.48\textwidth}
        \centering
        \includegraphics[width=0.46\linewidth]{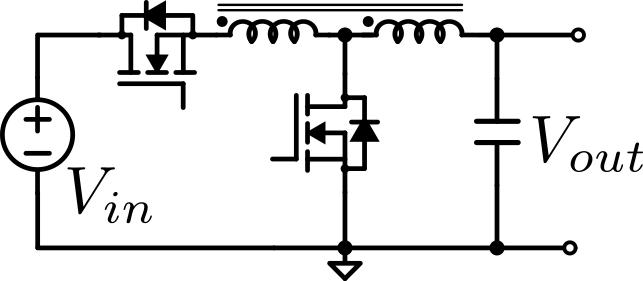}
        \caption{}
    \end{subfigure}
    \hfill
    \begin{subfigure}{0.48\textwidth}
        \centering
        \includegraphics[width=0.6\linewidth]{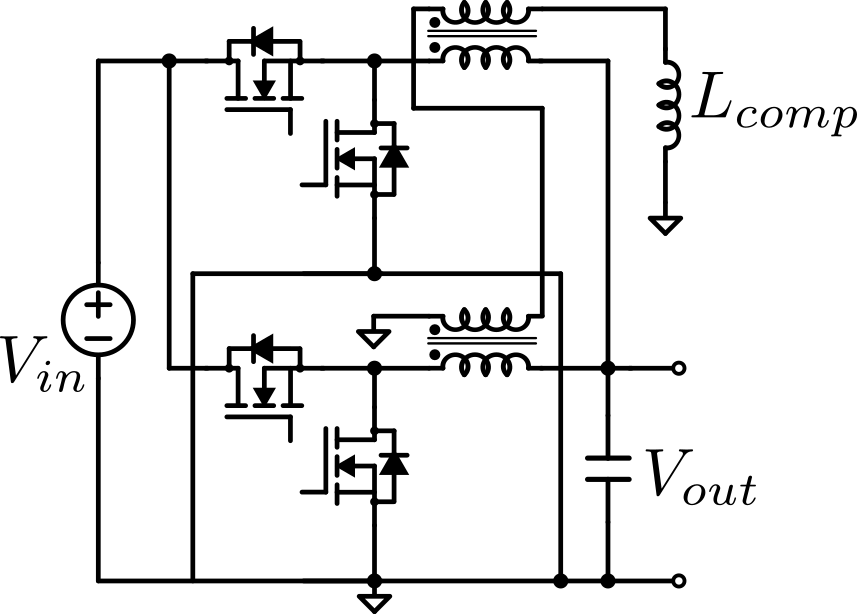}
        \caption{}
    \end{subfigure} 
    \hfill
    \begin{subfigure}{0.48\textwidth}
        \centering
        \includegraphics[width=0.55\linewidth]{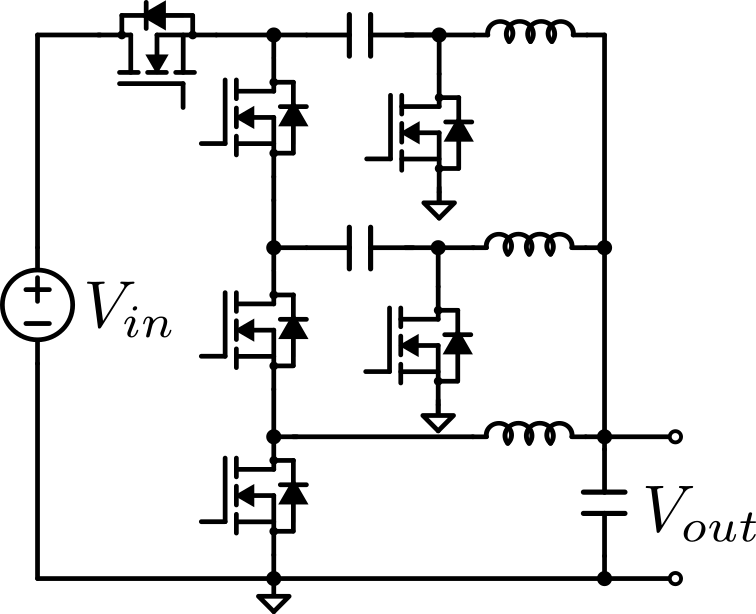}
        \caption{}
    \end{subfigure}
    \caption{Three different types of buck-derived converter topologies. (a) Buck converter employing tapped or coupled inductor. (b) Trans-inductor voltage regulator with fast dynamics. (c) Series capacitor buck converter.}
    \label{fig:buck_derived}
\end{figure}

The turns ratio of the tapped or coupled inductor enables a significantly higher voltage conversion ratio compared with that of a conventional buck converter. However, this class of buck converters suffers from large voltage spikes, necessitating the use of clamping circuits. Furthermore, these topologies inherently introduce a right-half-plane zero in the converter dynamics (i.e., non-minimum-phase behavior), which makes stability analysis and control design more challenging \cite{Tapped_Ind_challenge_1,Tapped_Ind_challenge_2,series_cap_buck_TI}.

In addition to achieving a high voltage conversion ratio, DC--DC converters in on-tray PDNs must exhibit excellent dynamic performance owing to the abrupt load transients typically observed in xPUs \cite{Minjie_VR_review}. Although buck or buck-derived converters can provide acceptable dynamic performance through PWM operation at high switching frequencies, the intrinsic dynamic response of buck-derived converters is fundamentally limited by the output inductor \cite{Minjie_VR_review}. Reducing the output inductance enables faster responses to load transients; however, it also results in large peak-to-peak current ripple, leading to increased switching losses, AC conduction losses, and magnetic core losses \cite{Minjie_VR_review}. 

To overcome this inherent tradeoff without introducing additional control complexity, a topology derived from the multiphase buck converter with coupled inductors has been proposed \cite{TLVR_1,TLVR_2,TLVR_3,TLVR_4}. Because this topology is commonly used as a voltage regulator, it is often referred to as the trans-inductor voltage regulator (TLVR). In this topology, the effective transient inductance is dynamically reduced during load steps, enabling faster dynamic response. The additional compensation inductance ($L_{\mathrm{comp}}$) actively induces current redistribution among phases, significantly enhancing scalability and manufacturability for industrial implementation \cite{Minjie_VR_review}. As a result, TLVR breaks the long-standing theoretical bandwidth limitation and achieves superior transient performance without compromising steady-state characteristics. 
\\

\subsubsection{Series Capacitor Buck Converter}
The second approach to achieving a high voltage conversion ratio in buck converters is to introduce a series capacitor. This class of buck-derived topologies is commonly referred to as the series capacitor buck converter \cite{series_cap_buck_2005,series_cap_buck_2006,series_cap_buck_2007,10263626}. The pioneering work was presented in \cite{series_cap_buck_2005}, originally termed the double step-down buck, and its extension to multiphase operation was discussed in \cite{series_cap_buck_2007}. The series capacitor acts as a DC voltage source, reducing the voltage stress on the switching devices and effectively increasing the achievable voltage conversion ratio \cite{series_cap_buck_2005,series_cap_buck_TI}. The reduced voltage stress on the switching devices leads to lower switching losses and improved efficiency. Because the inductor behaves as a current source, the series capacitor can be charged without current spikes, a feature commonly referred to as soft charging \cite{series_cap_buck_2005,series_cap_buck_TI}. Furthermore, \cite{series_cap_buck_TI} reports that the series capacitor buck converter exhibits reduced inductor current ripple and inherent current sharing among phases.

More recent series capacitor buck topologies tend to combine coupled inductors with series capacitors. This hybrid approach integrates the advantages of both tapped- and coupled-inductor buck converters and series capacitor buck converters, enabling very high voltage conversion ratios, reduced voltage stress, and inherent current balancing among phases \cite{SC+CL_1,SC+CL_2,SC+CL_3,SC+CL_4,SC+CL_6}. By jointly exploiting magnetic and capacitive components, soft switching can also be achieved \cite{SC+CL_1}. In addition, a hybrid topology combining the series capacitor buck converter with the TLVR concept has been reported in \cite{series_cap_TLVR}, achieving both a high voltage conversion ratio and fast dynamic response.

\subsection{Switched-Capacitor Network-Based Hybrid Converter}

Switched-capacitor (SC) networks can achieve high voltage conversion ratios and very high power density compared with those of magnetic-component-based solutions because the power density of capacitors is significantly higher than that of inductors \cite{Cap_vs_Ind_Pilawa}. Magnetic components generally dominate the size of a power converter, and thus eliminating or minimizing magnetic components offers substantial potential for power density improvement. However, pure SC networks typically support only fixed voltage conversion ratios \cite{Cap_vs_Ind_Pilawa}, which limits their output voltage regulation capability. Furthermore, due to the parallel connection of multiple capacitors at different voltage levels, charge redistribution occurs during capacitor charging and discharging. The losses associated with this charge redistribution within SC networks are referred to as charge-sharing losses \cite{Charge_sharing_losses_2025}. To mitigate charge-sharing losses, capacitors must be oversized or the switching frequency must be increased, both of which lead to suboptimal power conversion performance. As a result, small magnetic components are commonly introduced into SC networks, giving rise to hybrid SC converters.

The inclusion of small magnetic components in hybrid SC converters enables soft charging and improved voltage regulation \cite{Pilawa_Merged_SC_2008,Pilawa_Merged_SC_2012}. Owing to their high voltage conversion ratio and very high power density, hybrid SC converters have been extensively investigated for on-tray PDN applications. The following sections review three types of hybrid SC converters reported in the literature that are suitable for PDN applications (e.g., IBCs and VRs in Fig.~\ref{fig:conv}[b]).\\

\subsubsection{Cascaded Converter}
\begin{figure}[!t]
    \centering
    \begin{subfigure}{0.48\textwidth}
        \centering
        \includegraphics[width=0.6\linewidth]{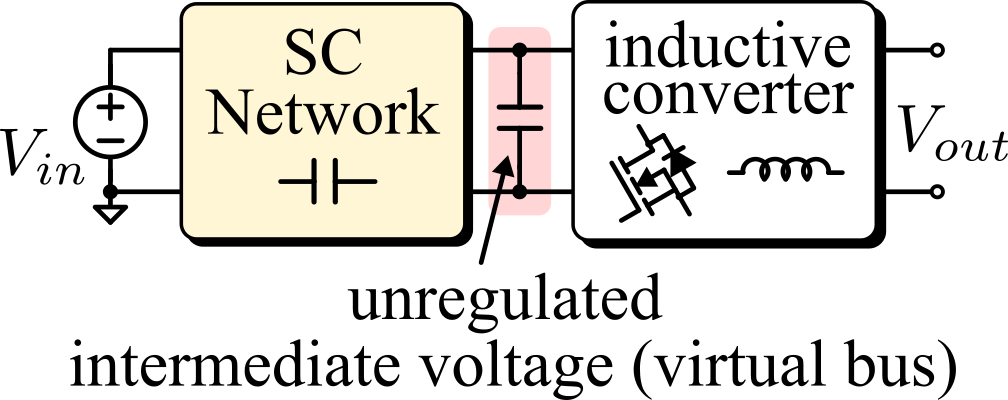}
        \caption{}
    \end{subfigure}
    \hfill
    \begin{subfigure}{0.48\textwidth}
        \centering
        \includegraphics[width=0.45\linewidth]{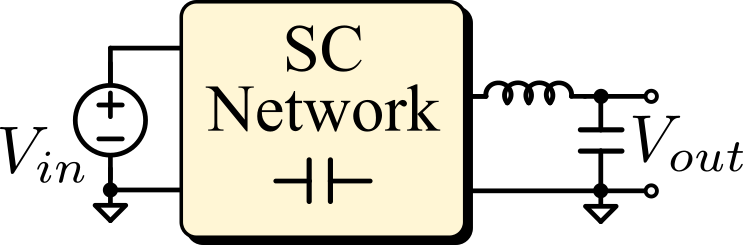}
        \caption{}
    \end{subfigure} 
    \hfill
    \begin{subfigure}{0.48\textwidth}
        \centering
        \includegraphics[width=0.4\linewidth]{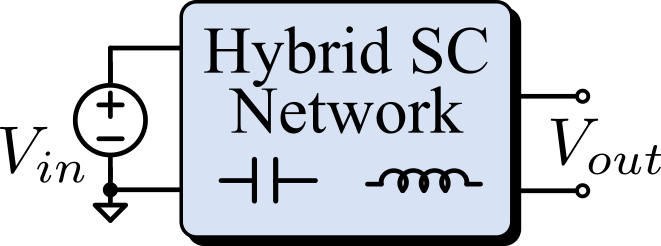}
        \caption{}
    \end{subfigure}
    \caption{Three types of hybrid SC converters. (a) Merged (or cascaded) with an inductive converter. This hybrid SC-converter topology typically has an unregulated IBC. (b) SC network with an input/output inductor. (c) Hybrid SC network with \textit{LC} resonant tanks.}
    \label{fig:hybrid_switched_cap_converters}
\end{figure}

The first approach to develop a hybrid SC converter is to merge the SC network with an inductive converter (e.g., buck converter), as shown in Fig. \ref{fig:hybrid_switched_cap_converters}(a). The first stage is the SC network that achieves high voltage step-down. Owing to the fixed voltage conversion ratio of the SC network, the first stage outputs are typically unregulated, forming an unregulated intermediate bus voltage \cite{Pilawa_Merged_SC_2008,Pilawa_Merged_SC_2012}, as shown in the red shaded area in Fig. \ref{fig:hybrid_switched_cap_converters}(a). The buck converter or other types of magnetic-component-based converters receives the unregulated intermediate bus voltage and converts it into tightly regulated voltage output.\\

\subsubsection{Direct Connection to Inductor or Inductive Load}
Although cascading an SC network with inductive converters (e.g., buck-derived converters) can achieve high voltage conversion ratios, soft charging of the capacitors, and tight output voltage regulation, this type of topology is equivalent to two-stage architecture, leading to additional losses and suboptimal power density. Instead of cascading the SC network with an inductive converter, it is possible to directly connect magnetic components at the input or output \cite{SC_single_inductor_resonant_proof,8867895,SC_output_inductor_1,SC_output_inductor_2} (or both \cite{SC_input_output_inductor_1,SC_input_output_inductor_2}) of the SC network, as shown in Fig. \ref{fig:hybrid_switched_cap_converters}(b) \cite{SC_single_inductor_resonant_proof,8867895}. Even with a single inductor, the hybrid SC network can have the soft-charging and resonant (i.e., soft switching) feature that can improve the efficiency of the converter \cite{SC_single_inductor_resonant_proof}. However, because the inductor should conduct the full load current, the thermal design and volume reduction of the inductor are challenging \cite{STC_cascaded_voltage_divider_JESTPE,HVCR_review}. Furthermore, owing to the centralized nature of the inductor in these topologies, the soft charging of all capacitors in the SC network is challenging, limiting the scalability of this type of topology \cite{STC_TPEL}. Note that the inductor in this type of topology can be replaced with a transformer \cite{SC+Transformer} or an inductive load \cite{resonant_FC_2015}.\\

\subsubsection{Hybrid Switched-Capacitor Network with Distributed Inductors}
The other type of hybrid SC converter topology distributes the small inductors in the SC network \cite{Distributed_SC_PFZ,Distributed_SC_1, Distributed_SC_2,Distributed_SC_3}, forming a hybrid SC network, as shown in Fig. \ref{fig:hybrid_switched_cap_converters}(c). The distributed nature of the inductor makes soft charging easier to achieve for all capacitors in the converter, improving the scalability of the topology \cite{STC_TPEL}. However, the performance of classical hybrid SC converters is sensitive to component nonidealities and has high voltage stress \cite{HVCR_review}. A more advanced topology called the \textit{switch tank converter} is proposed \cite{STC_ECCE,STC_TPEL} and optimized \cite{need_buck_for_start_up,STC_2}. The switch tank converter combines two building blocks (i.e., a clamping capacitor block and resonant tank block) to scale up the topology, ensuring the scalability of the topology \cite{STC_TPEL}. Also due to the clamping capacitor block in this topology, the voltage stress in the switching devices can be minimized \cite{STC_TPEL}. One challenge with this topology is that it cannot self-start \cite{need_buck_for_start_up}. Without a proper start-up method, the devices in the switch tank converter can undergo high voltage and current stress. One potential solution is to use an additional buck converter as the first stage of the switch tank converter \cite{need_buck_for_start_up}, at the cost of an increased number of power conversion stages.\\

\subsection{Isolation Transformer-Based Converters}

In the conventional rack with an ELV (48 V) bus, the ELV is considered to be touch-safe, and galvanic isolation is not required. However, transformer-based isolated DC--DC converters are a widely accepted solution in the industry sector for high voltage conversion ratios. Typically, active clamp forward converters as well as half- and full-bridge-based isolated DC--DC converters are employed as the DC--DC converter on the on-tray PDN in data center applications \cite{Minjie_VR_review, STC_TPEL,Isolated_DCDC_review_TIA_data_center_TIA}. \\

\subsubsection{Nonresonant Converters}
\begin{figure}[!t]
    \centering
    \begin{subfigure}{0.48\textwidth}
        \centering
        \includegraphics[width=0.5\linewidth]{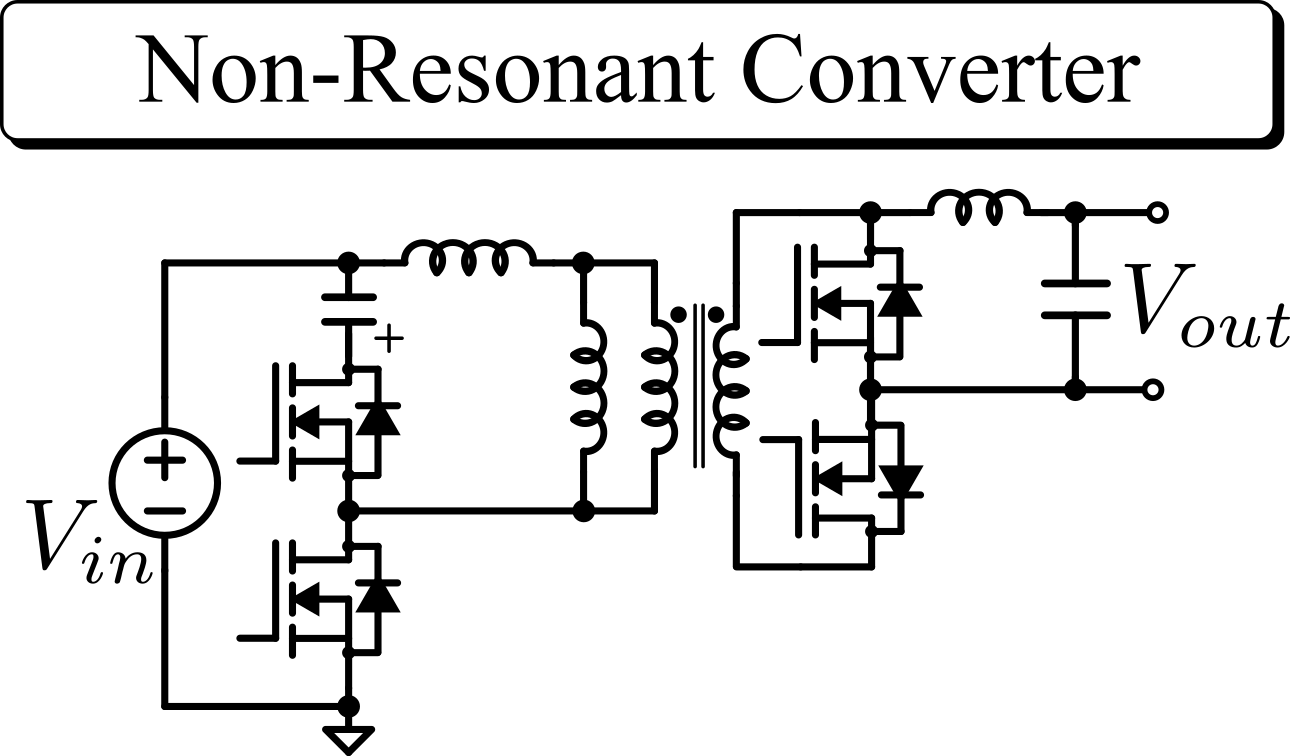}
        \caption{}
    \end{subfigure}
    \hfill
    \begin{subfigure}{0.48\textwidth}
        \centering
        \includegraphics[width=0.4\linewidth]{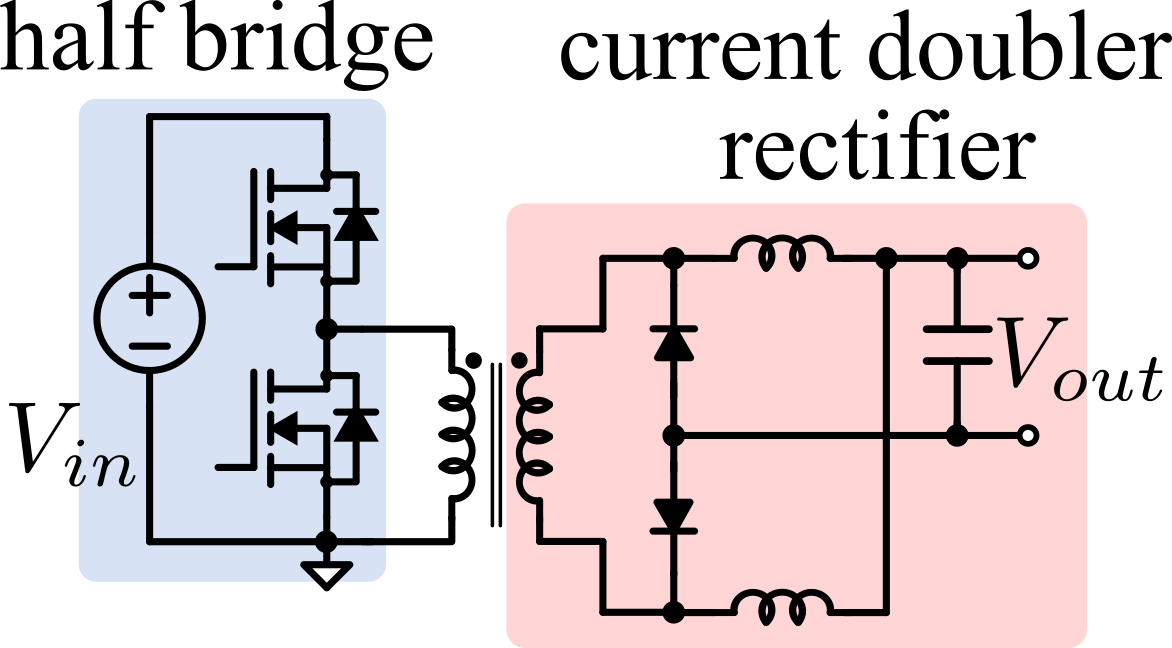}
        \caption{}
    \end{subfigure} 
    \hfill
    \begin{subfigure}{0.48\textwidth}
        \centering
        \includegraphics[width=0.45\linewidth]{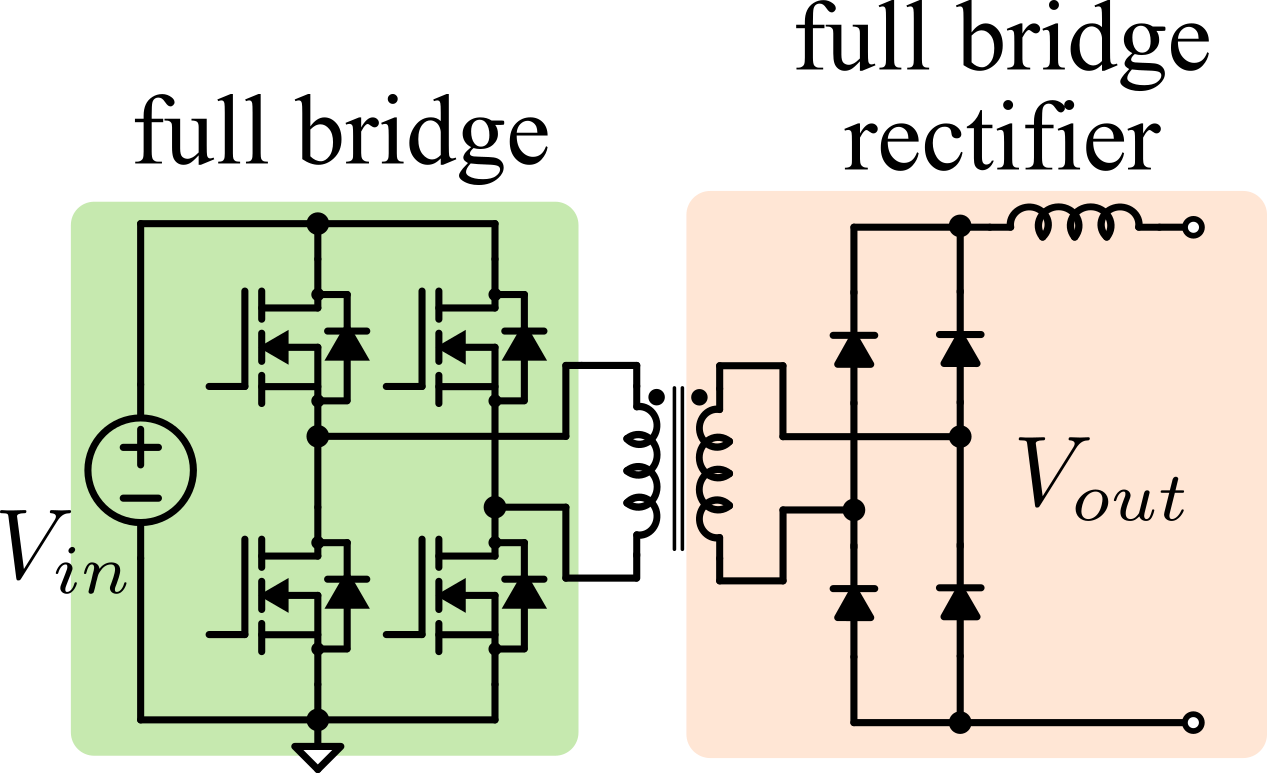}
        \caption{}
    \end{subfigure}
    \caption{Different types of nonresonant isolated converters. (a) Active clamp forward converter. (b) Primary: half-bridge converter (blue area) and secondary: current doubler rectifier (red area). (c) Primary: full-bridge converter (green area) and secondary: full-bridge rectifier (orange area).}
    \label{fig:non_resonant_isolated_DC_DC_conv}
\end{figure}

The simplest active clamp forward converter is shown in Fig. \ref{fig:non_resonant_isolated_DC_DC_conv}(a). The active clamp forward converter and its derivative topologies have been preferred in the industry because of the topology's simple structure, wide duty-cycle operation, and zero-voltage switching (ZVS) capability. Also, the active clamp forward converter topology has low secondary-side conduction losses owing to the presence of an output inductor, which is critical for converters with high voltage conversion ratios \cite{APM_review_2023,800V_14V_good_review}. However, its primary switches experience high voltage stress equal to the sum of the input voltage and the clamp capacitor voltage. Also, the active clamp forward converters suffer from a large DC offset current in the transformer, increasing magnetic core loss and potentially transformer size \cite{ACFW_DC_offset_current}. To address these limitations, improved active clamp forward converter topologies with auxiliary switches and innovative transformer designs have been discussed \cite{Modified_ACFW,ACFW_DC_offset_current}.

The family of half- and full-bridge-based isolated DC--DC converters are shown in Figs. \ref{fig:non_resonant_isolated_DC_DC_conv}(b) and (c). As well summarized in \cite{APM_review_2015, APM_review_2023}, the majority of isolated DC--DC converters with high voltage conversion ratios can be designed using the half- or full-bridge converter connected to the primary side of the transformer. The secondary side of the transformer is typically a current doubler rectifier (see the red shaded area in Fig. \ref{fig:non_resonant_isolated_DC_DC_conv}[b]), full-bridge rectifier (see the orange shaded area in Fig. \ref{fig:non_resonant_isolated_DC_DC_conv}[c]), center-tapped rectifier (see the yellow shaded area in Fig. \ref{fig:resonant_DC_DC_conv}[a]), or more advanced topologies derived from these two rectifiers \cite{APM_review_2015, APM_review_2023}. The diodes in the secondary side rectifier can be replaced with active devices to improve the performance, as shown in the center-tapped rectifier case in Fig. \ref{fig:resonant_DC_DC_conv}(a) \cite{FB_active_secondary}. The topology in Fig. \ref{fig:non_resonant_isolated_DC_DC_conv}(c) is also referred to as the phase-shift full-bridge converter \cite{classic_PSFB}. This topology is known for the zero transformer DC offset current and low voltage stress on the switches \cite{800V_14V_good_review}. The circuit topologies shown in Fig. \ref{fig:non_resonant_isolated_DC_DC_conv} do not operate under resonance and are classified as nonresonant converters. As a result, this class of converters suffers from limited soft-switching range.\\

\subsubsection{Resonant Converters}
\begin{figure}[!t]
    \centering
    \begin{subfigure}{0.48\textwidth}
        \centering
        \includegraphics[width=0.5\linewidth]{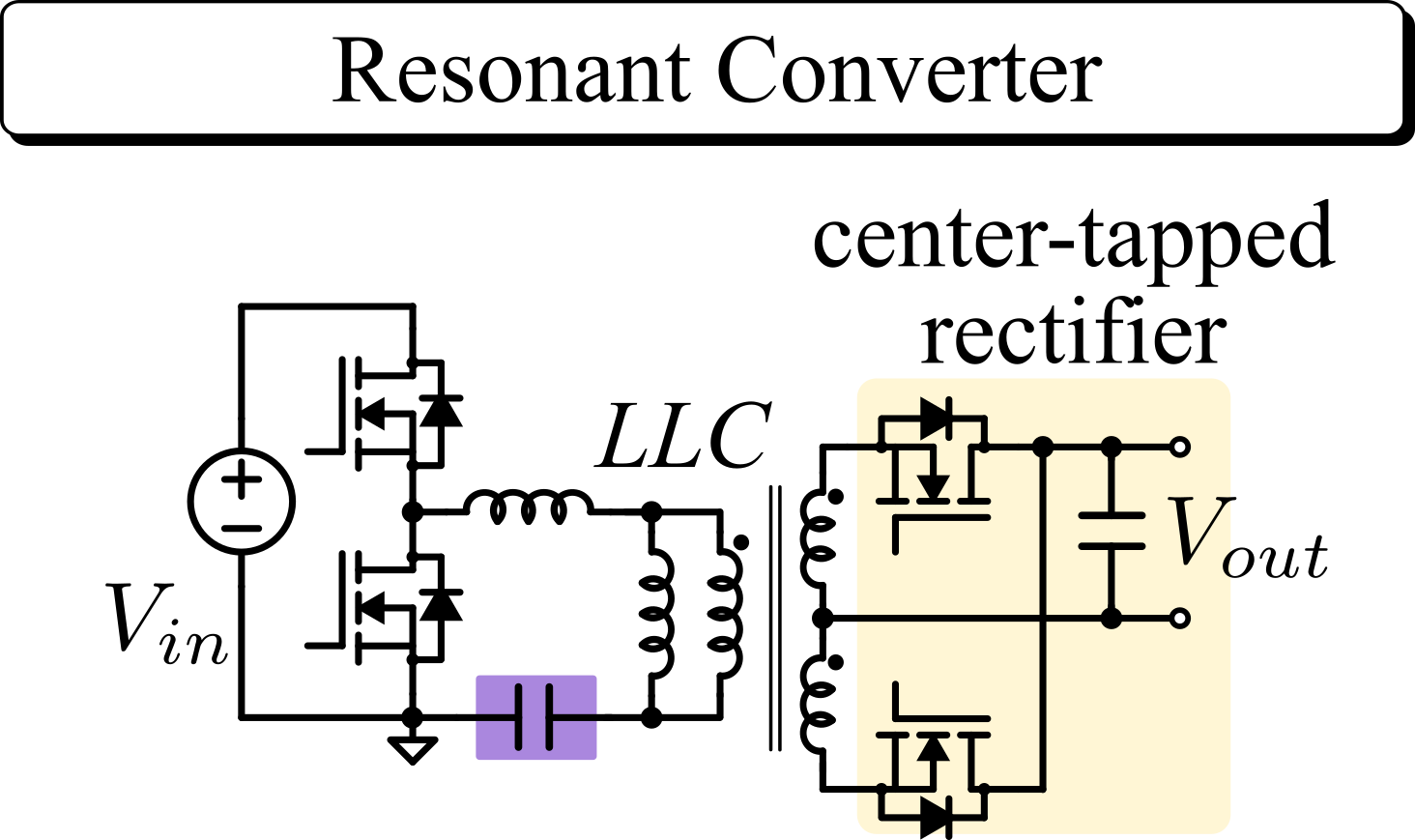}
        \caption{}
    \end{subfigure}
    \hfill
    \begin{subfigure}{0.48\textwidth}
        \centering
        \includegraphics[width=0.52\linewidth]{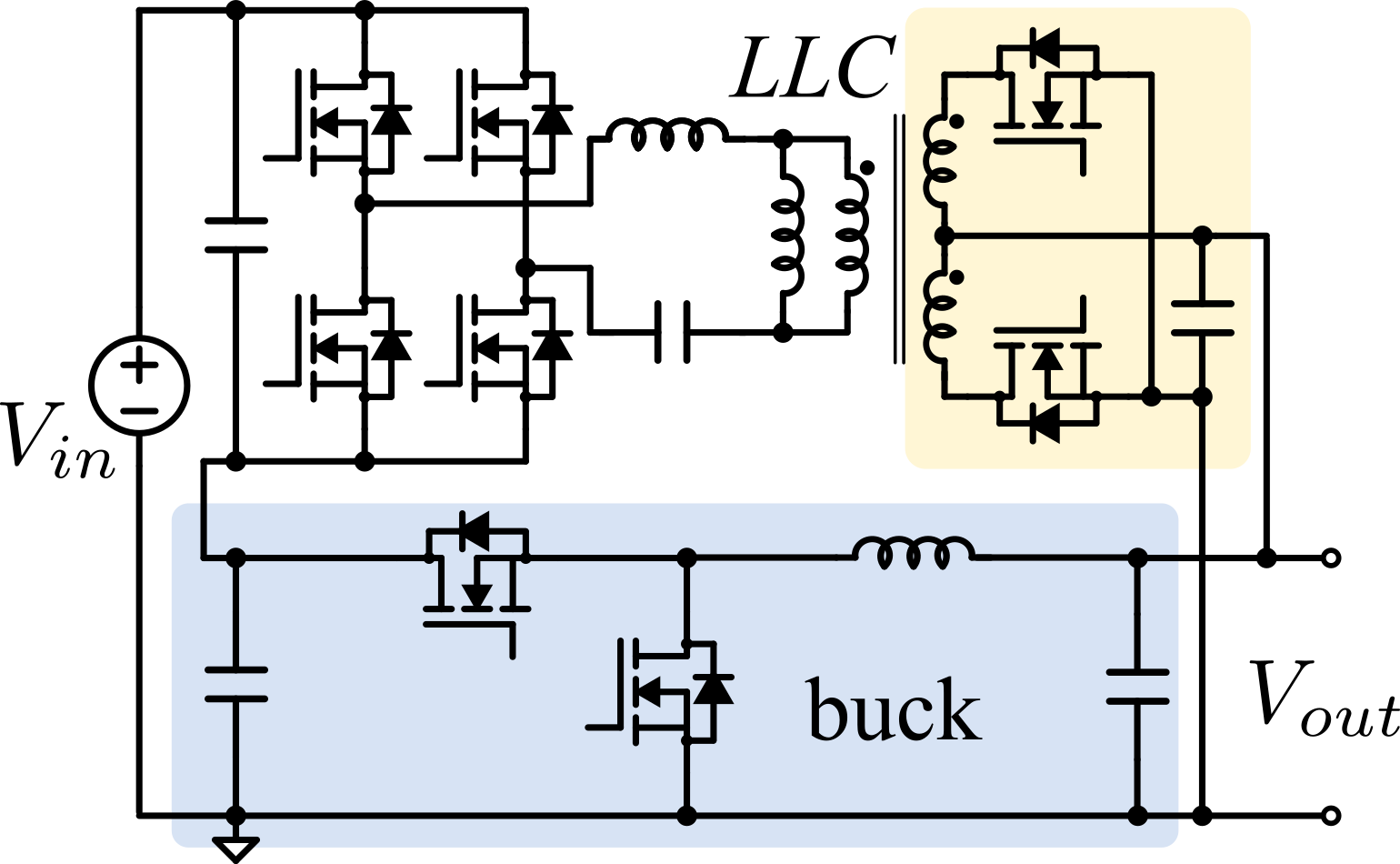}
        \caption{}
    \end{subfigure} 
     \caption{Load resonant DC--DC converters. (a) \textit{LLC} converter. (b) Sigma converter where the \textit{LLC} converter (yellow area) and buck converter (blue area) are connected in an input-series-output-parallel configuration \cite{sigma_converter_LLC}.}
    \label{fig:resonant_DC_DC_conv}
\end{figure}

A resonant converter is a converter with an \textit{LC} resonant tank, enabling a wide soft-switching range \cite{LLC_1,InfineonLLC2012}. One of the most popular resonant converters is the \textit{LLC} converter shown in Fig. \ref{fig:resonant_DC_DC_conv}(a) \cite{InfineonLLC2012}. The resonant capacitor (see the purple shaded area in Fig. \ref{fig:resonant_DC_DC_conv}[a]) is connected in series to the isolation transformer, causing a series resonance with the leakage inductance of the transformer. The family of \textit{LLC} converters is attractive because of its inherent ZVS capability for all primary switches, zero DC offset current in the transformer, and low voltage stress on the clamped input voltage \cite{800V_14V_good_review}. However, this topology is sensitive to input voltage fluctuations, and the converter is forced to operate in the suboptimal state when the input voltage deviates significantly \cite{InfineonLLC2012}. As a result, a major limitation of the \textit{LLC} converter is its poor output voltage regulation (sensitive to input voltage), which can make it unsuitable for IT loads in data center applications. A potential solution is to add a buck converter with high switching frequency, as shown in Fig. \ref{fig:resonant_DC_DC_conv}(b). In this circuit, named as sigma converter \cite{sigma_converter_LLC}, the \textit{LLC} converter (see the yellow shaded area in Fig. \ref{fig:resonant_DC_DC_conv}[b]) processes the majority of the power, and the buck converter (see the blue shaded area in Fig. \ref{fig:resonant_DC_DC_conv}[b]) regulates the output voltage $V_{out}$.

Although the \textit{LLC} converter is one of the most popular converters with an isolation transformer, it is challenging to employ for applications with higher voltages. When the resonant capacitor (see the purple shaded area in Fig. \ref{fig:resonant_DC_DC_conv}[a]) undergoes resonance, it can experience voltage that is multiple times higher than the input voltage $V_{in}$. As the input voltage $V_{in}$ scales up, the voltage rating of the resonant capacitor should also increase, leading to limited component options and increased cost. Also, capacitors are one of the weakest points in power electronics \cite{Cap_weak_motivate_DAB}, leading to reliability concerns in the family of resonant converters. As an alternative, the family of dual active bridge (DAB) converters, first proposed by De Doncker \cite{DAB_1988}, is gaining more attention for high-voltage and high-power applications \cite{HVCR_DAB_400VDC_Kolar,HVCR_DAB,DAB_NPC_half_bridge,DAB_Transformer_ISOP,DAB_Converter_ISOP_2016}. \\

\subsubsection{Dual Active Bridge Converter}
\begin{figure}[!t]
    \centering
    \begin{subfigure}{0.49\textwidth}
        \centering
        \includegraphics[width=0.55\linewidth]{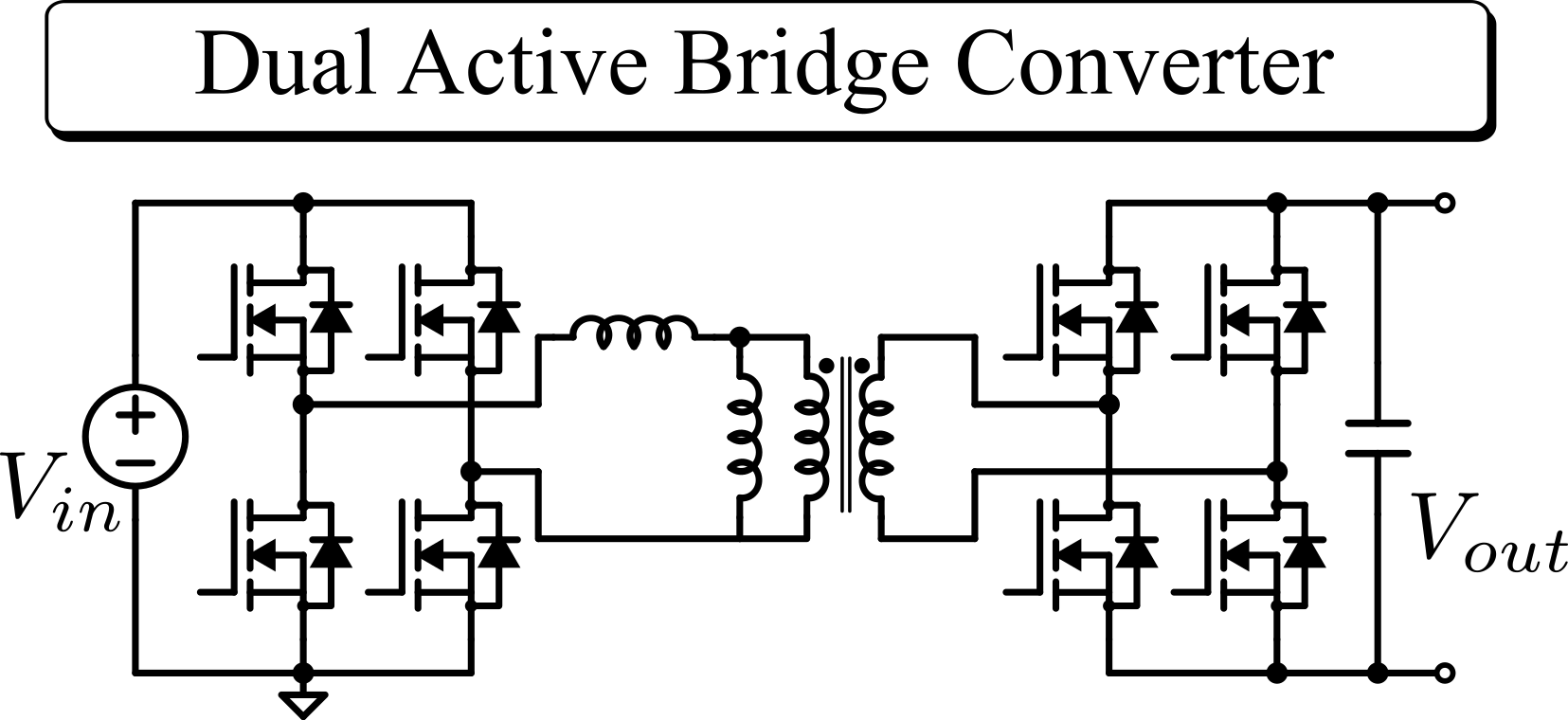}
        \caption{}
    \end{subfigure}
    \hfill
    \begin{subfigure}{0.49\textwidth}
        \centering
        \includegraphics[width=0.57\linewidth]{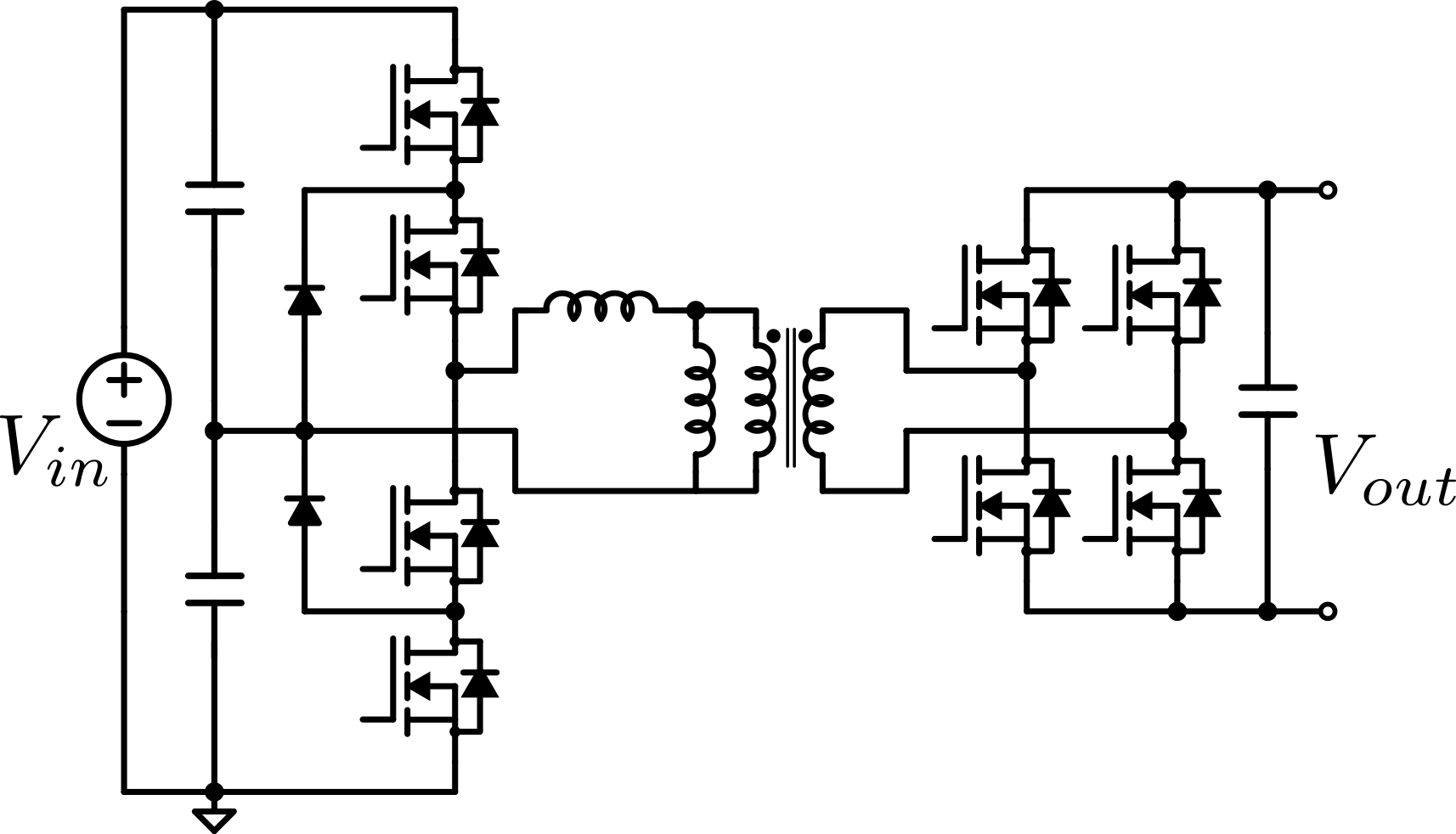}
        \caption{}
    \end{subfigure} 
     \caption{DAB converters. (a) Schematic of a DAB converter. (b) DAB converter with multilevel primary side converter (i.e., half-bridge neutral-point clamped converter) \cite{DAB_NPC_half_bridge}.}
    \label{fig:DAB}
\end{figure}

The schematic of a typical DAB converter is shown in Fig. \ref{fig:DAB}(a). Because a DAB converter does not operate under resonance, the voltage and current stress on the components are relatively low \cite{DAB_low_RMS_and_circulating_current}. This feature makes the DAB converter attractive for applications requiring high power \cite{compare_LLC_DAB}. DAB converters also have superior controllability at fixed frequencies because the main control variable is the phase-shift angle \cite{DAB_control_review}. The DAB converter allows bidirectional power flow. When the primary side voltage is very high, multilevel topologies, such as the neutral-point clamped converter (NPC) shown in Fig. \ref{fig:DAB}(b), can be used. However, DAB converters tend to lose ZVS under light load conditions, leading to lower efficiency \cite{compare_LLC_DAB}. The limited soft-switching range under low load conditions can become an issue for data center applications expecting dynamic loads \cite{Dynamic_AI_work_load}. DAB converters suffer from inherent circulating current (or reactive power), causing large peak current and additional losses \cite{DAB_Mi}. To mitigate this issue, more advanced control approaches, such as the dual phase-shift control, should be used \cite{DAB_Mi,DAB_control_review}.\\

\subsubsection{Transformer-Enabled Converter Configurations}

One interesting feature of transformer-based converter topologies is their flexibility of configurations and modularity, which is extremely advantageous for converters with high voltage conversion ratios. In high voltage step-down applications, the output current is extremely high, motivating the use of rectifier topologies that reduce the current stress (e.g., current doubler rectifier in Fig. \ref{fig:non_resonant_isolated_DC_DC_conv}[b] and center-tapped rectifier in Fig. \ref{fig:resonant_DC_DC_conv}[a]), or connect the output of the converter in parllel on the transformer's secondary side. The parallel connection reduces the high current stress on each converter. 

\begin{figure}[!t]
    \centering
    \begin{subfigure}{0.48\textwidth}
        \centering
        \includegraphics[width=0.5\linewidth]{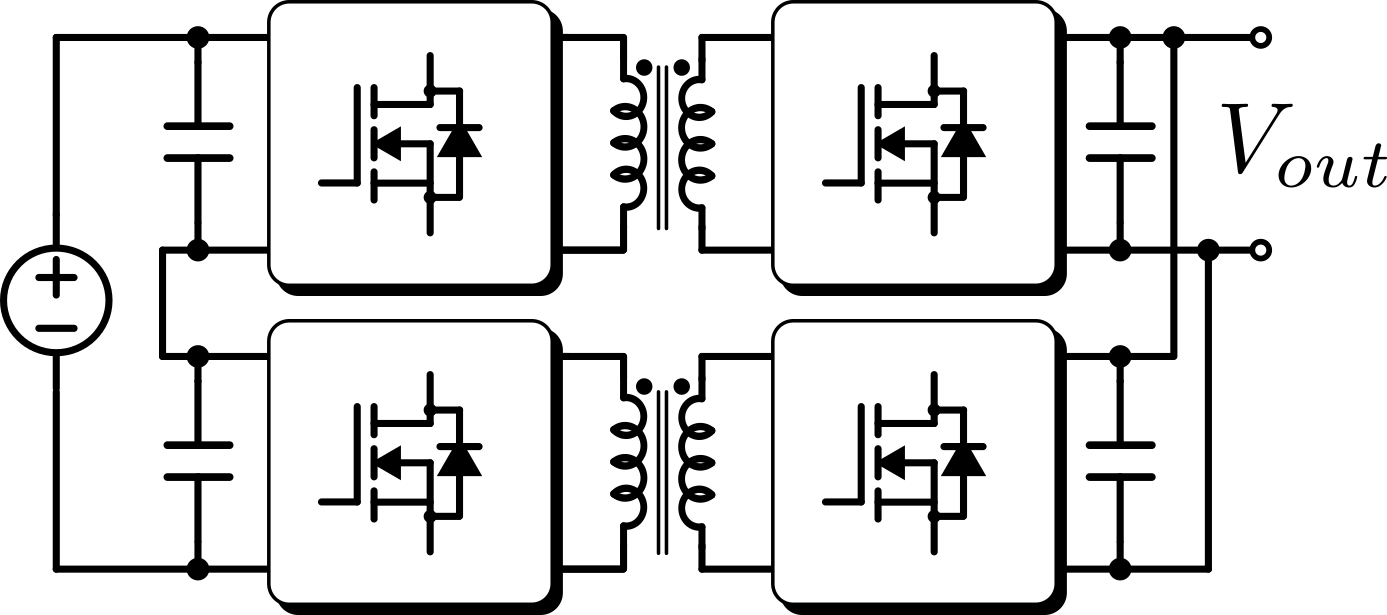}
        \caption{}
    \end{subfigure}
    \hfill
    \begin{subfigure}{0.48\textwidth}
        \centering
        \includegraphics[width=0.5\linewidth]{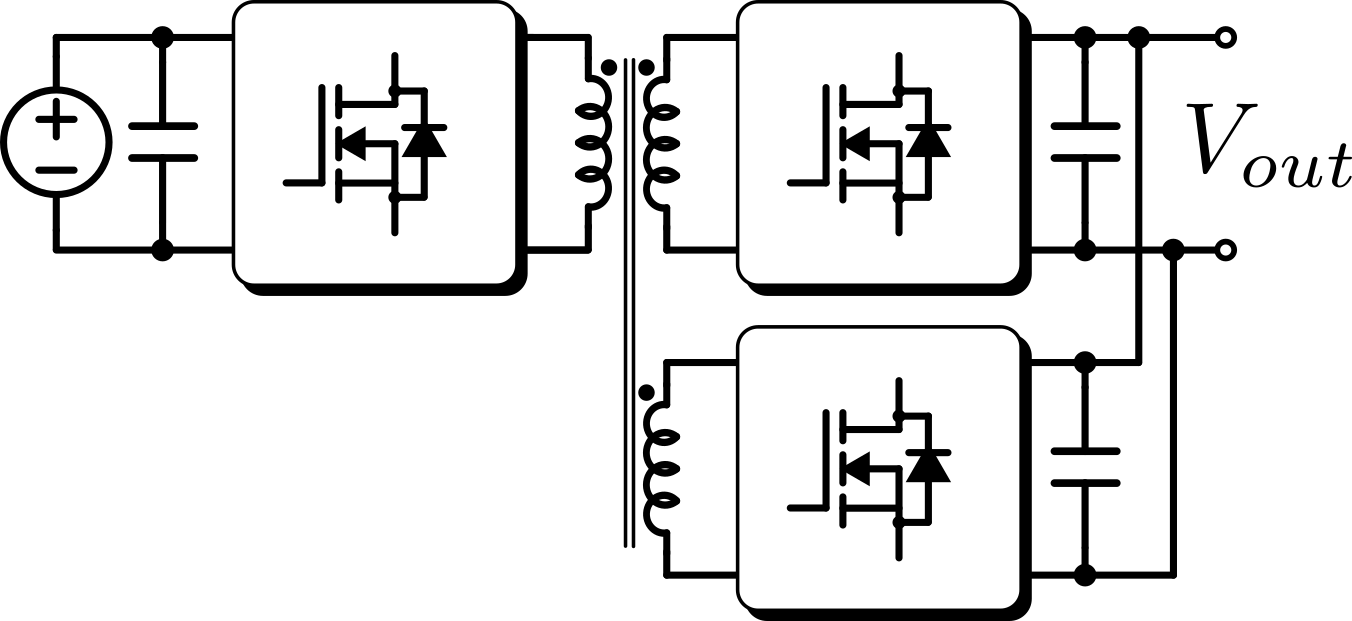}
        \caption{}
    \end{subfigure} 
    \hfill
    \begin{subfigure}{0.48\textwidth}
        \centering
        \includegraphics[width=0.5\linewidth]{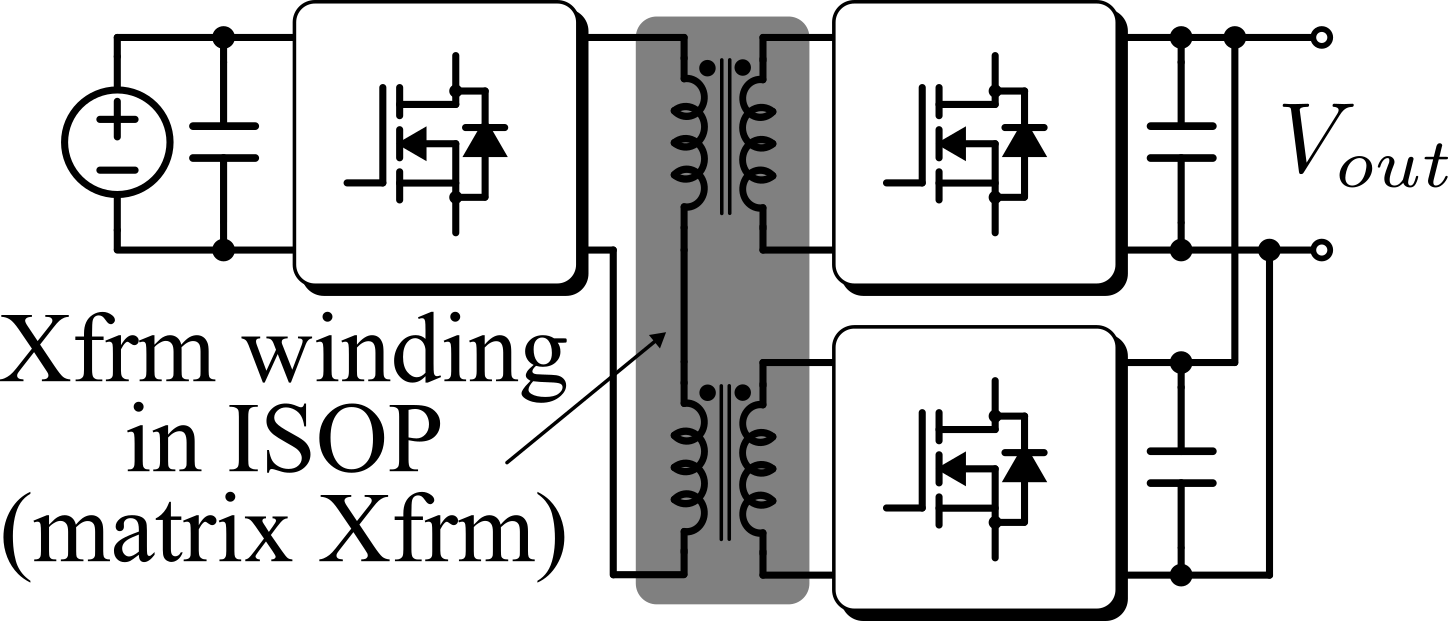}
        \caption{}
    \end{subfigure}
    \caption{Various output parallel configurations of isolation transformer-based converters. (a) Configuration 1: converter-level ISOP connection. (b) Configuration 2: single primary converter and parallel secondary converters with coupled secondary transformer windings. (c) Configuration 3: single primary converter and parallel secondary converters with transformer windings in the ISOP connection. The secondary windings are magnetically decoupled by using separate cores.}
    \label{fig:output_parallel_config}
\end{figure}

Different output-parallel configurations are shown in Fig. \ref{fig:output_parallel_config}. Whereas all three configurations connect the output of the secondary converter in parallel, the configuration of the primary windings of the primary converters is different. The first configuration (see Fig. \ref{fig:output_parallel_config}[a]) is the input-series-output-parallel (ISOP) configuration, in which the input to the primary converters is connected in series and the output of the secondary converters is connected in parallel. The ISOP configuration not only shares the high output current but also splits the high input voltage among the primary side converters connected in series, reducing the voltage stress on individual converters on the primary side. Reference \cite{DAB_Converter_ISOP_2016} demonstrated a high voltage step-down converter for data center applications by configuring multiple DAB converters. The 800-to-12.5 V DC--DC converter for data center on-tray PDN is proposed in \cite{EPC_LLC_ISOP}. This converter is based on an \textit{LLC} converter in the ISOP configuration. 

The ISOP configuration has high modularity and scalability because it is a converter-level configuration. The transformer of the isolated converter does not need to be redesigned, and the conventional isolated converters can be used as the fundamental building block. However, this configuration requires a high number of components, which can lead to higher volume, weight, and cost. The ISOP configuration is suitable for high voltage step-down applications, where high modularity and high redundancy are required.

The second configuration employs a modified transformer design, as shown in Fig. \ref{fig:output_parallel_config}(b). Unlike the ISOP configuration shown in Fig. \ref{fig:output_parallel_config}(b), there is only one primary side converter, significantly simplifying the overall system. The secondary windings are coupled to a common magnetic core, requiring minimum modification in the transformer design. However, because of the magnetic coupling between the secondary windings, this configuration suffers from circulating currents when a mismatch occurs between the paralleled secondary side converters, or the switching of the parallel-connected secondary converters is not synchronized \cite{DAB_Transformer_ISOP}. As a result, this configuration requires sophisticated control to suppress the circulating current \cite{DAB_Transformer_ISOP}.

\begin{figure}[!t]
     \centering
      \includegraphics[width=0.55\linewidth]{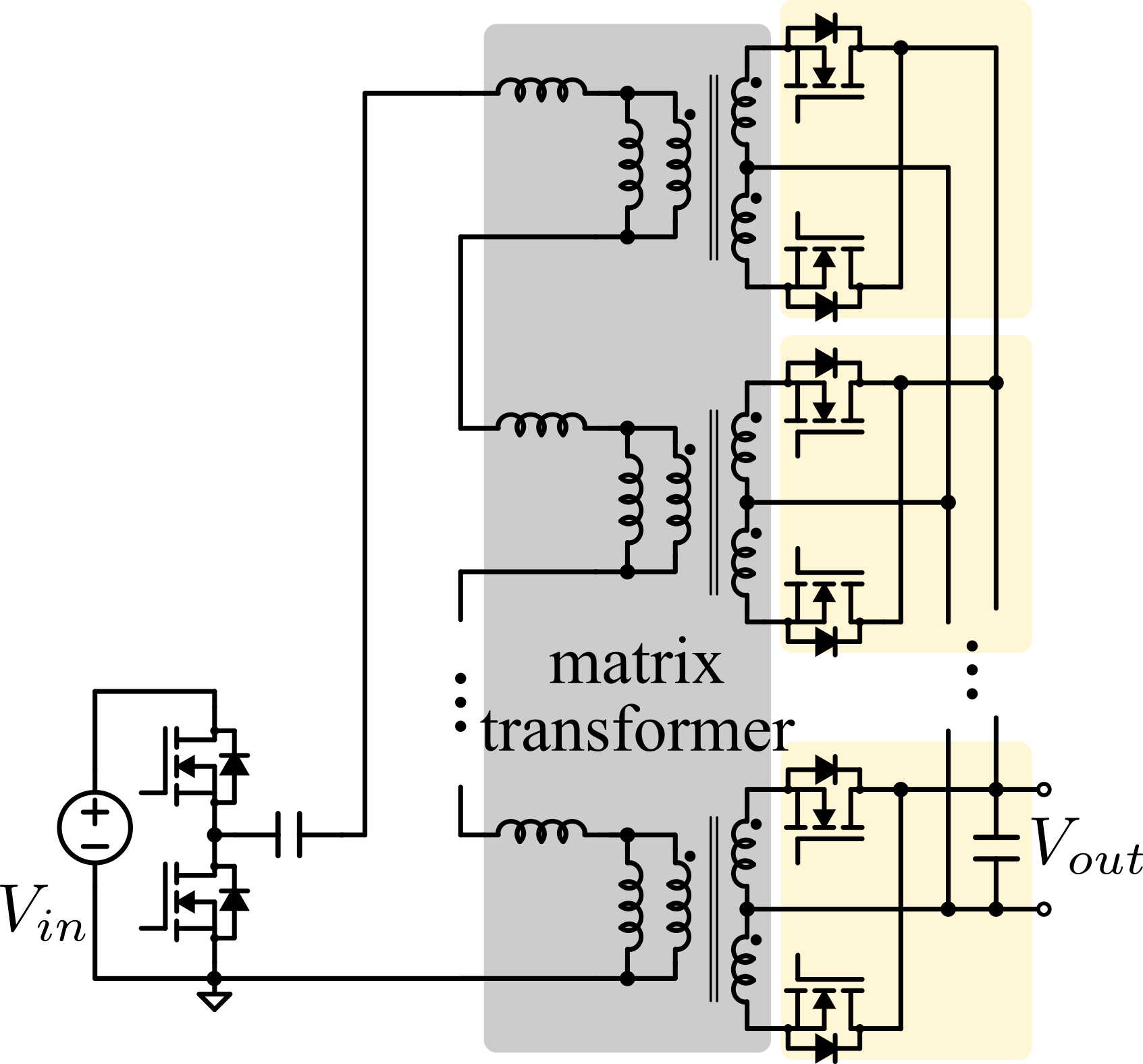}
     \caption{High step-down voltage \textit{LLC} converter in Configuration 3 \cite{LLC_Matrix_VT_2014}. This configuration has the secondary windings magnetically decoupled.}
     \label{fig:config_3_LLC}
 \end{figure}

The third configuration avoids the circulating current between the paralleled secondary converters by decoupling the secondary windings, as shown in Fig. \ref{fig:output_parallel_config}, by using separate cores. Such a modification of the transformer is pioneered in \cite{LLC_Matrix_VT_2014} for an \textit{LLC} converter with high voltage step-down, as shown in Fig. \ref{fig:config_3_LLC}. The design and optimization of this type of transformer are well documented in \cite{LLC_Matrix_1,LLC_Matrix_2}. Because the secondary windings are magnetically decoupled, it is possible to operate paralleled secondary converters asynchronously, providing more control flexibility and potentially reducing the output voltage ripple using carrier phase-shifted modulation \cite{config_3_carrier_phase_shift}. This transformer configuration is often referred to as the \emph{matrix transformer} \cite{Matrix_transformer_1990_book}. \\

\begin{table*}[!t]
\centering
\caption{Comparison of different types of high voltage step-down converters}
\label{tab:HVCR_topologies}
\renewcommand{\arraystretch}{1.25}
\begin{tabular}{lccc}
\hline
Class of topology                   & Buck-derived  & Hybrid SC                     & Transformer-based       \\
\hline
\textbf{Efficiency}                 & moderate      & low (charge sharing loss)     & high (soft-switching)   \\
\textbf{Power density}              & low           & very high                     & moderate                \\
\textbf{Control performance}        & very high     & low (fixed ratio)             & moderate                \\
Voltage conversion ratio            & moderate      & very high                     & high                    \\
Galvanic isolation                  & no            & no                            & yes                     \\
\hline
\end{tabular}
\end{table*}

\subsubsection{Matrix Transformer and Integrated Magnetics}

\begin{figure}
    \centering
    \begin{subfigure}{0.43\linewidth}
        \centering
        \includegraphics[width=1\linewidth]{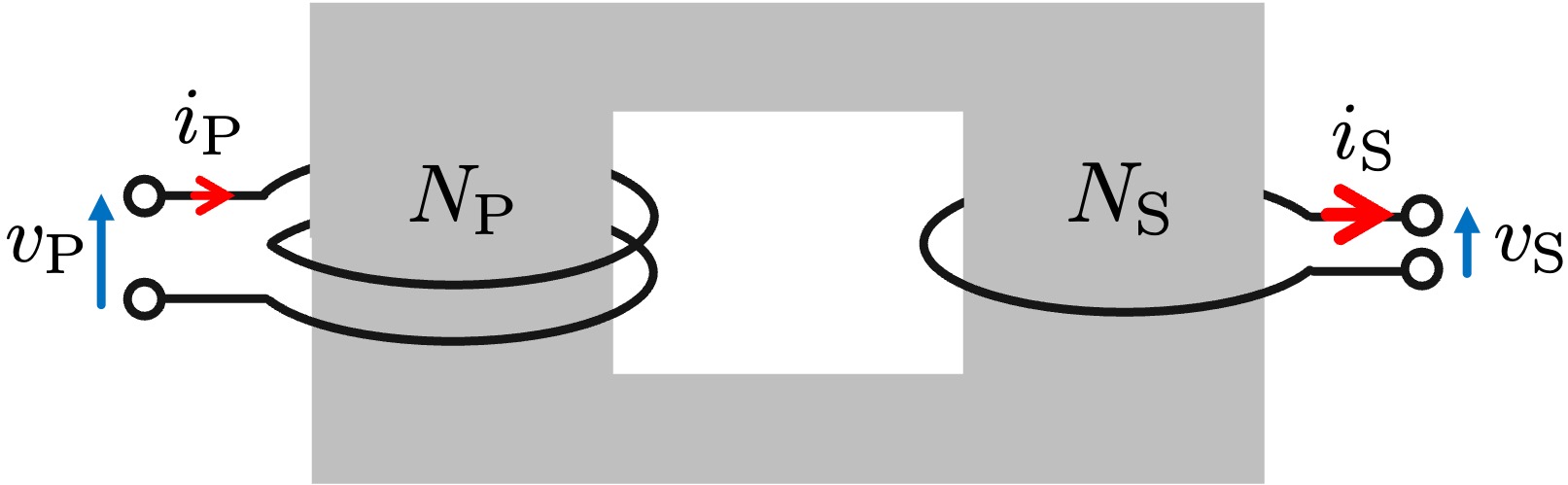}
        \caption{}
        \label{fig:conv_tx}
    \end{subfigure}
    \hfill
    \begin{subfigure}{0.55\linewidth}
        \centering
        \includegraphics[width=1\linewidth]{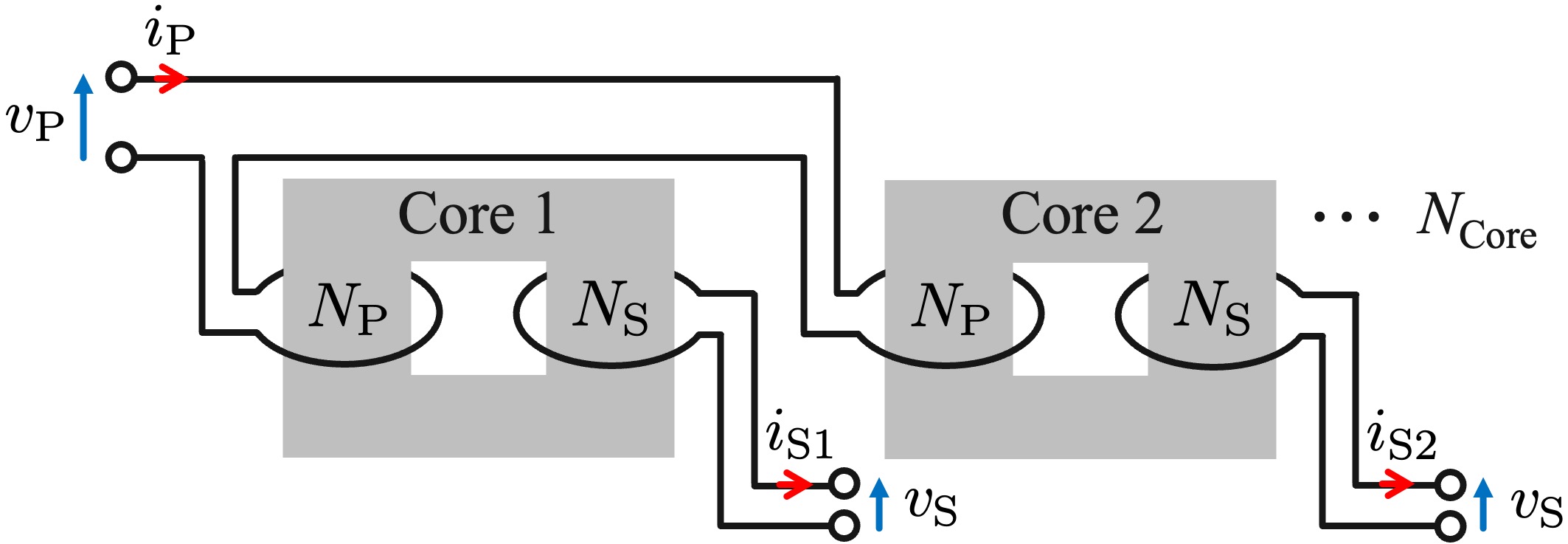}
        \caption{}
        \label{fig:matrix_tx}
    \end{subfigure}
    \caption{Height comparison of (a) conventional and (b) matrix transformers.}
    \label{fig:tx_height_comp}
\end{figure}

As introduced previously, the matrix transformer (Configuration 3 in Fig.~\ref{fig:output_parallel_config}[c]) uses series-connected primaries and parallel-connected, magnetically decoupled secondaries. 
This structure ensures equal current sharing and avoids circulating currents between secondary converters. 
Figure~\ref{fig:tx_height_comp} shows that the matrix transformer also achieves a lower profile than that of a conventional design.

In a conventional transformer, voltage and current follow
\begin{equation}
v_{\mathrm S} = \frac{N_{\mathrm S}}{N_{\mathrm P}} v_{\mathrm P}, \qquad
i_{\mathrm S} = \frac{N_{\mathrm P}}{N_{\mathrm S}} i_{\mathrm P}.
\end{equation}
In a matrix transformer (Fig.~\ref{fig:matrix_tx}), $N_{\mathrm{Core}}$ elemental transformers share a series-connected primary winding. The primary voltage divides equally, so each element sees $v_{\mathrm P}/N_{\mathrm{Core}}$. With turns ratio $N_{\mathrm P}:N_{\mathrm S}$ per element, the secondary voltage of each element is
\begin{equation}
v_{\mathrm {S}k} = \frac{N_{\mathrm S}}{N_{\mathrm P}} \cdot \frac{v_{\mathrm P}}{N_{\mathrm{Core}}} = \frac{N_{\mathrm S}}{N_{\mathrm{Core}} N_{\mathrm P}} v_{\mathrm P}.
\end{equation}
With identical elements and ideal coupling, each secondary carries
\begin{equation}
i_{\mathrm {S}k} = \frac{N_{\mathrm P}}{N_{\mathrm S}} i_{\mathrm P}, \quad k = 1, 2, \ldots, N_{\mathrm{Core}}.
\end{equation}
Because the secondaries are connected in parallel at equal voltage $v_{\mathrm S} = v_{\mathrm{S}k}$, the total output current is
\begin{equation}
i_{\mathrm {S}} = \sum_{k=1}^{N_{\mathrm{Core}}} i_{\mathrm {S}k} = \frac{N_{\mathrm{Core}} N_{\mathrm P}}{N_{\mathrm S}} i_{\mathrm P}.
\end{equation}
The effective turns ratio becomes $N_{\mathrm{Core}} N_{\mathrm P} : N_{\mathrm S}$, achieving a high step-down ratio without increasing physical turns on any single element. This reduces winding layers and transformer height.

Splitting the magnetic core distributes secondary current among multiple elements, making matrix transformers well suited for applications requiring hundreds of amperes at 12\,V or below. 
To further improve power density, matrix transformers often employ integrated core structures that eliminate dead space between separate cores, reduce component count, and minimize winding loss \cite{huang_tpel_2014, nabih_matrix_lowprofile}.

Integrated PCB-winding designs report $>\!98.8\%$ efficiency with significantly reduced winding loss \cite{nabih_matrix_lowprofile,cpes_matrix_term}. 
Resistance-matrix modeling allows conductor geometry and interleaving to be optimized for minimum AC resistance \cite{chen_sullivan_planar_model}.

Higher switching frequency shrinks magnetics but increases core loss nonlinearly, making geometry optimization critical. 
Pillar shape, window utilization, and gap placement strongly affect flux distribution and eddy loss. 
Round-pillar or ``snake-core'' PCB structures shorten winding length and spread flux more evenly, reducing both copper and core losses at megahertz frequencies \cite{knabben_gan_server,kasper_planar_vs_litz}. 
Splitting into multiple cores, however, can raise total core loss unless cross-section and flux density are re-optimized \cite{cpes_1mhz_llc,nabih_matrix_lowprofile}.

For resonant converters needing precise inductance for soft switching, matrix transformers also enable integration: (i) intentional leakage in multileg or modified-symmetry cores provides resonant inductance without a separate component \cite{cpes_1mhz_llc}; (ii) \emph{matrix inductors} distribute inductance laterally into multiple elements, lowering per-layer AC stress and improving thermal uniformity \cite{nabih_matrix_lowprofile}; (iii) integrated stacks placing transformer and resonant inductors on a shared core/PCB cut loop area, interconnect loss, and z-height \cite{nabih_matrix_lowprofile}.

\subsection{Discussions and Future Directions for Designing LV-PDN for Next-Generation Data Centers}
Previous subsections reviewed different LV-PDN architectures and state-of-the-art DC--DC converter topologies found in literature for high voltage step-down applications. The comparison of three classes of DC--DC converter topologies is shown in Table \ref{tab:HVCR_topologies}.

The design of LV-PDN and its converter will be achieved by combining these different classes of converters in an innovative way, maximizing their advantages and minimizing their limitations. For example, \cite{1500V_48V_switch_cap_DAB} combines the hybrid SC network with a DAB converter (i.e., a type of transformer-based DC--DC converter class) achieving a 1,500-to-48 V voltage conversion. The hybrid SC network down-converts the input voltage, and the cascaded DAB converter tightly regulates the output voltage. This DC--DC converter leverages the effective voltage conversion ratio of the hybrid SC network as well as the control and galvanic isolation features of the DAB converter. Various combinations and configurations of these three fundamental classes of voltage step-down converters should be explored for the LV-PDN converter.

\section{In-Facility DC Power Distribution}
\label{section:DC_Distribution}

The second architectural shift identified in Section~\ref{section:3_stages} is the transition from facility-level AC power distribution to LV DC distribution in AI data centers. This shift seeks to reduce the number of power conversion stages, improve end-to-end efficiency, and enable tighter integration with on-site ESS and backup resources \cite{400VDC_dis_INTELEC_07,Kumar}. As rack-level power delivery evolves toward LV in-rack DC architectures, the facility-level distribution system must likewise adapt to support higher power density, coordinated protection, and safe operation under DC fault conditions. 

Although LV-IBCs enable higher-voltage compute racks, their full benefits cannot be realized if the facility continues to rely on traditional 480 V AC distribution. Consequently, the architectural transition extends beyond the rack to the building-wide power system, where LV DC distribution has been proposed as a viable alternative for future AI data centers (Fig.~\ref{fig:stage2_3})~\cite{NVIDIA800VDC, ocp_diablo400}. A key advantage of LV DC distribution is the elimination of multiple conversion stages, as both IT loads and the majority of on-site ESS inherently interface with DC. Despite these potential benefits, AC distribution remains the dominant approach in current data center facilities~\cite{2025_OCP_GOOGLE, NVIDIA800VDC, EVatScale}. 

This section reviews in-facility LV DC distribution systems as the enabling technology for the second architectural shift in AI data centers. Emphasis is placed on typical DC busway architectures, grounding schemes, and protection challenges that directly influence safety, reliability, and scalability in high-power AI data center environments. The discussion highlights key design trade-offs and technical challenges that distinguish LV DC distribution in data centers from conventional AC facilities. 

\subsection{Selection of DC Distribution Architectures}

\begin{figure*}[!t]
    \centering
    \begin{subfigure}{0.46\textwidth}
        \centering
        \includegraphics[width=0.9\linewidth]{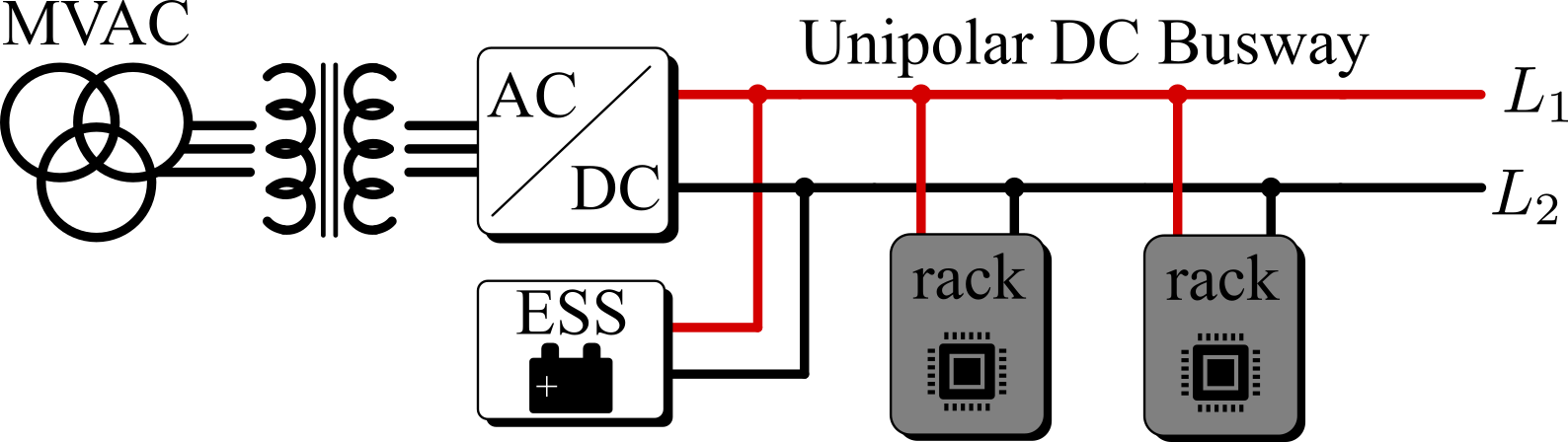}
        \caption{}
    \end{subfigure}
    \hfill
    \begin{subfigure}{0.51\textwidth}
        \centering
        \includegraphics[width=0.9\linewidth]{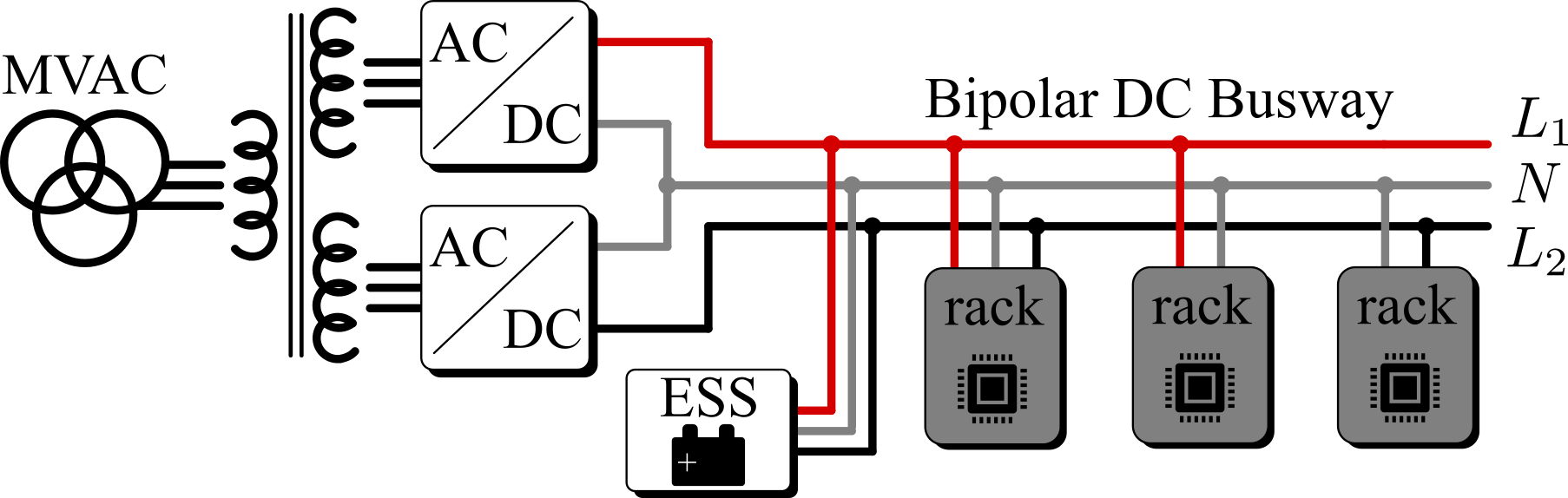}
        \caption{}
    \end{subfigure}     
    \hfill
    \begin{subfigure}{0.60\textwidth}
        \centering
        \includegraphics[width=0.9\linewidth]{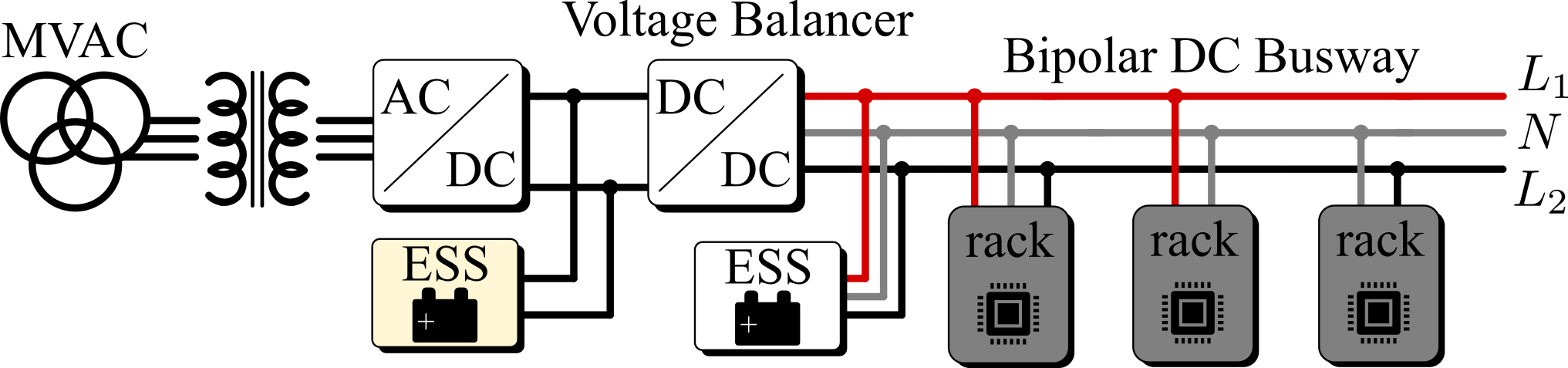}
        \caption{}
    \end{subfigure} 
    \caption{In-facility DC distribution architectures. (a) Unipolar DC busway architecture. (b) Bipolar DC busway architecture with a cascaded AC--DC converter. (c) Bipolar DC busway architecture with a single AC--DC converter and voltage balancer.}
    \label{fig:DC_uni_bi}
\end{figure*}

Two types of in-facility LV DC busway architectures are actively pursued for future AI data centers \cite{NVIDIA800VDC, ocp_diablo400, 2025_OCP_GOOGLE}. The first is the unipolar LV DC busway, in which a single line-to-line DC voltage is delivered to each compute rack, as shown in Fig. \ref{fig:DC_uni_bi}(a) \cite{NVIDIA800VDC, Cuzner_DC_ground, Kumar, Tomi_part2}. The second is the bipolar LV DC busway (e.g., ±400 V or ±750 V) with a neutral pole, as illustrated in Figs. \ref{fig:DC_uni_bi}(b) and (c) \cite{ocp_diablo400, 2025_OCP_GOOGLE, Cuzner_DC_ground, Kumar, Tomi_part2}. 

The unipolar architecture is the simplest in-facility DC distribution approach and is widely adopted in practice \cite{Tomi_part2}. For example, \cite{NVIDIA800VDC} proposes an 800 V unipolar DC busway for future AI data centers, which interfaces directly with compute racks through an in-rack LV DC (e.g., 800 V DC) distribution system. A BESS is typically connected to the busway to support stable operation, as shown in Fig. \ref{fig:DC_uni_bi}(a). Compared with a conventional AC busway (Fig. \ref{fig:conv}[a]), the DC busway architecture (Fig. \ref{fig:DC_uni_bi}[a]) eliminates an additional AC–-DC conversion stage and enables a more direct, tightly integrated BESS connection. 

Despite these advantages, the unipolar architecture offers no redundancy, which limits system-level reliability. It also supports only a single voltage level, even though different loads may require different DC input voltages. In short, though the unipolar DC busway is simple and cost-effective, it provides limited reliability and voltage flexibility \cite{Kumar}. 

The bipolar DC busway architecture addresses the limitations of its unipolar counterpart. It consists of three poles—positive ($L_{1}$, red in Figs. \ref{fig:DC_uni_bi}[b] and [c]), negative ($L_{2}$, black), and neutral ($N$, gray)—and provides higher redundancy, improved reliability, and more flexible voltage options for the load \cite{Kumar,Bipolar_ex1}. Owing to these advantages, \cite{2025_OCP_GOOGLE} proposes an in-facility ±750 V DC distribution architecture in which the busway’s line-to-line voltage is 1,500 V DC, and compute racks are supplied from the line-to-neutral 750 V DC. 

The bipolar busway can be realized in two ways. The first is a natural extension of the unipolar architecture, in which two AC–DC converters are cascaded to form two independently regulated DC voltages (i.e., $L_{1}-N$ and $N-L_{2}$), connected in series, as shown in Fig. \ref{fig:DC_uni_bi}(b) \cite{Bipolar_review,Bipolar_ex1}. Independent control of these converters eliminates challenges associated with unbalanced loading on the bipolar busway. However, this approach requires redesigning the LFT secondary into a split winding (see Fig. \ref{fig:DC_uni_bi}[b]) to withstand any DC offset caused by load imbalance \cite{Bipolar_review}.

Alternatively, the bipolar output can be generated using NPC AC–DC converters, which inherently produce a bipolar DC voltage using two split capacitors \cite{Cuzner_DC_ground}. A key challenge with these topologies is regulating the neutral-point voltage (i.e., maintaining balanced capacitor voltages at the DC output). Several methods have been proposed to address this issue \cite{NPC_hw, NPC_MPC, NPC_carrier, 3186927, SL_ITEC}, but none can guarantee neutral-point balance under all operating conditions \cite{Cuzner_DC_ground, NPC_balance_limit}. Consequently, an additional voltage-balancing circuit is typically required, as shown in Fig. \ref{fig:DC_uni_bi}(c) \cite{Bipolar_review, Cuzner_DC_ground, Bipolar_legend, Kumar}. Comprehensive reviews of voltage balancer circuits are provided in \cite{Bipolar_review, Voltage_balancer_review}. Depending on the balancer implementation, it is also possible to use an AC–DC converter that is not NPC-based \cite{Cuzner_DC_ground}. 

The ESS can be connected either to the intermediate DC link between the AC–DC converter and the voltage balancer (yellow ESS in Fig. \ref{fig:DC_uni_bi}[c]) \cite{Bipolar_legend} or directly to the bipolar DC busway (white ESS in Fig. \ref{fig:DC_uni_bi}[c]) \cite{2025_OCP_GOOGLE}. 

\subsection{Different Grounding Schemes of DC Busway Architecture}
Grounding enhances safety, minimizes electrical hazards, reduces EMI, and improves overall system reliability \cite{IEC60364-1_2005, Cuzner_DC_ground}. This is especially critical for future AI data centers using LV DC distribution in both the facility and compute racks (e.g., 800 V DC in \cite{NVIDIA800VDC}), where users face a higher likelihood of exposure to LV DC compared with conventional architectures. However, no widely accepted grounding configuration has been established for future AI data centers. Accordingly, this subsection reviews and compares different grounding configurations suitable for LV DC distribution. 

According to IEC 60364-1, DC grounding configurations fall into three categories: Terre–Terre (\textit{TT}, Fig. \ref{fig:DC_TT}), Terre–Neutre (\textit{TN}, Fig. \ref{fig:DC_TN-S}, Fig. \ref{fig:DC_TN-C}, and Fig. \ref{fig:DC_TN-C-S}), and Isolated–Terre (\textit{IT}, Fig. \ref{fig:DC_IT}) \cite{IEC60364-1_2005, Cuzner_DC_ground}. Note that \textit{IT} refers to a grounding configuration and is distinct from the IT acronym used for information technology. The first letter (\textit{T} or \textit{I}) denotes how the DC source (e.g., a substation or the DC output of a power converter forming the DC busway voltage) is grounded: \textit{T} indicates direct grounding to earth, whereas \textit{I} indicates isolation from earth. The second letter (\textit{T} or \textit{N}) indicates how exposed conductive parts at the load (e.g., metal enclosures of compute racks) are grounded: \textit{T} denotes direct connection to earth, whereas \textit{N} denotes grounding through the system neutral.

The following subsections provide an overview of these grounding configurations. A performance comparison across four categories—personal safety, equipment safety, EMC, and fault tolerance—is summarized in Fig. \ref{fig:grounding_spider_plot} \cite{Cuzner_DC_ground}.\vspace{11pt}

\subsubsection{DC Grounding Configuration \textit{TT}}
\begin{figure}[!t]
    \centering
    \begin{subfigure}{0.25\textwidth}
        \centering
        \includegraphics[width=1\linewidth]{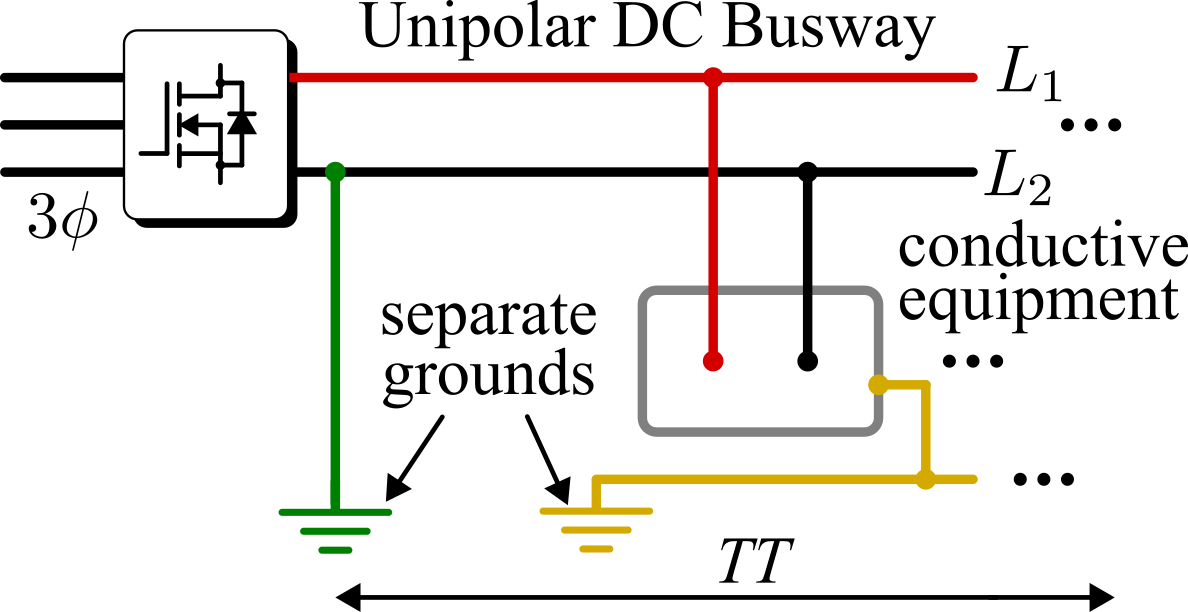}
        \caption{}
    \end{subfigure}
    \hfill
    \begin{subfigure}{0.23\textwidth}
        \centering
        \includegraphics[width=1\linewidth]{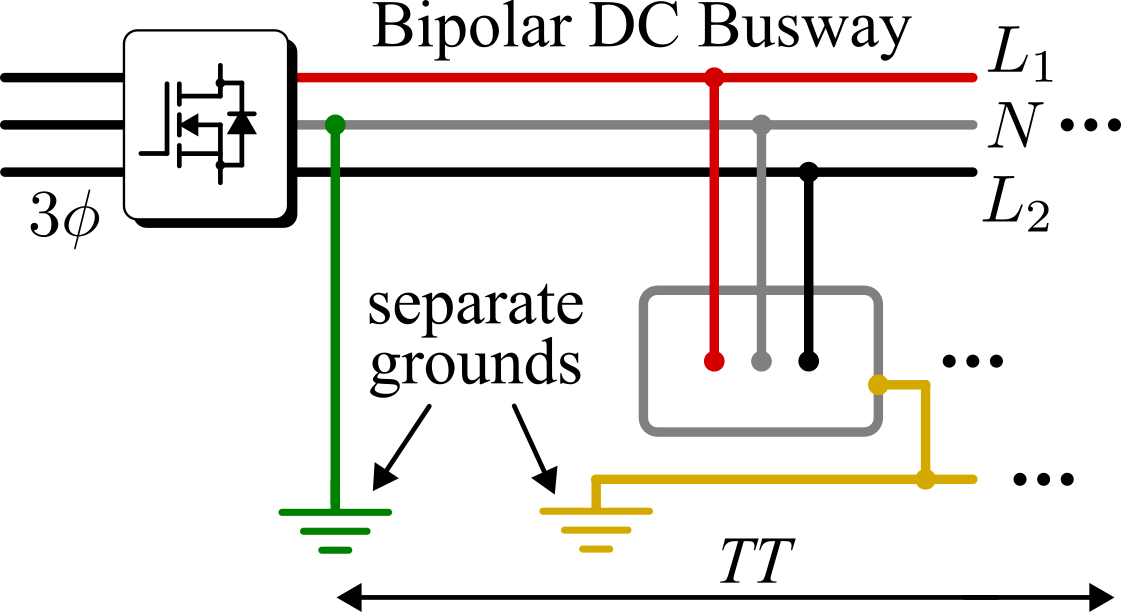}
        \caption{}
    \end{subfigure} 
    \caption{Schematic of \textit{TT} ground configuration. (a) Unipolar DC busway. (b) Bipolar DC busway.}
    \label{fig:DC_TT}
\end{figure}

The schematic of the \textit{TT} grounding configuration is shown for both the unipolar (Fig. \ref{fig:DC_TT}[a]) and bipolar (Fig. \ref{fig:DC_TT}[b]) DC busways. Because the first letter of \textit{TT} is \textit{T}, the DC output of the AC–DC conversion system that forms the DC busway (i.e., the DC source) must be grounded to earth. In the unipolar case, either the positive or negative pole may be grounded (green grounding in Fig. \ref{fig:DC_TT}[a]) \cite{positive_pole_ground, positive_pole_ground_IEEE}. For the bipolar busway, the neutral pole of the DC source is typically grounded, as illustrated in Fig. \ref{fig:DC_TT}(b) \cite{positive_pole_ground, positive_pole_ground_IEEE}. 

The second letter of \textit{TT} is also \textit{T}, indicating that exposed conductive parts on the load side (e.g., metallic components of the IT racks) must likewise be grounded. In this configuration, a separate grounding network (yellow ground in Fig. \ref{fig:DC_TT}) is installed within the facility. This separate ground maintains exposed conductive parts at the same potential, thereby protecting users from electric shock. Because of this protective function, it is commonly referred to as the protective earth (PE). 

The \textit{TT} configuration is straightforward to install, and the separation of grounding paths helps prevent faults from propagating to other parts of the DC busway. However, this same separation can permit circulating currents and introduce significant voltage stress—one of the primary disadvantages of the \textit{TT} configuration in DC busway architectures \cite{TT_circulating_current}.
\vspace{11pt}
\subsubsection{DC Grounding Configuration \textit{TN}}
\begin{figure}[!t]
    \centering
    \begin{subfigure}{0.24\textwidth}
        \centering
        \includegraphics[width=1\linewidth]{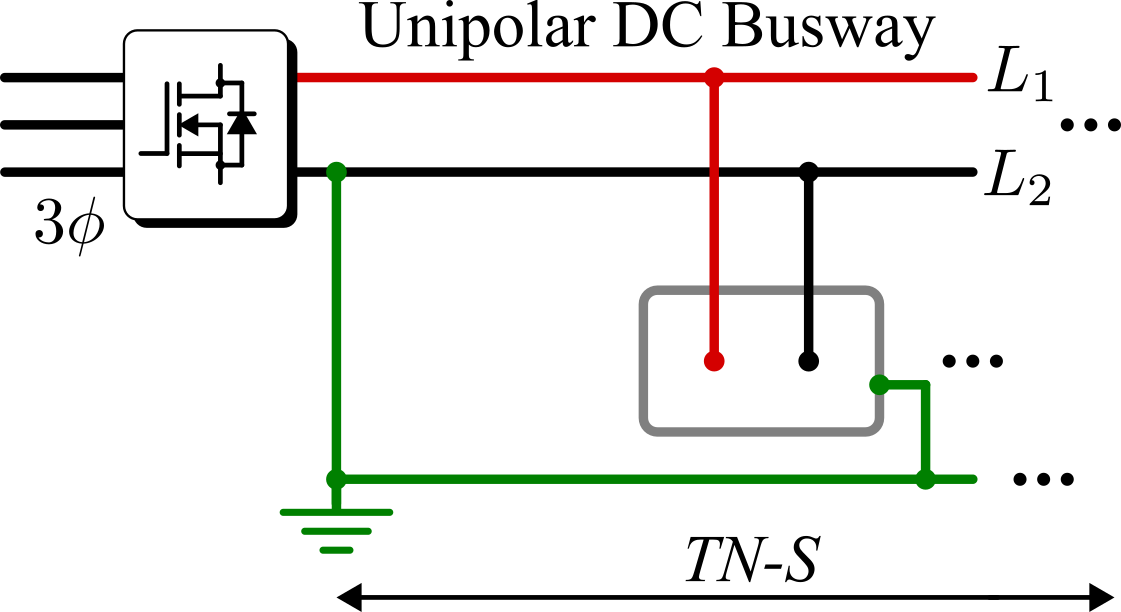}
        \caption{}
    \end{subfigure}
    \hfill
    \begin{subfigure}{0.24\textwidth}
        \centering
        \includegraphics[width=1\linewidth]{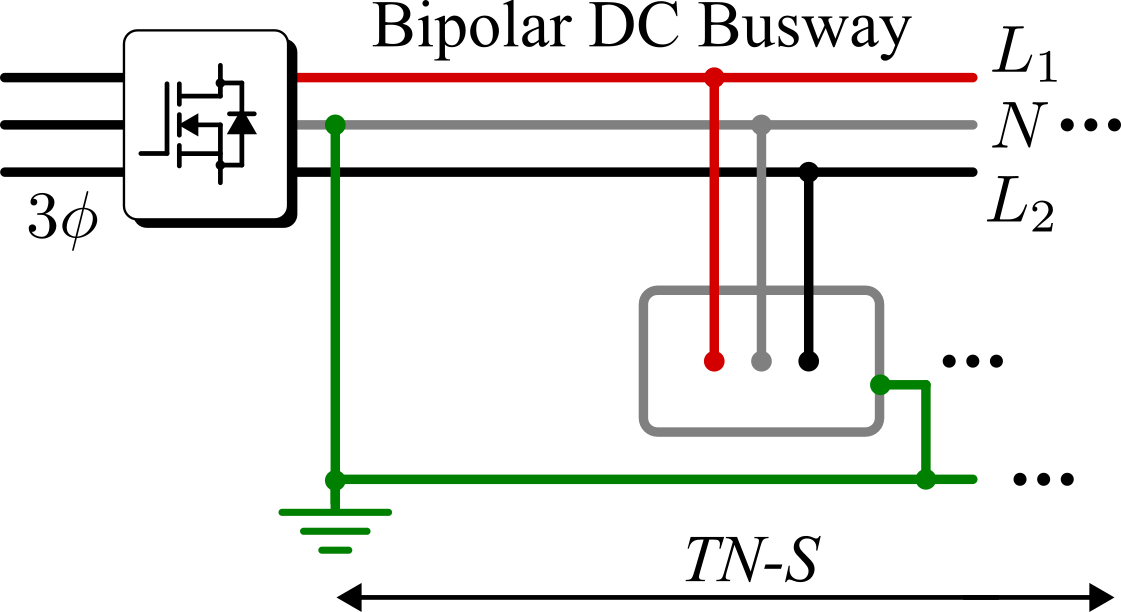}
        \caption{}
    \end{subfigure} 
    \caption{Schematic of \textit{TN-S} ground configuration. (a) Unipolar DC busway. (b) Bipolar DC busway.}
    \label{fig:DC_TN-S}
\end{figure}

\begin{figure}[!t]
    \centering
    \begin{subfigure}{0.24\textwidth}
        \centering
        \includegraphics[width=1\linewidth]{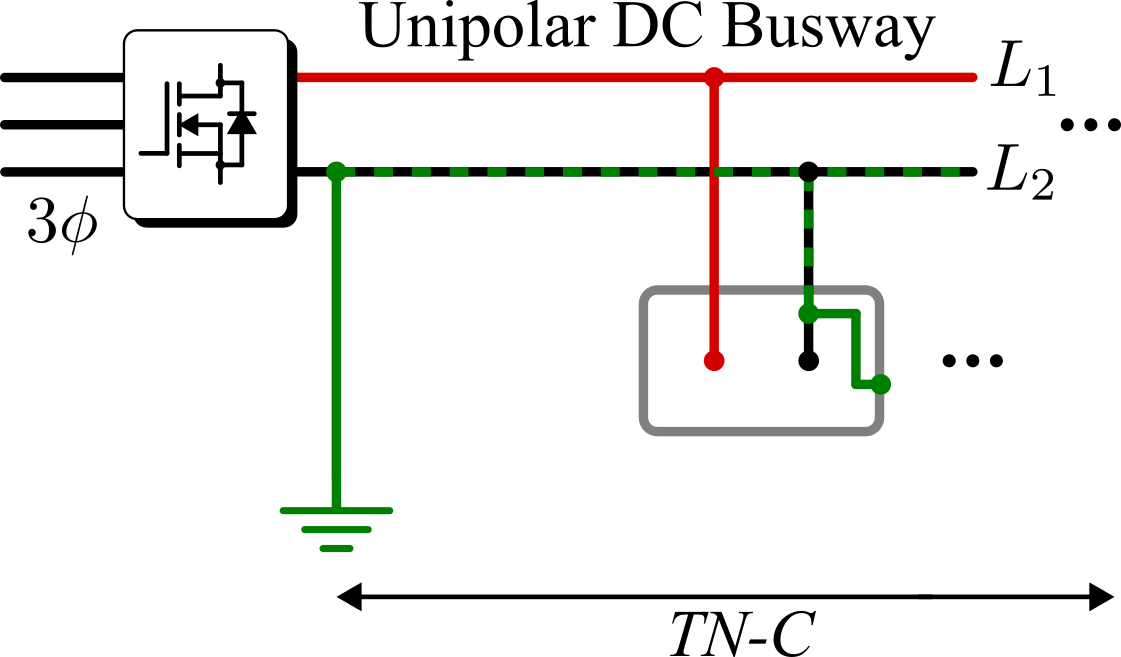}
        \caption{}
    \end{subfigure}
    \hfill
    \begin{subfigure}{0.24\textwidth}
        \centering
        \includegraphics[width=1\linewidth]{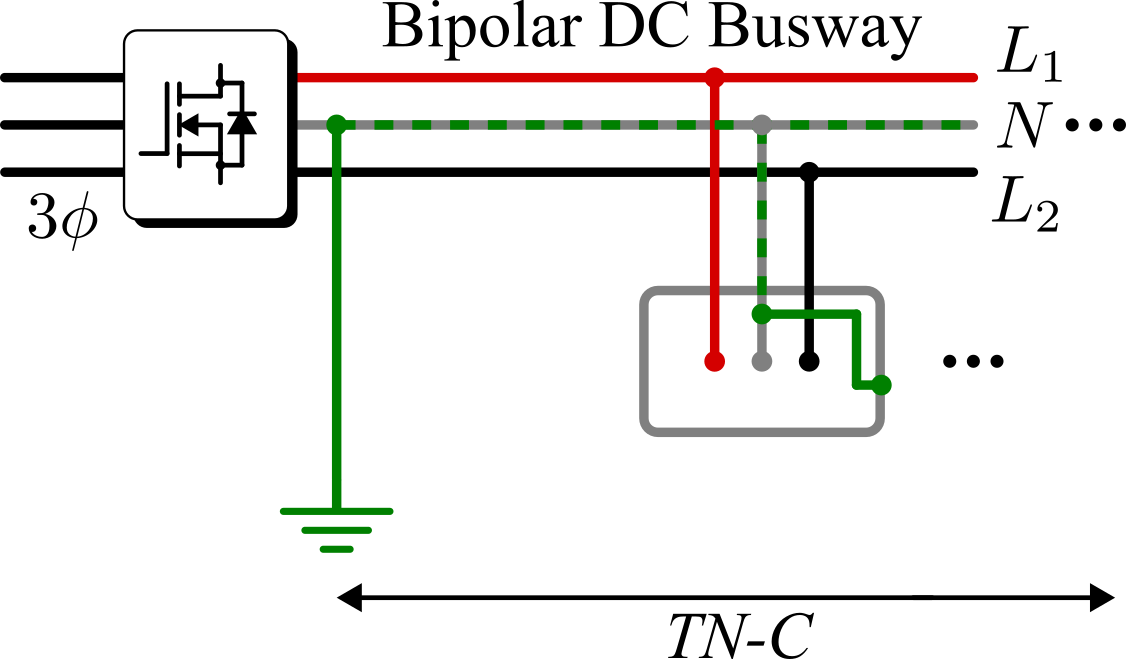}
        \caption{}
    \end{subfigure} 
    \caption{Schematic of \textit{TN-C} ground configuration. (a) Unipolar DC busway. (b) Bipolar DC busway.}
    \label{fig:DC_TN-C}
\end{figure}

\begin{figure}[!t]
    \centering
    \begin{subfigure}{0.24\textwidth}
        \centering
        \includegraphics[width=1\linewidth]{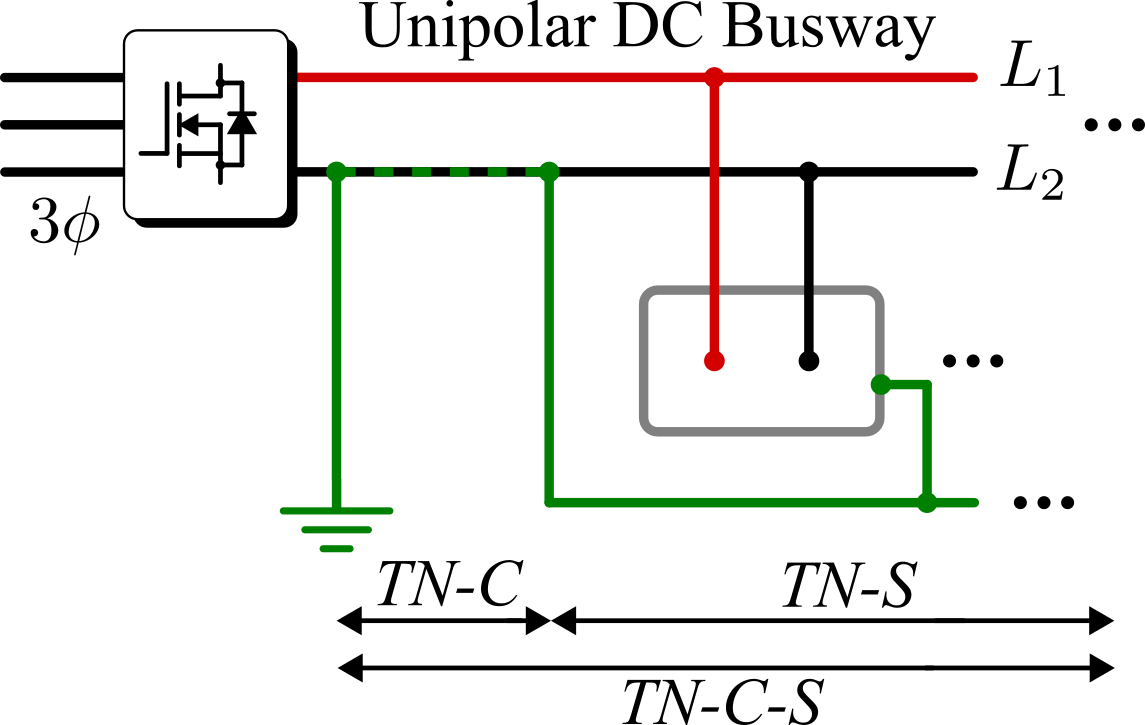}
        \caption{}
    \end{subfigure}
    \hfill
    \begin{subfigure}{0.24\textwidth}
        \centering
        \includegraphics[width=1\linewidth]{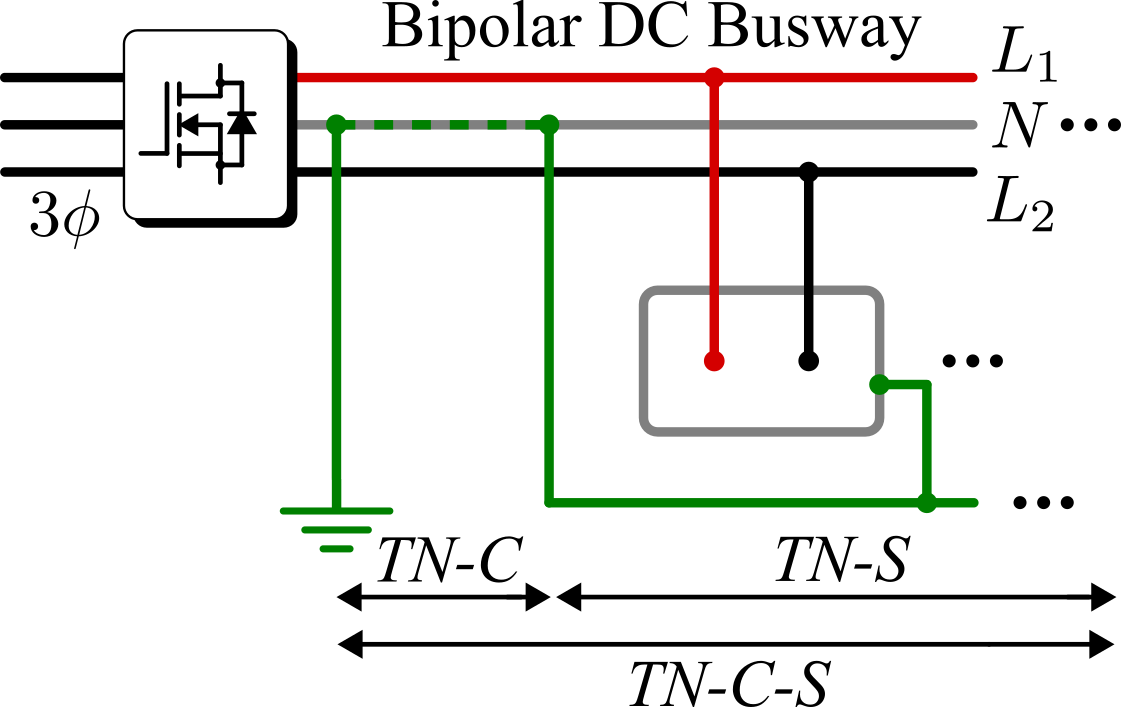}
        \caption{}
    \end{subfigure} 
    \caption{Schematic of \textit{TN-C-S} ground configuration. (a) Unipolar DC busway. (b) Bipolar DC busway.}
    \label{fig:DC_TN-C-S}
\end{figure}

The IEC 60364-1 standard defines three types of \textit{TN} grounding configurations: \textit{TN-S} (Fig. \ref{fig:DC_TN-S}), \textit{TN-C} (Fig. \ref{fig:DC_TN-C}), and \textit{TN-C-S} (Fig. \ref{fig:DC_TN-C-S}). In all \textit{TN} configurations, the DC source is grounded to earth, and the exposed conductive parts on the load side are connected to the neutral. Although the unipolar DC busway does not include a neutral pole, the pole that is grounded at the DC source is treated as the neutral from the load perspective. Unlike the \textit{TT} configuration, the \textit{TN} configurations use a \textit{single} grounding system shared by both the DC source and the load (green grounds in Fig. \ref{fig:DC_TN-S}, Fig. \ref{fig:DC_TN-C}, and Fig. \ref{fig:DC_TN-C-S}). How this shared ground is delivered to the load distinguishes the three \textit{TN} variants—\textit{TN-S}, \textit{TN-C}, and \textit{TN-C-S}. 

In the \textit{TN-S} configuration, the “S” denotes “separate,” meaning that although the DC source and load share a common grounding point, \textit{separate conductors} are used to connect them, as shown in Fig. \ref{fig:DC_TN-S}. Because the two grounding conductors (each connected to the same earth point) remain physically separated, \textit{TN-S} offers the best EMC performance among all \textit{TN} types, as shown in Fig. \ref{fig:grounding_spider_plot} \cite{Cuzner_DC_ground}. The redundant grounding conductor also improves personal safety, making \textit{TN-S} one of the most favorable grounding options for applications with sensitive equipment (e.g., data centers and telecoms) \cite{Cuzner_DC_ground}. The primary drawback is the additional installation cost associated with the extra grounding conductor on the load side. 

The \textit{TN-C} configuration addresses this cost issue by combining the neutral and PE conductors into a single conductor, as shown in Fig. \ref{fig:DC_TN-C}. The “C” denotes “combined.” This conductor—commonly referred to as the protective earth–neutral (PEN)—is represented by the dashed green line overlapping the black line in Fig. \ref{fig:DC_TN-C}(a) for the unipolar busway and by the dashed green line overlapping the gray line in Fig. \ref{fig:DC_TN-C}(b) for the bipolar busway. The PEN conductor connects to exposed conductive parts on the load side while also serving as the return path for DC current (e.g., the negative pole in Fig. \ref{fig:DC_TN-S}). Although cost-effective, the \textit{TN-C} configuration exhibits poor EMC performance (see Fig. \ref{fig:grounding_spider_plot}), limiting its suitability for sensitive equipment such as data-center loads \cite{Cuzner_DC_ground}. 

The \textit{TN-C-S} configuration is a hybrid in which part of the system uses \textit{TN-C} and part uses \textit{TN-S}, as shown in Fig. \ref{fig:DC_TN-C-S}. The \textit{TN-C} portion reduces installation cost, whereas the \textit{TN-S} portion improves grounding reliability. 
\vspace{11pt}
\subsubsection{DC Grounding Configuration \textit{IT}}
\begin{figure}[!t]
    \centering
    \begin{subfigure}{0.24\textwidth}
        \centering
        \includegraphics[width=1\linewidth]{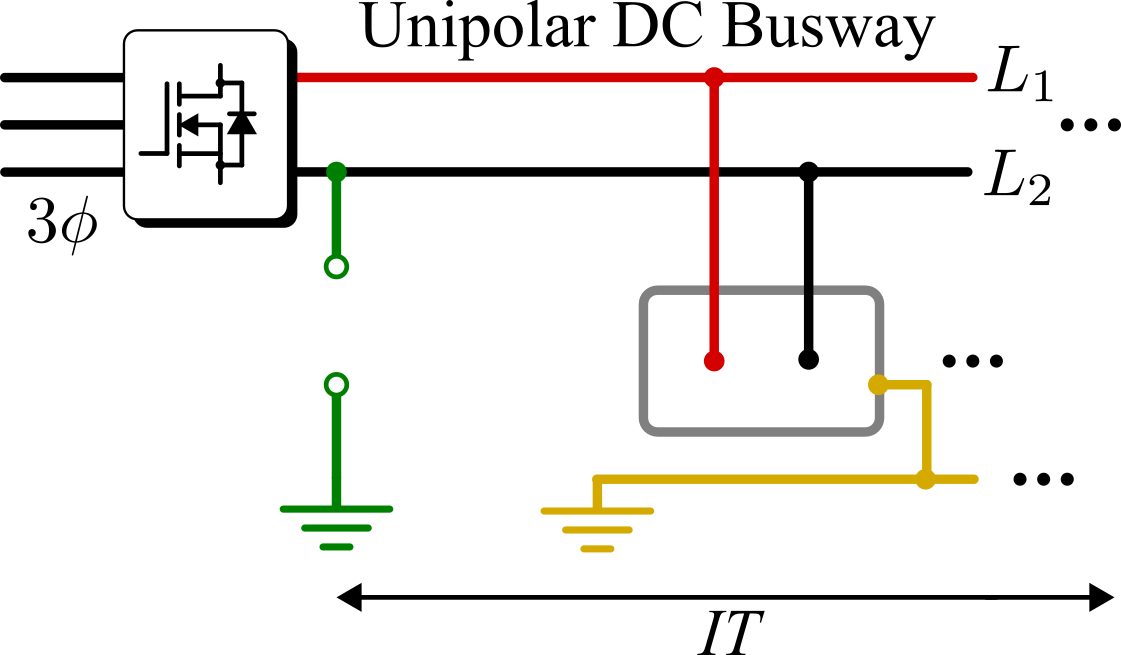}
        \caption{}
    \end{subfigure}
    \hfill
    \begin{subfigure}{0.24\textwidth}
        \centering
        \includegraphics[width=1\linewidth]{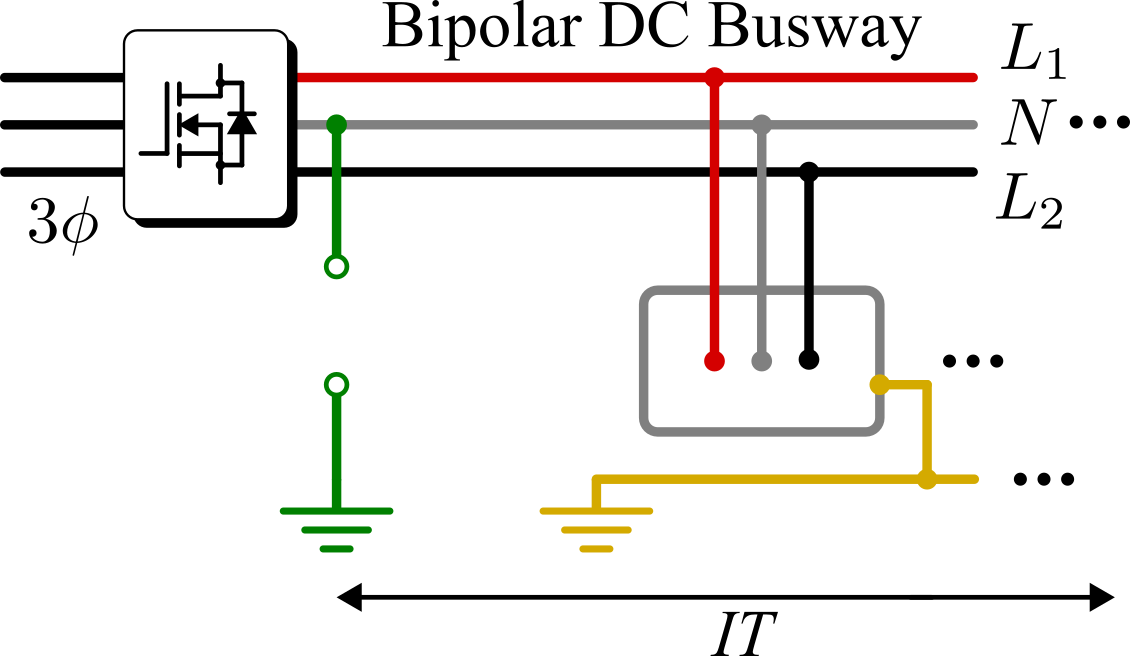}
        \caption{}
    \end{subfigure} 
    \caption{Schematic of \textit{IT} ground configuration. (a) Unipolar DC busway. (b) Bipolar DC busway.}
    \label{fig:DC_IT}
\end{figure}

The final DC grounding configuration defined in IEC 60364-1 is the \textit{IT} configuration. The first letter, \textit{I}, indicates that the DC source is isolated from earth, as shown in Fig. \ref{fig:DC_IT}. The second letter, \textit{T}, specifies that exposed conductive parts on the load side must be grounded to earth (yellow ground in Fig. \ref{fig:DC_IT}). 

A key advantage of the \textit{IT} configuration is its high fault tolerance (see Fig. \ref{fig:grounding_spider_plot}). The system can continue operating through a first line-to-ground fault, maintaining power availability \cite{Cuzner_DC_ground}. However, if a second line-to-ground fault occurs, the resulting fault-current path becomes unpredictable and difficult to detect \cite{Cuzner_DC_ground}. Another important advantage of the \textit{IT} configuration is its strong compatibility with different AC grounding schemes. Although future AI data centers are expected to adopt in-facility DC distribution (Fig. \ref{fig:stage2_3}[a] and Fig. \ref{fig:stage2_3}[b]), they will still interface with the MV AC grid. As discussed in \cite{Kumar}, not all DC grounding configurations are compatible with all AC grounding configurations. Because the \textit{IT} configuration isolates the DC source from earth while maintaining a separate load-side earth ground, it can be interfaced with AC systems using arbitrary grounding schemes. Notably, the interaction between AC and DC grounding configurations remains an important and underexplored topic—particularly for future AI data centers that demand extremely high availability. 
\vspace{11pt}
\subsubsection{Summary and Comparison of DC Grounding Schemes}
\begin{figure}[!t]
     \centering
      \includegraphics[width=0.55\linewidth]{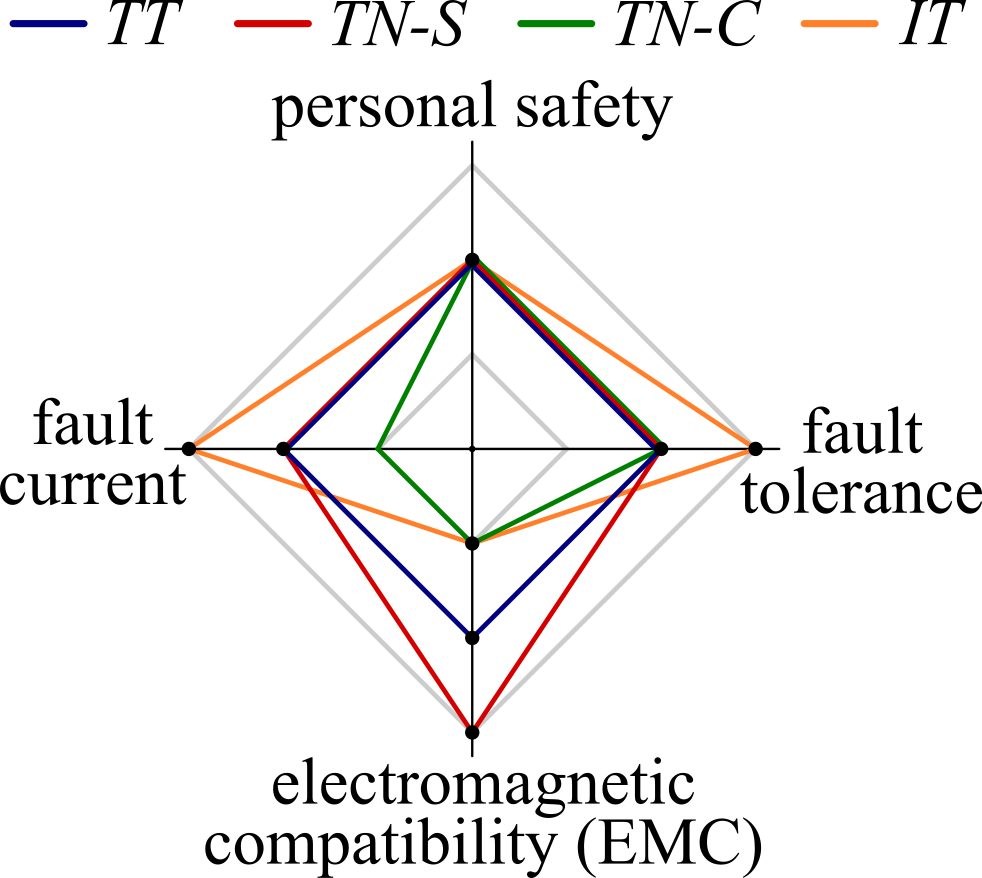}
     \caption{Comparison of different DC grounding schemes.}
     \label{fig:grounding_spider_plot}
 \end{figure}

Grounding is a key design consideration for future AI data centers adopting LV DC distribution, directly affecting safety, EMC performance, and system availability. The grounding configurations reviewed earlier—\textit{TT}, \textit{TN}, and \textit{IT}—present distinct trade-offs for in-facility DC busways, and their comparative performance in terms of personal safety, equipment protection, EMC, and fault tolerance is summarized in Fig.~\ref{fig:grounding_spider_plot}. 

The \textit{TT} configuration grounds the DC source and exposed conductive parts through separate earth paths. Its simplicity and isolation help limit fault propagation but can introduce circulating currents and voltage stress, reducing suitability for large-scale DC distribution in data centers. 

The \textit{TN} configurations (\textit{TN-S}, \textit{TN-C}, and \textit{TN-C-S}) share a common grounding system. \textit{TN-S} uses separate neutral and PE conductors, providing strong EMC performance and high personal safety at the cost of additional wiring. \textit{TN-C} combines these conductors into a PEN conductor, reducing cost but degrading EMC performance. \textit{TN-C-S} offers an intermediate approach, balancing cost and grounding reliability. 

The \textit{IT} configuration isolates the DC source from earth while grounding load-side conductive parts. It provides high fault tolerance, enabling continued operation through the first line-to-ground fault, and is compatible with a broad range of AC grounding schemes—an important advantage as LV DC architectures interface with MV AC utility systems. However, second line-to-ground faults create unpredictable current paths and require effective detection mechanisms. 

As data centers transition toward high-power LV DC distribution (see Fig. \ref{fig:stage2_3}), grounding strategy will remain fundamental to achieving safe, reliable, and EMC-compliant operation. Further research is needed to identify grounding configurations best suited for future AI-driven facilities, particularly as power densities and availability requirements continue to rise. In addition, the interaction between LV DC grounding schemes and the grounding practices of the MV AC grid warrants deeper investigation, especially under fault conditions, to ensure robust and predictable system behavior. 

\subsection{Challenges in Adopting LV DC Distribution: Protection and Standards}

Despite the numerous advantages of LV DC busway architectures, conventional AC distribution remains the dominant approach in data centers \cite{2025_OCP_GOOGLE, NVIDIA800VDC, EVatScale}. A primary barrier to broader adoption of DC distribution is the challenge of protection. The impedance of a DC busway is predominantly resistive, and DC fault currents can rise with extremely high $di/dt$ \cite{Cuzner_DC_ground}. Moreover, unlike AC systems, DC distribution systems lack natural current zero-crossings, making traditional fuses and circuit breakers—whose interruption mechanisms rely on current zero-crossing—ineffective \cite{2243478}. These inherent characteristics significantly complicate fault interruption and remain a major bottleneck for widespread deployment of DC architectures in future AI data centers \cite{2243478, Cuzner_Mag, NVIDIA800VDC}. 

Protection challenges are further compounded by DC-specific safety hazards, particularly those related to arc faults and arc-flash events. DC arcs are more persistent than AC arcs because they are not extinguished by periodic current zero-crossings, necessitating dedicated mitigation strategies \cite{Cuzner_Mag}. Recent research has explored overload and short-circuit protection using semiconductor-based or solid-state circuit breakers (SSCBs) \cite{6374505, 5606638, 5685956, 7855145}, which offer fast interruption capability and the potential for selective coordination. Complementary efforts have focused on developing arc-free connectors, sockets, and load-break DC disconnects suitable for data center operating environments \cite{6374522, 6803624}. Though promising, SSCBs still face challenges related to cost, conduction loss, thermal management, and long-term reliability, all of which must be addressed before large-scale commercial adoption \cite{Cuzner_Mag}. 

In addition to protection, the lack of widely accepted standards for LV DC distribution represents another critical obstacle \cite{EVatScale}. Existing standards have historically been developed for AC systems, and the industry has not yet converged on uniform design practices, protection philosophies, grounding requirements, or equipment ratings for high-power LV DC distribution in data centers. The industrial shift toward LV DC distribution in data centers (see Fig. \ref{fig:stage2_3}) highlights the urgency of establishing such standards \cite{NVIDIA800VDC,EVatScale,2025_OCP_GOOGLE}. Emerging initiatives are beginning to define guidelines for DC busway ratings, connector design, arc-flash mitigation, grounding compatibility, and coordination with MV AC feeders \cite{NVIDIA800VDC,2025_OCP_GOOGLE}. However, these efforts remain in early stages, and substantial work is needed to ensure interoperability, safety uniformity, and long-term reliability across vendors and deployment scales. 

Overall, protection and standardization remain the two most significant challenges hindering the adoption of LV DC distribution in future AI data centers \cite{Cuzner_Mag,NVIDIA800VDC,EVatScale}. Addressing these challenges will require coordinated advancements in fault-interruption technologies, arc mitigation, equipment design, and comprehensive standard development. As power densities continue to rise and DC architectures gain traction, overcoming these barriers will be essential for enabling safe, resilient, and scalable LV DC distribution systems. 

\section{MV-SST Design Challenges for Next Generation Data Center}
\label{section:MV-SST}

The third architectural shift identified in Section~\ref{section:3_stages} is the direct interfacing of data centers with the MV grid, for which MV-SSTs have emerged as a promising enabling technology \cite{NVIDIA800VDC,ocp_diablo400}. Conventional LFT–based MV AC–LV conversion remains a robust and highly mature solution, particularly in facilities employing LV AC distribution. However, as data center architectures evolve toward LV DC distribution (Shift~2) and reduced conversion-stage count (Shift~3), limitations associated with LFT-based substations become increasingly pronounced. In particular, the large footprint and low power density of LFTs pose challenges as facility power scales and space utilization becomes critical. By replacing line-frequency magnetics with medium-frequency isolated conversion, MV-SSTs offer substantially higher power density while enabling enhanced controllability, flexible power routing, and advanced protection capabilities. Reported studies further indicate that, for MV AC–LV DC conversion, MV-SSTs can achieve significantly reduced losses as well as one-third the weight and volume compared with LFT-based solutions\cite{SST_motivation_Proc_IEEE,SST_motivation_IES_mag,SST_motivation_IES_mag_2025,2014_SST_Kolar,SST_advantage}.

However, adopting MV-SSTs for data center infrastructure introduces significant challenges related to the following: 

\begin{itemize}
    \item Power electronics scaling for MV interface
    \item Insulation design
    \item Data-center-rated reliability    
\end{itemize}
Unlike many industrial power-electronic applications, data centers operate continuously under mission-critical workloads and are subject to stringent service-level agreements that permit little or no unplanned downtime. Consequently, power-delivery components are required to achieve reliability and availability levels comparable to, or exceeding, those of conventional utility infrastructure while simultaneously accommodating highly dynamic load profiles and frequent operational transients. In this context, MV-SSTs form the primary electrical interface between the MV grid and the internal power distribution system, making reliability-oriented design a first-order consideration rather than a secondary objective. Accordingly, the following sections review the primary technical challenges associated with MV-SST deployment in AI data centers and summarize representative research directions aimed at enabling their reliable and widespread adoption.



\subsection{Scaling SST for Medium Voltage}

\subsubsection{Equivalent MV-Rated Power Device}
\begin{figure}[!t]
     \centering
      \includegraphics[width=0.8\linewidth]{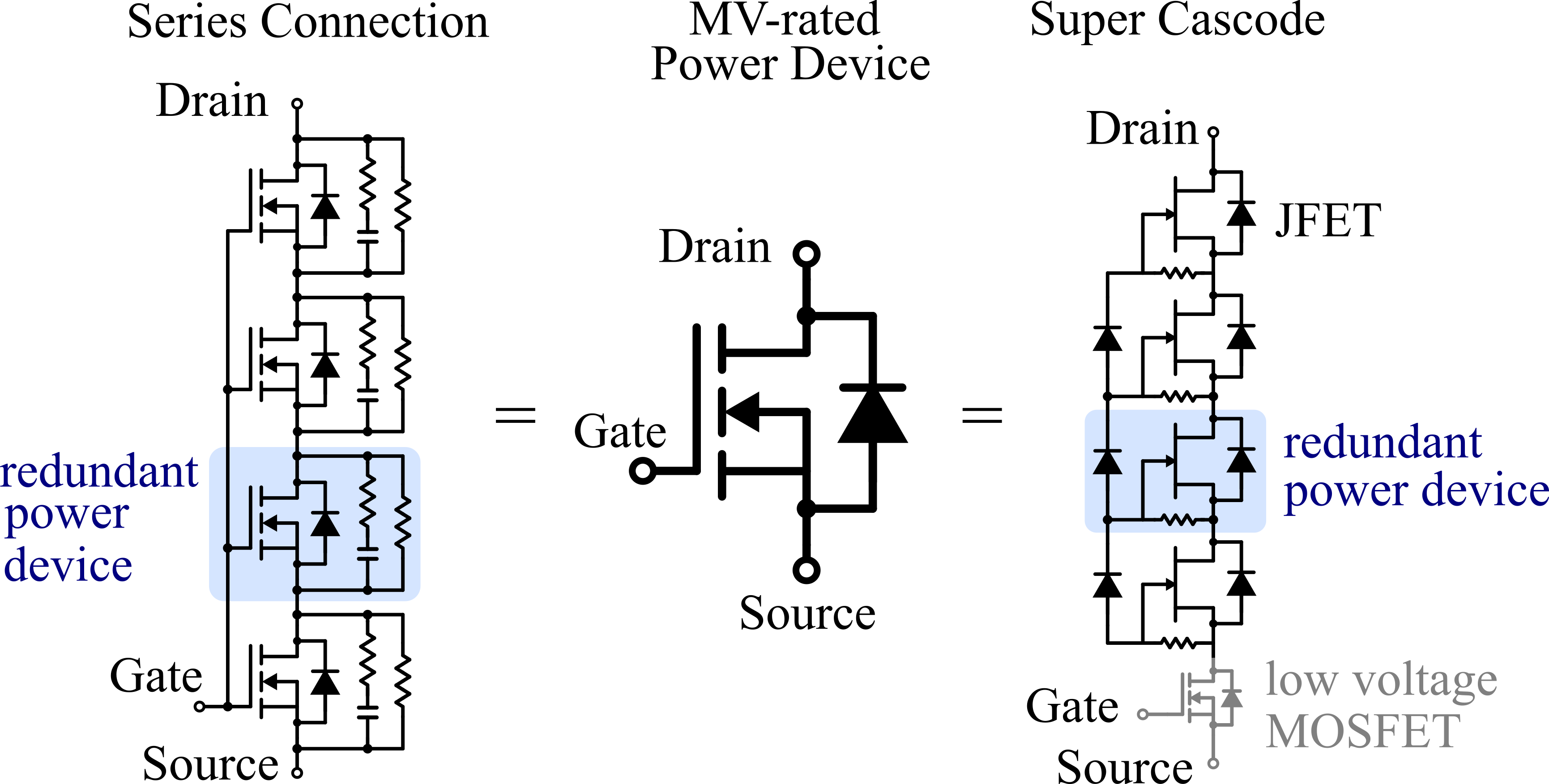}
     \caption{Implementation of equivalent MV-rated power devices.}
     \label{fig:eqvl_MV_device}
 \end{figure}

The first challenge in developing MV-SSTs is scaling power-conversion hardware for direct interfacing with the MV grid. Owing to the limited availability of MV–rated power devices, a class of nonmodular approaches has been widely investigated, in which equivalent MV-rated power devices are constructed at the device or module level and then applied within conventional three-phase converter topologies. The most straightforward nonmodular approach is to realize an equivalent MV-rated power device by stacking multiple lower-voltage devices either through series connection or by employing super-cascode configurations, as illustrated in Fig. \ref{fig:eqvl_MV_device}. These equivalent MV-rated devices can be readily integrated into standard three-phase converter topologies, including two-level, NPC, and matrix-type converters \cite{MV_two_level,7420219,2197867,2258472,NPC_SST_NCSU_2024,2019_Kolar_single_phase_10kV_SiC_PFC}. When three-phase topologies are used, such MV-SST implementations benefit from relatively simple control and do not suffer from pulsating power issues. However, the series-connected nature of the device stack inherently introduces voltage-balancing challenges, requiring additional balancing circuits to ensure uniform voltage sharing among devices, as shown in Fig. \ref{fig:eqvl_MV_device} \cite{SuperMOS, SuperCascode, SuperCascode_15kV_SiC_IGBT,40kV_kolar}.

Beyond voltage balancing, a critical limitation of nonmodular, device-stacked approaches lies in redundancy and fault tolerance. Because the equivalent MV-rated device is formed by a series string of lower-voltage devices, the failure of a single device can compromise the entire stack. To address the stringent reliability requirements of data center applications, redundancy is often introduced through modest overdesign using additional devices within the series string (Fig. \ref{fig:eqvl_MV_device}). In such configurations, fault-tolerant operation fundamentally relies on each lower-voltage device exhibiting a stable short-circuit failure mode (SCFM) upon failure, thereby maintaining current conduction until scheduled maintenance, which may occur on the order of months to a year \cite{Rudy_2023,voltage_bal,fail_short_2023}. SCFM is particularly desirable because it suppresses internal arcing and mitigates catastrophic device rupture associated with open-circuit failures under high-voltage stress \cite{Rudy_2023,SiC_fail_short_2017,HVDC_design_book_2016,fail_short_2019}.

In practice, SCFM is most commonly achieved using press-pack power module technology, where the semiconductor die is directly clamped between metallic interconnects that serve as both electrical terminals and heat sinks \cite{SiC_fail_short_2017,Press_pack_2017}. In silicon-based devices, SCFM arises when device heating leads to silicon melting and the formation of a conductive eutectic alloy between silicon and adjacent metals, typically silver or aluminum \cite{voltage_bal,press_pack_not_explode}. The absence of solder layers and bond wires eliminates dominant failure mechanisms such as delamination and bond-wire lift-off, thereby improving robustness under high surge currents \cite{voltage_bal}. However, these advantages are largely confined to silicon-based technologies. For wide-bandgap (WBG) devices, particularly SiC, the much higher thermal stability, reduced die and contact-pad area, and shorter short-circuit withstand time significantly complicate the realization of reliable and repeatable SCFM behavior \cite{Rudy_2023,SiC_fail_short_2017,fail_short_2017,press_pack_patent}. Although die fracture–induced conductive paths have been observed experimentally, such mechanisms lack predictability and long-term reliability, leaving SCFM implementation for WBG and ultrawide-bandgap (UWBG) devices as an open research challenge.

In parallel with series-connected stacks, recent research has explored advanced nonmodular solutions based on higher-voltage-rated devices and enhanced packaging concepts. Examples include MV-SSTs employing 10 kV–class SiC devices for direct MV-grid interfacing, as well as NPC-based converters combining 10 kV and 15 kV devices for operation at 13.8 kV \cite{10kV_SiC_VT,10kV_SiC_WS}. Beyond this range, 20 kV SiC insulated gate bipolar transistors and UWBG devices with even higher voltage ratings are under active development \cite{10kV_SiC_SST,15kV_SiC_SST}. Although such MV-rated devices can significantly reduce device count and simplify system architecture, their practical adoption is constrained by severe dv/dt and $di/dt$ stresses, which introduce substantial challenges in parasitic control, EMI mitigation, gate-drive design, electric-field management, thermal performance, and reliability assessment \cite{15kV_SiC_SST,20kV_SiC_IGBT}. To address these issues, advanced packaging strategies—such as stacked direct bonded copper substrates, spring-loaded interconnects, cavitated structures, and common-mode noise mitigation techniques—have been proposed, demonstrating measurable reductions in electric-field stress and common-mode noise emission \cite{Tomi_part2,Bipolar_ex1,Bipolar_review}.

\vspace{11pt}
\subsubsection{Modular Approach}
The second approach to scale the SST is to cascade multiple submodules (SMs) with lower voltage ratings. This modular approach distributes the voltage and power among multiple SMs, successfully scaling the voltage of the SST without the need for MV-rated or equivalently MV-rated devices. The modular designs offer inherent redundancy and improved fault containment, which are attractive properties for data center applications requiring high fault tolerance.

\begin{figure}[!t]
     \centering
      \includegraphics[width=0.65\linewidth]{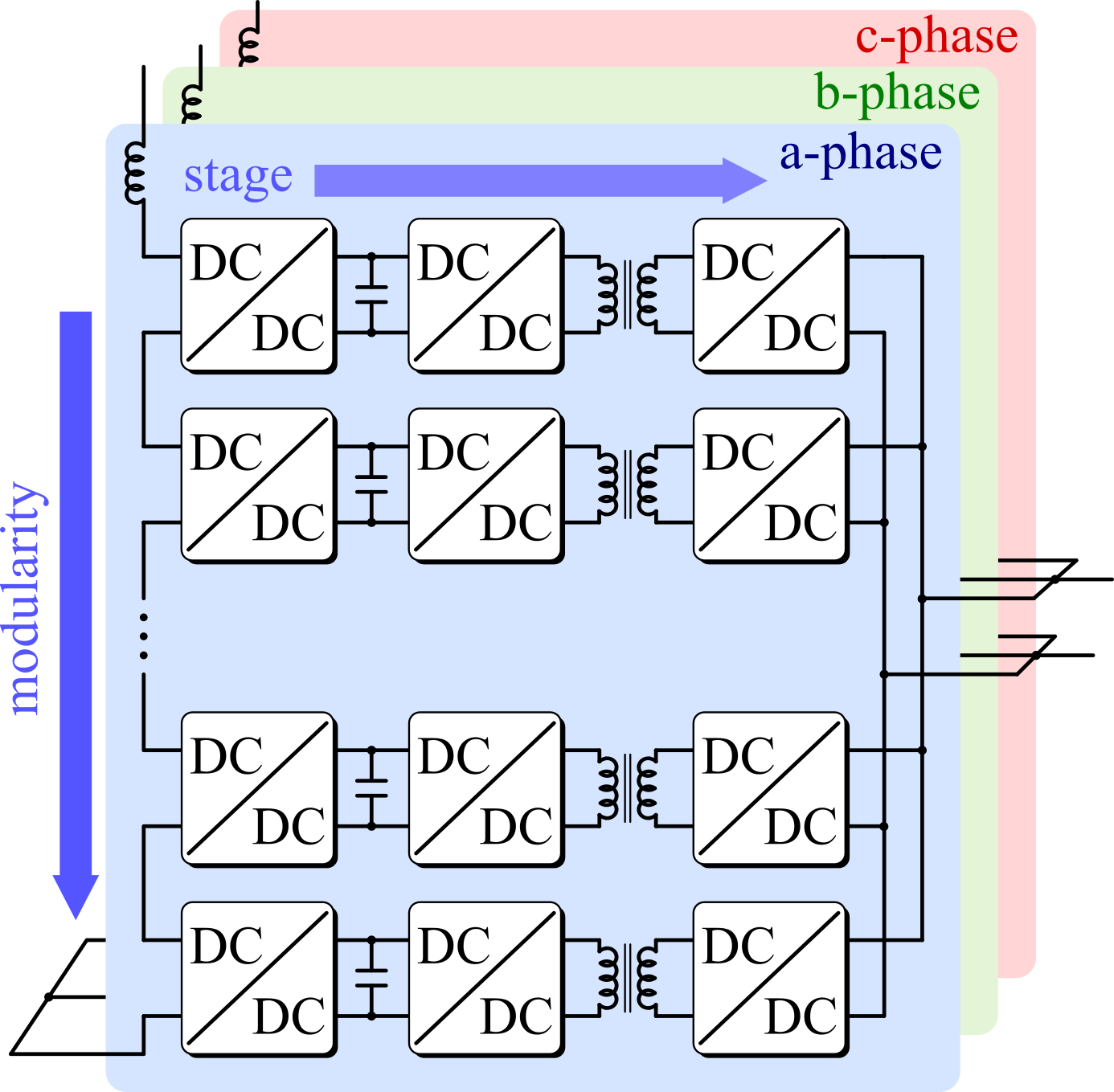}
     \caption{Cascaded H-bridge converter in ISOP configuration for MV-SST application.}
     \label{fig:CHB_ISOP}
 \end{figure}

A classic example of a series connection of converter cells, as shown in Fig. \ref{fig:CHB_ISOP}, is the cascaded H-bridge (CHB) converter, which was patented in the 1970s \cite{McMurray1970}. In principle, one can accommodate virtually any grid voltage by increasing the number of cascaded SMs. Implementing that isolation with high-frequency–transformer-based isolated DC–-DC converters and then tying all of their LV DC outputs together forms an ISOP configuration. This yields the representative modular MV AC-to-LV DC SST topology (e.g., the three-phase configuration illustrated in Fig. \ref{fig:CHB_ISOP}).

This modular configuration is one of the most well-accepted topologies in the industry when developing MV-SST. Delta Electronics, which developed a 13.2 kV/400 kW SST for XFC applications, reported that the system was designed based on this architecture (i.e., a modular, two-stage ISOP configuration) \cite{Delta_SST}. Similarly, Hitachi also announced the development of a 6.6 kV/350 kW SST for EV fast-charging, which was likewise based on the same architecture \cite{Hitachi2022}. In addition, \cite{3462920} presented the design, control, and protection of a 13.2 kV/1 MVA SST using the same approach—although laboratory constraints restricted experimental validation to a 200 kW load. Furthermore, \cite{Optimal_cell_number} investigated how to optimize the number of cells and device voltage ratings for efficiency, power density, and reliability, and \cite{10977184} provided a comparative analysis of the material usage of SSTs versus LFT-based solutions. Additionally, depending on the DC-link location within an ISOP architecture, the system can realize either an isolated front-end or an isolated back-end configuration \cite{Kolar_SST_2016}.

\begin{figure*}[!t]
    \centering
    \begin{subfigure}{0.31\textwidth}
        \centering
        \includegraphics[width=1\linewidth]{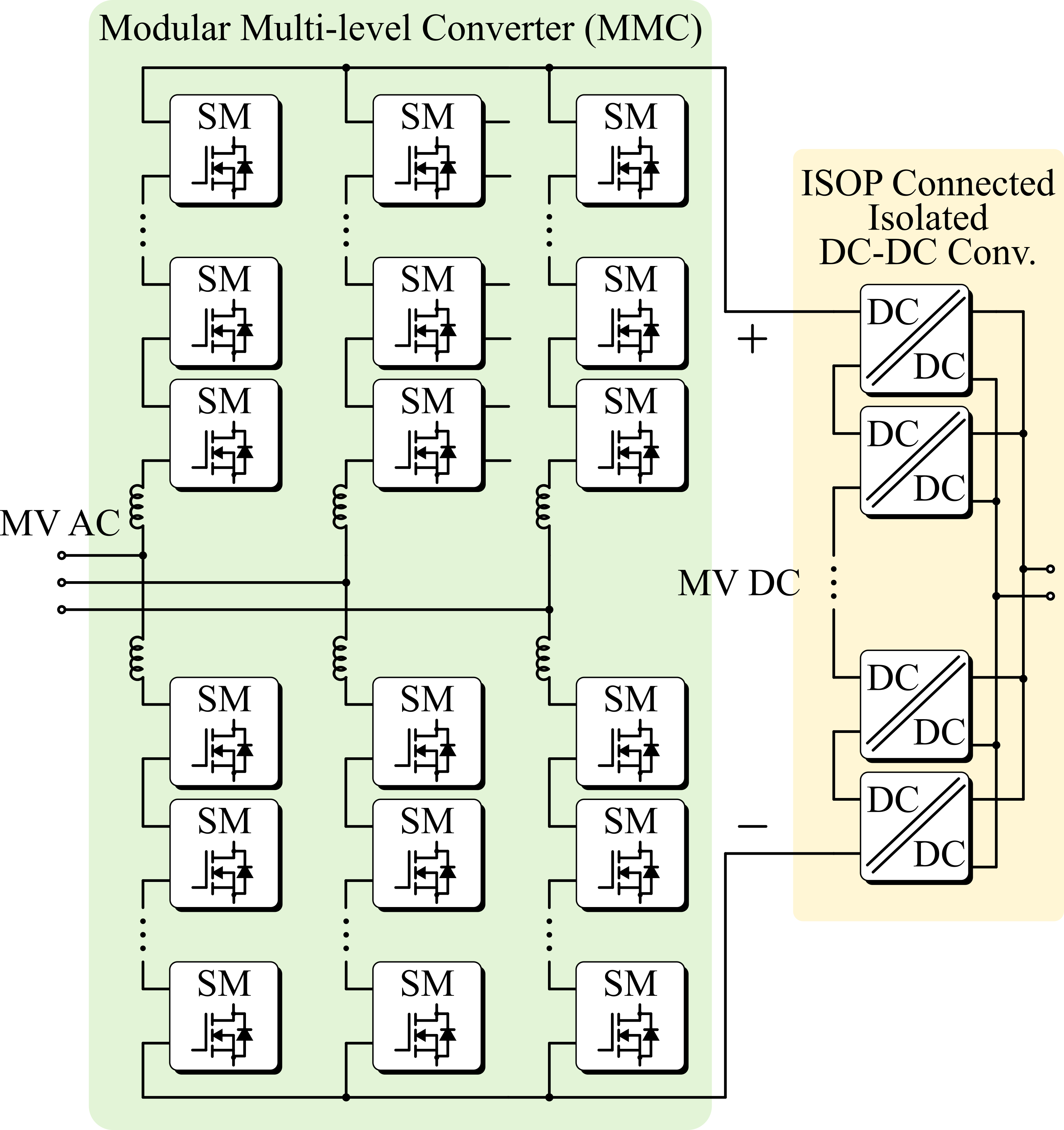}
        \caption{}
    \end{subfigure}
    \hfill
    \begin{subfigure}{0.35\textwidth}
        \centering
        \includegraphics[width=1\linewidth]{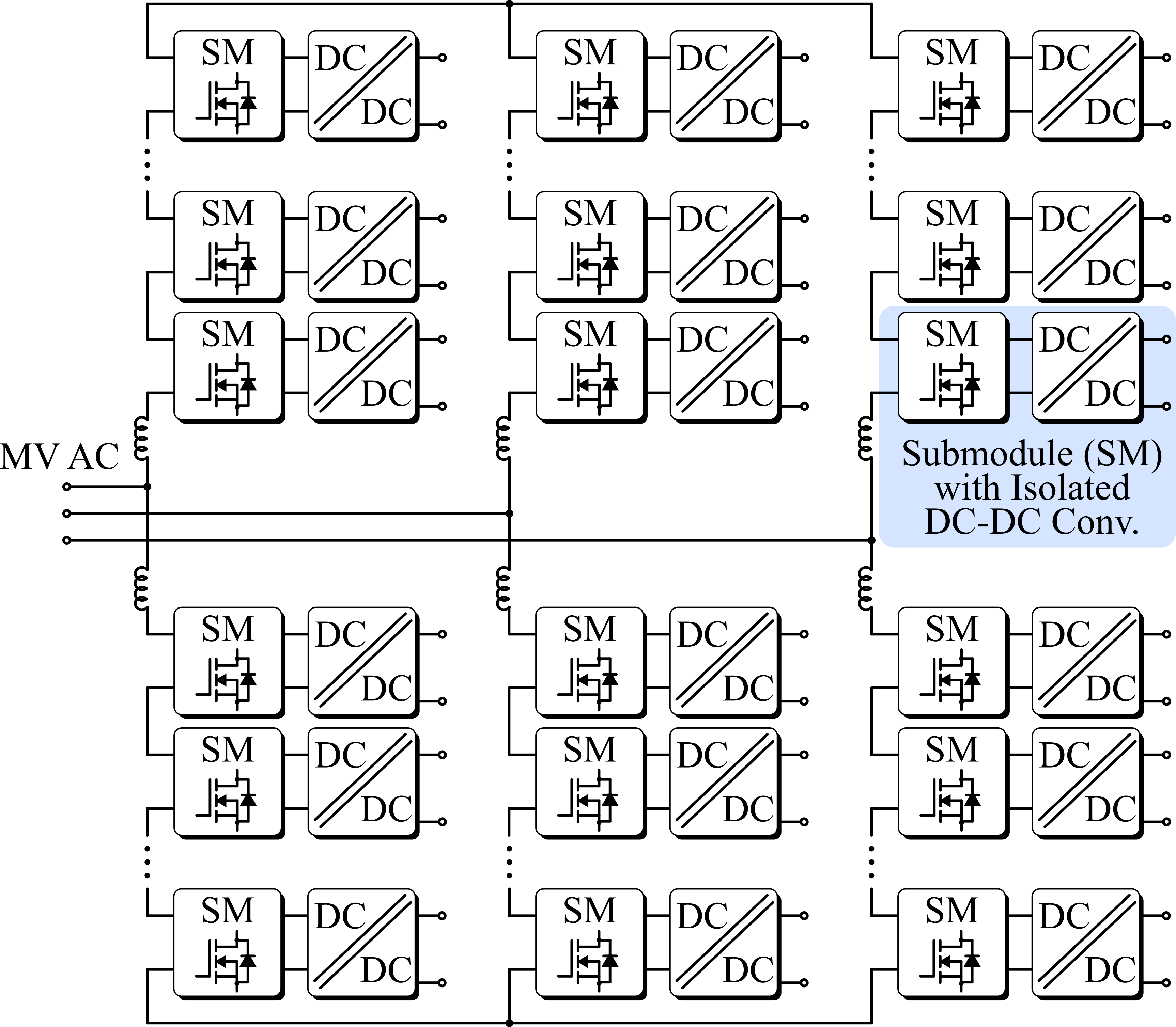}
        \caption{}
    \end{subfigure} 
    \hfill
    \begin{subfigure}{0.31\textwidth}
        \centering
        \includegraphics[width=1\linewidth]{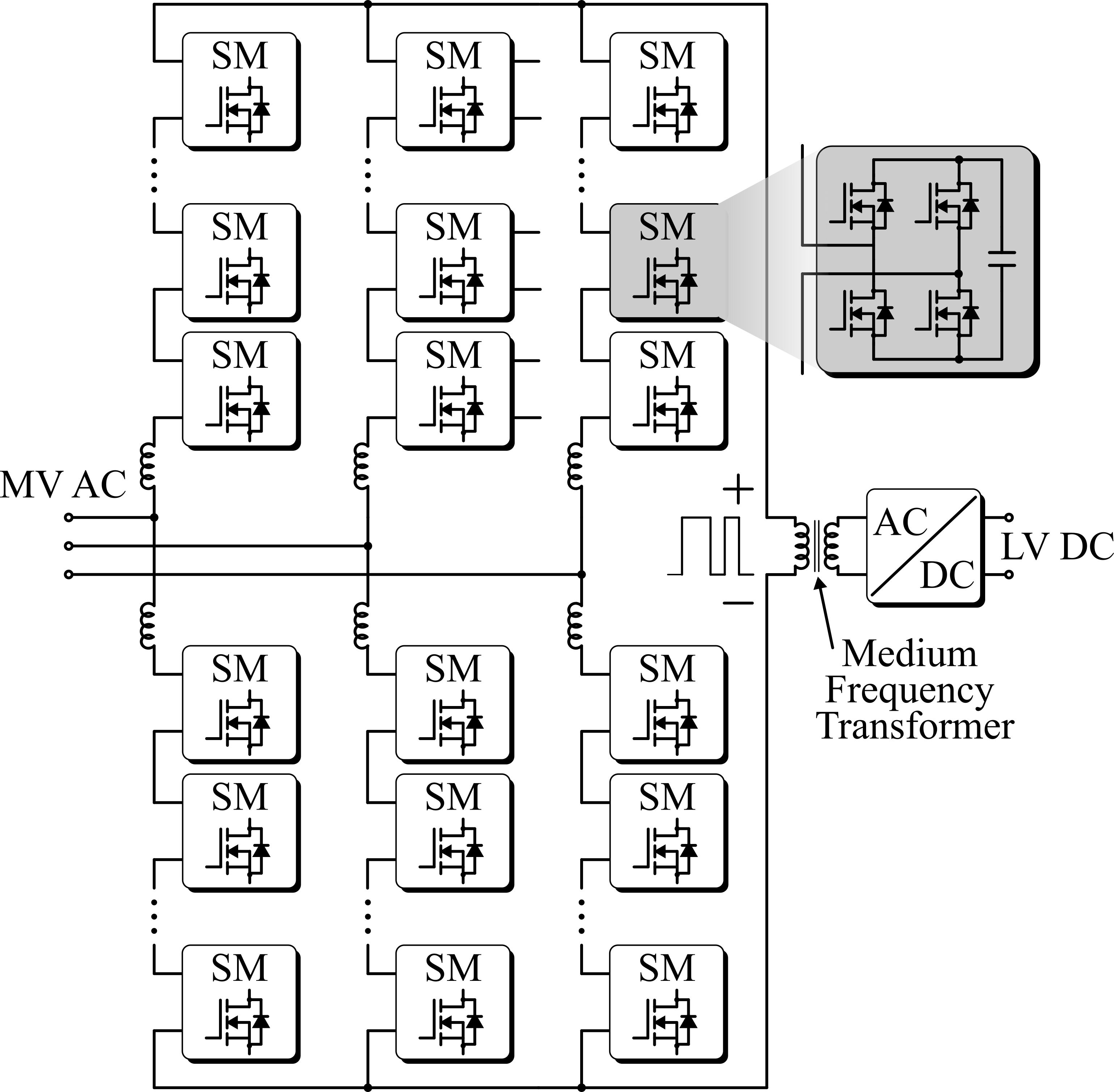}
        \caption{}
    \end{subfigure} 
    \caption{Various configurations of MMC-based MV-SSTs. (a) MMC-based MV-SST with MV DC intermediate bus voltage followed by a cascaded isolated DC--DC converter. (b) MMC-based MV-SST with isolated SMs. (c) MMC-based MV-SST with a single isolation point. Full-bridge SMs should be used for the MMC to achieve switching voltage at the primary side of the MFT.}
    \label{fig:MMC_MV_SST}
\end{figure*}

Aside from the CHB converter, the modular multilevel converter (MMC) also has been widely investigated for MV-SST \cite{MMC_Marquardt,10667423,9034866,6631810,2923355,3065235,8590678,Fault_tolerant_cell,3534427}. Because of the inherent modularity and flexibility of MMC-based MV-SSTs, various configurations can be considered. The simplest approach, as illustrated in Fig. \ref{fig:MMC_MV_SST}(a), is the two-stage approach in which the MMC converts MV AC to MV DC, and the isolated DC--DC converter in ISOP configuration converts the MV DC to LV DC, which is suitable for the data center busway \cite{10667423}, \cite{9034866}, and \cite{6631810}. Another approach is to integrate an isolated DC--DC converter into each MMC SM, as shown in Fig. \ref{fig:MMC_MV_SST}(b) \cite{2923355,3065235,8590678,Fault_tolerant_cell}.

In the case of MMCs with full-bridge SMs (see the gray shaded area in Fig. \ref{fig:MMC_MV_SST}[c]), a direct AC--AC conversion can be achieved, as shown in Fig. \ref{fig:MMC_MV_SST}(c). The AC output of the MMC is fed to a medium-frequency transformer (MFT), followed by a DC--AC converter, as depicted in Fig. \ref{fig:MMC_MV_SST}(c), enabling a more compact single-stage MMC-based MV-SST \cite{3534427,3504337,3343330,Kolar_13084}. However, in single-stage configurations, the coupling between the grid low-frequency and the MFT excitation limits the achievable switching frequency and increases current ripple \cite{3534427}. To mitigate this issue and verify efficient high-frequency transformer operation, \cite{3534427} developed a single-stage MMC-based charger rated at 10 kV/400 V and proposed a frequency-decoupled modulation combined with soft-switching control. 


\subsection{Insulation Design of MV-SST}
Based on existing studies, insulation-related challenges in MV-SSTs can be broadly classified into two categories: (i) insulation coordination under grid-originated overvoltages and (ii) insulation stress induced by converter-generated high dv/dt and mixed-frequency voltage waveforms. The first category involves insulation coordination in grid-connected MV-SSTs. According to IEEE C62.82.1, insulation coordination refers to the systematic selection of insulation strength to withstand expected system overvoltages while considering environmental conditions and protection-device characteristics \cite{insul_coordination_def,insul_coordination_divan}. In MV-SSTs, insulation must withstand both normal operation and fault or protective events. This task is complicated by the absence of a dedicated insulation-coordination framework for MV power-electronic converters, requiring designers to interpret and combine existing requirements to define insulation levels, clearances, creepage distances, and qualification tests \cite{insul_corrdination_Cuzner}.

A key difficulty arises from the presence of multiple internal insulation interfaces—such as module-to-ground, transformer primary-to-secondary, and terminal regions—in addition to exposure to grid-originated surges \cite{insul_coordination_divan}. Under lightning impulse conditions, internal insulation stress is governed not only by terminal peak voltage, but also by surge attenuation and redistribution through protective devices and parasitic elements. As a result, localized insulation overstress may occur even when terminal-level clamping appears adequate, particularly in cases where protection levels are mismatched with internal insulation strength \cite{insul_coordination_divan}. Verification further complicates this issue because conventional lightning impulse tests were developed for passive insulation systems and may not accurately reflect internal stress in protected MV-SSTs, motivating protection-aware insulation-coordination approaches based on internal stress modeling \cite{insul_coordination_divan}.

The second category of challenges originates from converter-induced insulation stress. The use of WBG devices introduces fast-rising, repetitive switching voltage waveforms with high dv/dt, which propagate through insulation systems in MFTs, busbars, and power-module interfaces \cite{PWM_MFT_insulation, MV_busbar, MV_PM_review}. Under such conditions, electric-field stress depends not only on voltage magnitude, but also on waveform shape and frequency content. During fast voltage transitions, voltage sharing within windings and insulation stacks becomes dominated by parasitic capacitances, leading to nonuniform electric-field distribution and localized field enhancement at terminations and interfaces. Studies on MV MFTs for SST applications show that peak electric-field stress is strongly influenced by winding geometry, shielding, and insulation layering rather than nominal insulation thickness alone \cite{MFT_insulation,3094674}. In addition, repetitive PWM excitation and mixed-frequency voltage stress affect partial discharge behavior and introduce dielectric losses that can contribute to insulation heating under continuous operation \cite{PWM_MFT_insulation, MV_dielc_loss_Guillod}.

To address these converter-induced challenges, recent research has emphasized structural electric-field control rather than limiting switching speed at the system level. Approaches such as electrostatic shielding, graded insulation layers, and controlled termination geometries have been proposed to reduce local field concentration under fast-switching excitation \cite{MFT_insulation,9887908}. Complementary optimization-based methodologies jointly consider insulation thickness and electric-field distribution, whereas waveform-aware evaluation techniques—including PWM-specific PD characterization and frequency-domain dielectric-loss modeling—enable more accurate assessment of insulation robustness under converter-representative operating conditions \cite{3094674,PWM_MFT_insulation,3566999,MV_dielc_loss_Guillod,10186238}.

Overall, insulation design in MV-SSTs is a multidimensional problem that requires simultaneous consideration of grid-originated overvoltages and converter-induced electric-field stress. Current research trends point toward integrated insulation-design frameworks that combine protection-aware insulation coordination, electric-field-controlled insulation structures, and waveform-aware evaluation methods to enable reliable, high-power-density MV-SST operation in data center environments.

\subsection{Protection of MV-SST}
Protection constitutes a critical challenge for MV-SSTs because these systems replace passive LFTs with semiconductor-based power conversion stages that exhibit limited fault withstand capability while simultaneously being required to meet the stringent availability and safety requirements of data center infrastructures. Unlike conventional transformers, which primarily rely on upstream protection and benefit from substantial thermal and magnetic fault tolerance, MV-SSTs must actively detect, limit, and interrupt fault currents within timescales compatible with semiconductor safe-operating limits. Consequently, protection becomes a first-order design consideration in MV-SSTs rather than an auxiliary system function \cite{SST_protection_2017_Kolar}.

For data center applications, this challenge is further intensified by the requirement to contain faults without unnecessary service interruption. MV-SSTs must handle severe grid-side contingencies while preventing internal faults from propagating into downstream LV DC distribution networks, where low impedance and large stored energy can significantly aggravate fault severity. Prior studies identified MV short-circuit faults, overcurrent events, and overvoltage disturbances as the dominant protection scenarios that fundamentally distinguish SST protection from that of passive LFT-based systems \cite{SST_protection_2017_Kolar}.

On the MV AC side, protection is constrained by the combination of high fault energy, rapid current rise, and converter-limited short-circuit withstand capability, rendering conventional breaker-based protection insufficient. This limitation has motivated coordinated protection strategies that integrate fast converter blocking, current limiting, and surge protection, and has driven system-level co-design of protection with converter topology and control \cite{SST_protection_2017_Kolar,MV_MFT_DongDong,3082033}. On the LV DC side, short-circuit faults are dominated by extremely fast current transients owing to low distribution impedance and distributed capacitance. Although MV-SSTs do not function as DC circuit breakers, their internal energy storage and control actions critically shape the initial fault response, indicating that LV DC fault behavior must be intentionally managed through MV-SST architecture and control rather than delegated entirely to downstream protective devices \cite{2575139,3056205}.

In addition, internal faults—such as semiconductor device failures and insulation breakdowns—introduce further protection challenges. In modular MV-SST architectures, effective fault containment relies on protection partitioning at the module or cell level to prevent cascading failures. Although such partitioning enables localized fault handling and controlled shutdown or partial operation, it requires careful coordination of protection thresholds and control responses across multiple modules \cite{MV_MFT_DongDong,3066908}.

Overall, effective protection in MV-SSTs for data center applications requires a multilayered approach that combines fast converter-level actions, controlled energy management, modular fault isolation, and coordination with upstream MV and downstream LV DC protection systems. Rather than being realized by a single protective device or algorithm, MV-SST protection emerges from the co-design of topology, control, and protection philosophy, with fault containment and service continuity as primary objectives.

\subsection{Fault Tolerance}
Fault tolerance in MV-SSTs refers to the ability to maintain acceptable power delivery after internal faults or selected external disturbances rather than relying solely on protective shutdown. This capability is particularly important for data center applications, where controlled degradation and rapid recovery are preferred over complete service interruption. Because MV-SSTs are typically realized using modular, multistage power-electronic architectures, fault tolerance is achieved through a combination of redundancy, topology-level fault containment, and fault-aware control rather than a single protective element.

Across the literature, fault-tolerant MV-SST design converges on three complementary approaches:

\begin{itemize}
    \item Redundancy-based architectures enabling post-fault reconfiguration \cite{2964950,MV_MFT_DongDong,3082033,10567148}
    \item Topology-embedded fault containment that limits fault propagation \cite{3153801,3056205}
    \item Control-based reconfiguration that preserves safe operation under degraded conditions \cite{2964950,3294186}
\end{itemize}

For data center applications, where availability, maintainability, and graceful degradation are paramount, effective fault-tolerant MV-SSTs are increasingly treated as a co-design problem spanning redundancy planning, converter topology, and fault-aware control.\\

\subsubsection{Redundancy-Based Fault Tolerance}
Redundancy is one of the most established approaches to fault tolerance in MV-SSTs. In cascaded or ISOP-based architectures, redundant modules enable post-fault reconfiguration in which remaining healthy cells redistribute voltage and power to maintain operation with limited derating. Prior studies demonstrated that effective fault tolerance depends on coordinated post-fault voltage and power balancing, whereas large-scale hardware demonstrations confirm that modular redundancy enables localized fault handling and system-level reconfiguration at MV ratings \cite{2964950,MV_MFT_DongDong,3082033}. However, redundancy introduces trade-offs among availability, cost, efficiency, and insulation overhead, motivating careful redundancy planning.\\

\subsubsection{Fault-Tolerant Submodule and Topology Design}
Fault tolerance can also be embedded directly into the converter topology or SM structure, reducing reliance on spare modules. In modular SSTs derived from MMC concepts, appropriate SM configurations can provide inherent fault-blocking or fault-isolation capability. Prior work demonstrated that topology-level features can enable uninterrupted power transfer or limit fault propagation during DC-side faults, simplifying fault management at the expense of increased conduction losses and control complexity \cite{3153801,3056205, FT_DAB}.\\

\subsubsection{Control-Based Fault Tolerance and Post-Fault Reconfiguration}

Independent of hardware redundancy, control reconfiguration plays a central role in fault-tolerant MV-SST operation. Following a module or cell fault, control strategies must redistribute voltage and power while maintaining device and insulation limits. Existing studies show that post-fault power-balancing control and modulation adaptation can preserve operation in cascaded SST architectures, reducing the need for dedicated bypass hardware and highlighting that fault tolerance must be co-designed with control rather than guaranteed by hardware alone \cite{2964950,3294186}.\\

\subsubsection{Fault Ride-Through as a Fault-Tolerance Mechanism}

Beyond internal faults, fault tolerance in MV-SSTs may include ride-through of selected external disturbances, provided semiconductor and insulation constraints are respected. Prior work demonstrated that dedicated fault-mode control can enable uninterrupted power transfer during MV DC short-circuit events and emphasized that ride-through strategies must be bounded by clearly defined electrical and thermal limits to avoid compromising long-term reliability \cite{3153801,3056205}. These results indicate that fault tolerance can be realized as a controlled operating mode rather than solely as a shutdown response.

\section{Remaining Challenges and Future Directions}
\label{section:Emerging}

The challenges discussed thus far reveal both the complexity and the opportunity inherent in future AI data center design. Several emerging technologies—including digital twins, advanced hot-swap systems, and cross-domain technology transfer—offer promising paths forward but require further investigation. Section \ref{section:Emerging} consolidates these forward-looking themes, highlighting directions where research and industry development can accelerate the readiness of next-generation power architectures.

\subsection{System Integration Challenges Across Architectural Shifts}
Whereas Sections \ref{section:LV-IBC}–\ref{section:MV-SST} examined the challenges associated with individual power electronics building blocks, the deployment of these technologies at scale introduces additional challenges that arise from their interaction across architectural layers. These challenges are not attributable to a single converter, distribution system, or grid interface; rather, they emerge from the integration of rack-level, facility-level, and grid-level power delivery architectures in future AI data centers. 

A central integration challenge is the coordination of protection and fault management across multiple voltage domains. As AI data centers adopt LV DC distribution at both rack and facility levels and incorporate actively controlled MV grid interfaces, fault behavior becomes distributed across interconnected subsystems. Ensuring selective fault isolation and predictable system response requires coordination among rack-level converters, facility DC protection devices, and grid-facing power-electronic interfaces. Such coordination cannot be achieved through isolated design of individual components and remains an open system-level problem. 

Another integration challenge stems from the propagation of highly dynamic AI workload-induced power transients across architectural boundaries. Rapid load changes originating at the IT level can influence facility-level DC buses and interact with grid-facing converters, particularly as data centers transition toward more actively controlled grid interfaces. Addressing these interactions requires control and energy-management strategies that consider the collective behavior of the power delivery chain rather than local regulation objectives alone. 

\begin{figure}[!t]
    \centering
    \includegraphics[width=1\linewidth]{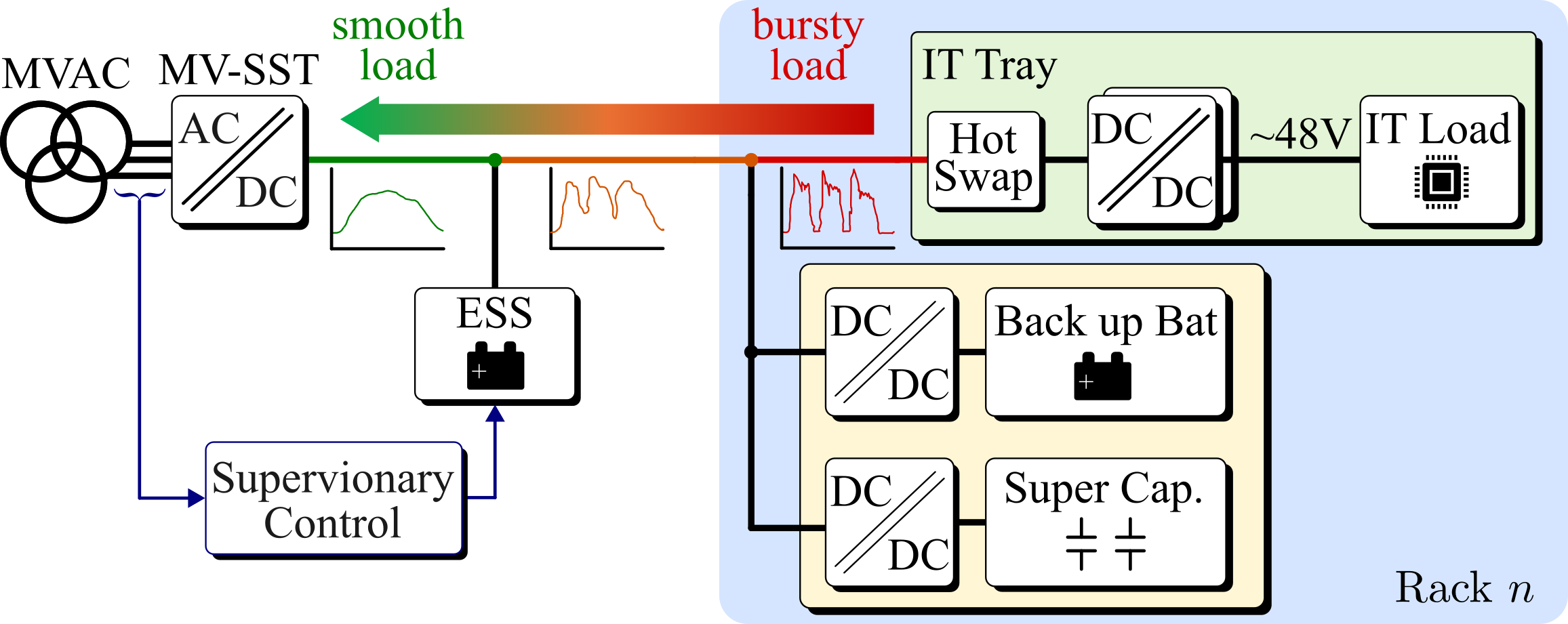}
    \caption{Supervisory control and load-smoothing features of next-generation AI data centers.}
    \label{fig:ESS_distribution}
\end{figure}

Thermal and reliability considerations further illustrate the need for cross-layer integration. Decisions made at one architectural level—such as voltage selection, converter placement, or cooling topology—can significantly influence thermal stress and lifetime margins elsewhere in the system. Without coordinated electrothermal design and operational strategies, local optimizations may inadvertently limit overall system reliability or scalability. 

Finally, the readiness of next-generation AI data center power architectures is constrained by the lack of unified standards and qualification practices spanning rack-, facility-, and grid-level technologies. Differences in grounding philosophy, protection requirements, and validation criteria across voltage domains complicate system integration and slow large-scale deployment, even when individual technologies demonstrate promising performance in isolation. 

These integration-driven challenges motivate the future research directions discussed in the following subsections, which focus on unresolved issues that emerge at the boundaries between architectural layers rather than within individual technology building blocks. 

\subsection{Hot-Swap and Protection at Rack--Facility Interface}
The transition toward LV in-rack DC distribution introduces new challenges for hot-swap operation that extend beyond the design of individual rack-level converters. Although hot-swap functionality has long been employed in conventional 48 V architectures, its implementation at elevated DC voltage levels fundamentally changes the interaction between rack-level power electronics and facility-level distribution and protection systems. 

In future AI data centers, hot-swappable compute trays equipped with LV-PDN converters interface directly with facility-level LV DC busways. As a result, hot-swap events—such as insertion, removal, or fault isolation of IT trays—can influence voltage stability, fault currents, and protection coordination beyond the rack boundary. Unlike traditional architectures, where hot-swap disturbances are largely confined to the rack, LV DC environments couple rack-level events more tightly to facility-level DC buses and upstream protective devices. 

A key integration challenge lies in coordinating rack-level hot-swap behavior with facility-level DC protection schemes. The higher stored energy associated with LV DC busways, combined with fast electronic fault-interruption devices, creates operating scenarios in which protection actions at different hierarchical levels may interact unintentionally. Ensuring selective isolation of faulty or removed racks without triggering upstream protection or propagating disturbances to adjacent racks remains an open system-level problem. 

Grounding and fault-reference considerations further complicate hot-swap operation at elevated DC voltages. As discussed in Section \ref{section:DC_Distribution}, different grounding philosophies for LV DC distribution can significantly influence fault paths and transient behavior. Hot-swap mechanisms must therefore be compatible not only with rack-level converter requirements but also with the grounding and protection strategy adopted at the facility level, reinforcing the need for coordinated design across architectural layers. 

Addressing these challenges requires re-examining hot-swap not as an isolated rack-level function, but as a coordinated interaction between rack-integrated power converters, facility-level DC distribution, and protection infrastructure. Future research directions include system-level modeling of hot-swap transients, co-design of rack- and facility-level protection schemes, and development of standardized interfaces that enable predictable and safe hot-swap behavior in high-power LV DC data center environments. 

\subsection{Advanced Packaging}

Conduction, switching, and resistive losses within power conversion, power distribution, and computation systems in data center facilities are primarily transformed into thermal energy. If thermal energy is not relocated effectively from its origin, device and system temperatures can rise significantly, limiting performance and negatively affecting reliability. To that end, the facility-wide cooling schedule for present-day and future data center architectures must account for each of these systems within the center. Each of these subsystems within the center will have differing thermal energy extraction strategies and needs depending on their power level and packaging architecture, increasing the complexity of the overall data center thermal management network. Furthermore, the incorporation of increased power density components into each respective system for data centers will make extracting thermal energy even more challenging because of the inherent increase in heat flux resulting from miniaturization \cite{W6JPD}. 

Packaging will play a critical role in transitioning to new power conversion and distribution architectures (new designs and adaptations updating legacy data centers) in addition to emerging topologies for computing within data center compute racks. The packaging strategies for power and compute systems within data centers will need to be an elegant blend of enabling new and evolving power schedules/loads (e.g., high voltage levels) and improving cooling capacity \cite{W6JPD}. For example, moving to higher voltages at power conversion and distribution sections of the data center facility can help to improve efficiency with compact designs \cite{10849971, LAD2023119726}. 

However, these design strategies can also increase the risk of partial discharge and dielectric breakdown \cite{10568122, 10770218, 10908700}. This problem is potentially worsened with the integration of WBG and UWBG devices because they enable reduced package sizes and high-voltage operation \cite{W6JPD}, introducing an architecture in which large voltage gradients can be present across shorter electrical isolation gaps. 

With these decreasing sizes and higher voltage gradients, traditional packaging materials will also likely need to be replaced or phased out of emerging designs. For example, traditional encapsulating silicone gels used in power modules have an increased partial discharge probability with increasing temperature and increased partial discharge frequency with decreasing gap distances, largely driven by material properties and defects within the silicone structure \cite{10770218}. Therefore, when integrating new technologies to enhance data center performance, power module packaging strategies enabling new layouts will also need to incorporate new materials and fabrication techniques to curb concerns about electrical-isolation-related failures while still promoting heat dissipation to curb concerns related to thermally accelerated failures. 

Additionally, packaging approaches that enable systems to become more compact (such as heterogeneous integration) co-locate heat-generating components, leading to difficulty extracting heat without the use of expensive materials (e.g., diamond) and/or complex strategies for heat dissipation and relocation (e.g., multimaterial substrates, thermal spreading augmentation features, multimaterial cooling assemblies, and highly integrated cooling) \cite{LAD2023119726, 9119164, 10387492, 10443930, 9236179, 10938321}. Many of these emerging thermal management and packaging strategies are in their developmental phase and show promising performance but have also specified the need for further development to reduce the risk of failure with respect to electrical isolation before system integration \cite{LAD2023119726}. These considerations emphasize the need for co-designing power packages and compute packages for electrical and thermal performance to mitigate reliability risks while maximizing power density. An example of optimizing many of these parameters while integrating emerging technologies has previously been executed for GaN HEMT devices \cite{9360269}.

\subsection{Thermal--Electrical Co-Design Across Rack-Level Power Delivery and IT Loads}
As AI data centers transition toward higher in-rack DC voltages and increased power density, thermal management challenges increasingly intersect with power electronics design decisions. Considering cooling technologies and heat-transfer mechanisms at the facility and rack levels, an unresolved challenge lies in the coordinated thermal–electrical co-design of rack-integrated power delivery components and IT loads. 

In next-generation 1 MW+ scale compute racks, power converters such as LV-IBCs are co-located with high-power xPUs and memory devices on tightly constrained IT trays. This physical proximity creates strong coupling between electrical operating conditions and local thermal environments. Electrical design choices—including switching frequency, voltage level, and power density—directly influence heat generation and spatial heat distribution; thermal conditions, in turn, affect converter efficiency, reliability margins, and insulation lifetime. These interactions cannot be fully addressed through independent optimization of power electronics or cooling subsystems. 

The adoption of liquid-based cooling strategies further amplifies the need for coordinated design. Shared or closely coupled cooling loops for IT loads and rack-level power electronics introduce new trade-offs among electrical performance, thermal stability, and fault containment. For example, thermal transients induced by dynamic AI workloads may simultaneously stress power converters and computing devices, requiring joint consideration of electrical control, thermal response, and reliability objectives at the rack level.

From a system-integration perspective, thermal–electrical co-design also influences facility-level scalability and operational flexibility. Decisions made at the rack level—such as allowable junction temperatures, cooling interface requirements, and power density targets—propagate upward to affect facility cooling capacity, redundancy planning, and maintenance practices. Without coordinated design methodologies, local improvements in rack-level power density may impose disproportionate burdens on facility-level thermal infrastructure or limit overall system availability.

Future research directions therefore include the development of integrated electrothermal modeling frameworks that capture interactions between rack-level power electronics and IT cooling systems, as well as control and design approaches that explicitly account for coupled electrical, thermal, and reliability constraints. Addressing these challenges is essential for enabling scalable, high-density AI compute racks while maintaining predictable thermal behavior and long-term reliability across the power delivery chain.

\subsection{Digital Twin of Data Center for System-Level Modeling}

As AI data center architecture evolves, operational challenges increasingly extend beyond static power converter or component design considerations. Although digital twins have been widely explored for data center energy optimization and thermal management, an emerging research direction lies in their application in power electronics-aware operation, control, and reliability assessment across architectural layers \cite{bb,cc,dd,ee}.

In future AI data centers, the power delivery system comprises tightly coupled rack-level converters, facility-level DC distribution, energy storage/generation systems, and grid-interfacing power electronics. The dynamic behavior of this system is strongly influenced by AI workload dynamics, fast power transients, and coordinated control actions across multiple voltage domains. Digital twins that integrate electrical, thermal, and control models provide a framework for capturing these interactions in real time, enabling predictive analysis that extends beyond steady-state efficiency metrics.

From a power-electronics perspective, digital twins can enable stress-aware operation by estimating the electrical and thermal loading of power converters under realistic AI data center workload conditions. This capability is particularly relevant for future AI data centers employing higher-voltage power delivery architectures in which insulation aging, thermal cycling, and switching-stress accumulation increasingly govern the long-term reliability of power-electronics components across the power delivery chain. By integrating physics-based models with real-time measurements and data-driven techniques, digital twins can support control strategies that explicitly balance performance objectives with reliability margins.

Digital-twin-assisted control also enables improved coordination between data centers and the power grid. As data centers increasingly adopt actively controlled grid interfaces and participate in grid-support functions, real-time visibility into internal power-electronics states becomes critical. Digital twins can support supervisory control layers that anticipate the system-level impact of workload changes, energy-storage dispatch, and grid disturbances, thereby improving operational predictability at both the facility and grid levels \cite{ee,ff,gg}.

Despite this progress, several challenges remain before digital twins can be fully integrated into AI data center power systems. These include the development of scalable and validated multidomain models, seamless integration with existing control and protection frameworks, and the definition of appropriate abstractions that balance model fidelity with computational tractability. Addressing these challenges is essential for enabling digital-twin-assisted, reliability-aware operation of next-generation AI data center power architectures.


\begin{figure}[!t]
    \centering

    \begin{subfigure}{\linewidth}
        \centering
        \includegraphics[width=\linewidth]{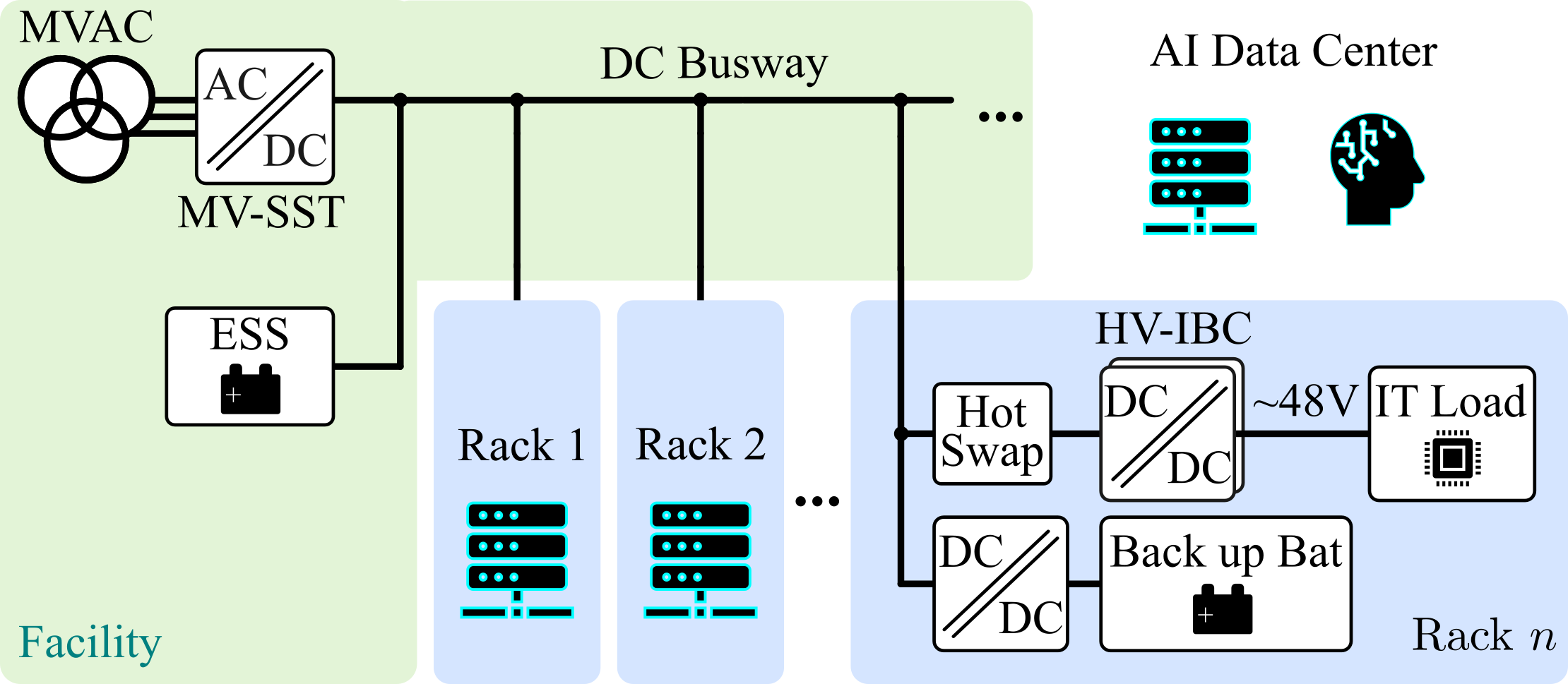}
        \caption{Next generation AI data center}
    \end{subfigure}

    \vspace{0.5em} 

    \begin{subfigure}{\linewidth}
        \centering
        \includegraphics[width=\linewidth]{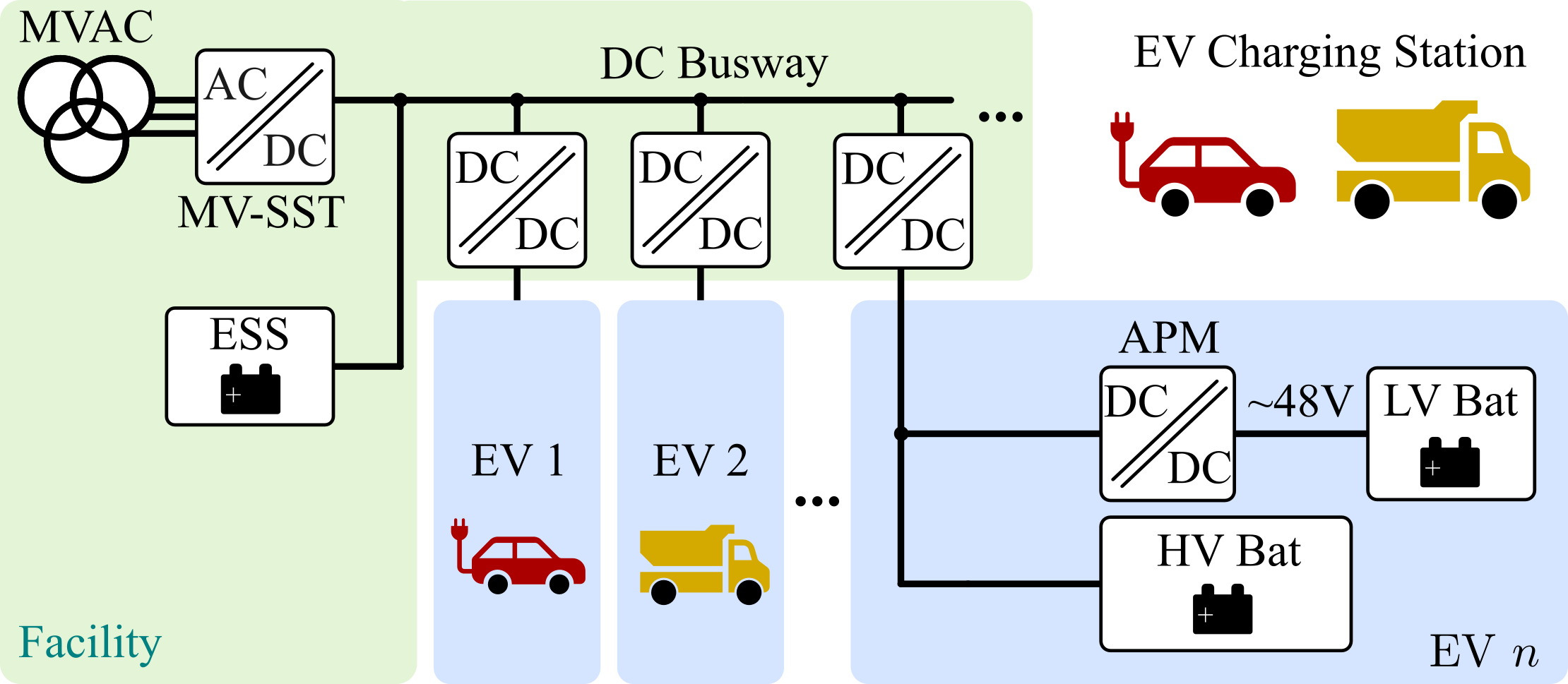}
        \caption{EV charging stations with in-facility LV DC busway}
    \end{subfigure}

    \caption{Architectural similarities between future AI data centers and EV charging stations.}
    \label{fig:AI_EV_comparison}
\end{figure}

\subsection{Cross-Domain Technology Transfer}
Section~\ref{section:3_stages} identified three key technology building blocks for future AI data centers: ultra-compact DC--DC converters with high voltage conversion ratios, in-facility DC distribution, and MV-SSTs. Although the corresponding architectural transitions (i.e., Shifts~1--3) are relatively new to the data center industry, comparable voltage levels, power ratings, and system functions have already been widely deployed and commercialized in other power-electronics-intensive domains. Leveraging cross-domain technology transfer therefore represents an important pathway for accelerating the development and deployment of next-generation AI data center power architectures.

Among potential reference domains, the transportation sector provides a particularly relevant example. Power technologies originally developed for EVs and fast-charging infrastructure operate at similar DC voltage levels and have matured rapidly in recent years. In particular, EV power train and charging systems have transitioned from 400~V~DC architectures toward 800~V~DC systems to enable higher power density, improved efficiency, and faster power delivery~\cite{800V_EV}. This evolution has fostered a robust technology ecosystem encompassing WBG power devices, high-frequency magnetics, connectors, protection components, and safety standards, many of which are directly applicable to future AI data centers~\cite{NVIDIA800VDC,ocp_diablo400}.

Fig.~\ref{fig:AI_EV_comparison} illustrates the architectural similarities between a future AI data center (Fig.~\ref{fig:AI_EV_comparison}[a]) and a state-of-the-art EV charging station (Fig.~\ref{fig:AI_EV_comparison}[b])~\cite{NVIDIA800VDC,ocp_diablo400,DC_EV_Charging_Station}. Both systems interface directly with the MV AC grid through MV-SSTs and employ in-facility LV DC distribution networks with storage/generation systems to smooth peak power demand and support grid interaction. In AI data centers, the in-facility LV DC busway voltage is stepped down through LV-PDN converters to supply IT loads, whereas auxiliary power modules perform an analogous function in EV charging systems.

Although many of the functional and technical building blocks required for next-generation AI data centers—including LV-PDN converters, DC distribution, and MV-SSTs—have already been explored in the context of EV fast-charging systems, direct technology transfer is not always straightforward. Data centers impose distinct requirements related to continuous operation, hot-swap capability, fault tolerance, EMC, and long-term reliability under sustained high usage. Consequently, assessing technology readiness for AI data center applications requires careful evaluation of which cross-domain solutions can be adapted with minimal modification and which demand application-specific redesign.

Overall, cross-domain technology transfer offers a valuable reference for accelerating innovation in AI data center power delivery, but it must be guided by a clear understanding of technology readiness and application-specific constraints. Continued cross-sector collaboration among the data center, transportation electrification, and power systems communities will be essential for translating mature power-electronics technologies into reliable, scalable infrastructure capable of supporting future AI workloads.

\section{Conclusion}
\label{Section:Conclusion}

The rapid emergence of AI workloads is transforming data center power delivery from both a technological and architectural perspective. Traditional 48 V rack systems, multistage AC distribution, and LFT interfaces are no longer adequate to support megawatt-scale compute racks and gigawatt-scale facilities. This review highlighted three foundational architectural shifts now reshaping next-generation AI data centers: the move toward high-voltage in-rack DC distribution, the adoption of facility-level LV DC busways, and the direct interfacing of data centers with the MV grid through MV-SSTs. These shifts establish clear technology building blocks—ultracompact isolated DC-DC converters, robust DC distribution systems, and reliable MV-SSTs—each of which presents unique challenges in efficiency, power density, safety, insulation, protection, and standardization.

At the rack level, the transition to 800 V-class architectures requires LV-IBCs with unprecedented voltage conversion ratios, power density, and EMI performance. Magnetics design, high-frequency operation, and isolation requirements remain the principal bottlenecks. At the facility level, LV DC distribution offers significant efficiency gains and tighter integration with battery energy storage, but its widespread adoption is hindered by the absence of mature DC protective devices, grounding schemes, and industry standards. At the grid interface, MV-SSTs promise transformative improvements in footprint, controllability, and efficiency; however, their long-term reliability—particularly under high dv/dt and medium-frequency insulation stress—must be substantially improved before replacing LFTs at scale.

Beyond these electrical considerations, AI data centers introduce new system-level behaviors shaped by highly dynamic IT power profiles. These fluctuations require coordinated operation of MV-SSTs, active front-end converters, rack- and facility-level energy storage, and supervisory control strategies capable of ensuring both internal power quality and external grid stability. As data centers evolve into active grid participants, they will increasingly provide ancillary services such as fast frequency response, peak shaving, and voltage support—further blurring the boundary between load and distributed energy resources.






\section{Acknowledgment}
This material is based upon work supported by the US Department of Energy (DOE), Office of Critical Materials and Energy Innovation, under contract number DE-AC05-00OR22725. The authors would like to thank Fernando Salcedo of DOE for his support and guidance.

\bibliographystyle{IEEEtran}
\bibliography{References,Ref_HV-IBC, Ref_DC_Distribution, Ref_MV_SST, Ref_Grid_Interactive}

\vspace{-5mm}
\begin{IEEEbiography}[{\includegraphics[width=1in,height=1.25in,clip,keepaspectratio]{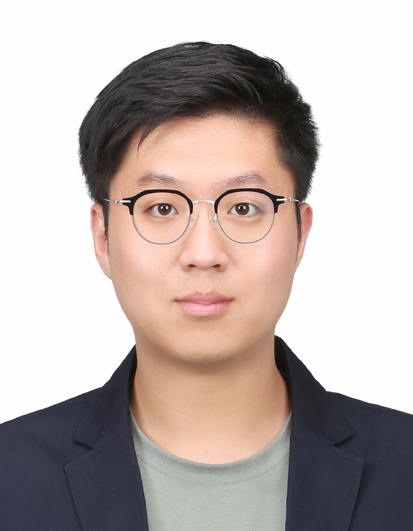}}]{Sangwhee Lee} $(\text{Member, IEEE})$ received a B.S.
degree in electrical engineering from the Korea University, Seoul, South Korea, in 2018 and M.S. and Ph.D. degrees in electrical and computer engineering from the University of Wisconsin-Madison, Madison, WI, USA, in 2020 and 2024, respectively.

Since 2024, he has been an R\&D Associate in the Energy Science and Technology Directorate at Oak Ridge National Laboratory, Oak Ridge, TN, USA. His research interests include power electronics and their advanced controls, reliability, grid-connected converters, and electromagnetic interference. 

Dr. Lee was a winner of the IEEE Power Electronics Society (PELS) Ph.D. Thesis Talk (P3 Talk) in 2025. He was the recipient of the Best Paper Award at the 2025 IEEE/AIAA Transportation Electrification Conference and Electric Aircraft Technologies Symposium (ITEC+EATS) in 2025, Novotny Power Engineering Award from the University of Wisconsin-Madison in 2023, and two IEEE student competition awards in 2021.
\end{IEEEbiography}

\vspace{-4mm}
\begin{IEEEbiography}[{\includegraphics[width=1in,height=1.25in,clip,keepaspectratio]{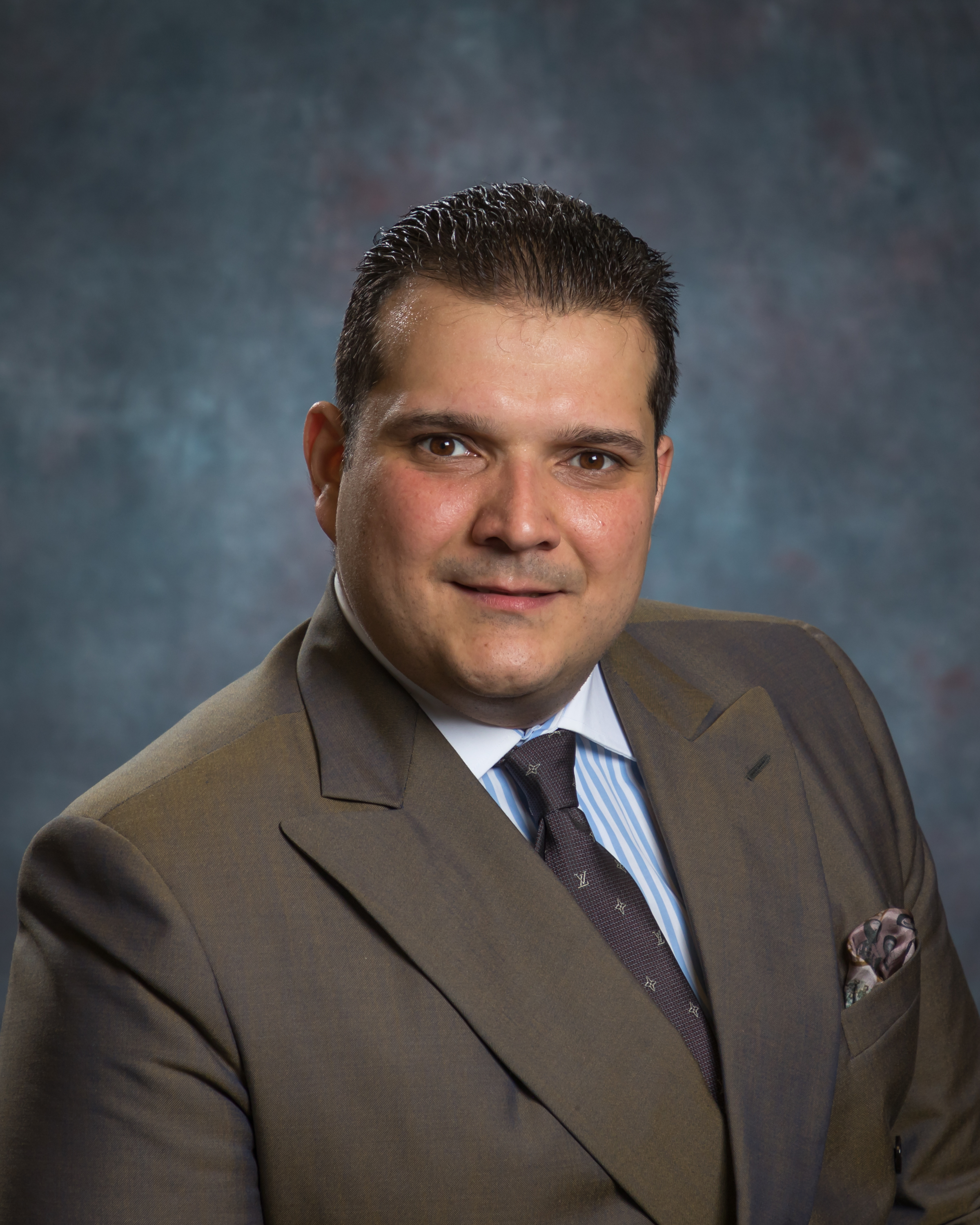}}]{Rafal P. Wojda} 
(Senior Member, IEEE) received B.S. and M.S. degrees (summa cum laude)
in electronics engineering from the Institute of Radioelectronics and Multimedia Technology at Warsaw University of Technology, Poland, in 2007 and 2009, respectively. 
He completed his Ph.D. in electronics at Wright State University, Dayton, OH, USA, in 2012, with a focus on microwave engineering, VLSI, and nanotechnology. 

From 2012 to 2018, Dr. Wojda was a Senior Scientist at the ABB Corporate Research Center in Krakow, Poland, where he led the development and optimization of magnetic components for power conversion systems. 

Since 2018, he has been a R\&D Staff Member at Oak Ridge National Laboratory (ORNL). His research focuses on high-frequency and medium-voltage magnetic components, with emphasis on loss modeling, thermal management, and insulation design. 

Dr. Wojda has authored numerous peer-reviewed publications and has received several distinctions, including the IET Power Electronics Premium Award for Best Paper in 2014. He is an alumnus of both the Fulbright Program and the US Department of State Exchange Program.

He is actively involved in the professional community as Chair of the IEEE East Tennessee PELS/PES Joint Chapter and serves as Secretary of the IEEE Power Electronics Society Technical Committee on Design Methodologies (TC10).
\end{IEEEbiography}

\vspace{-4mm}
\begin{IEEEbiography}[{\includegraphics[width=1in,height=1.25in,clip,keepaspectratio]{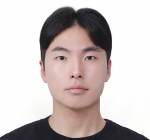}}]{Cheol-Hee Jo} $(\text{Member, IEEE})$ received B.S., M.S., and Ph.D. degrees in electrical engineering from Chonnam National University, Gwangju, South Korea, in 2019, 2021, and 2025, respectively. 

He is currently a Postdoctoral Research Associate with the Vehicle Power Electronics Research Group at Oak Ridge National Laboratory, Oak Ridge, TN, USA. His research interests include AC--DC and DC--DC power converters, wireless power transfer systems, and integrated wired/wireless charging architectures for electric vehicles. He also works on real-time simulation, hardware-in-the-loop simulation (HILS), and power hardware-in-the-loop (PHIL) techniques for renewable energy systems and power grid applications.
\end{IEEEbiography}

\vspace{-4mm}
\begin{IEEEbiography}[{\includegraphics[width=1in,height=1.25in,clip,keepaspectratio]{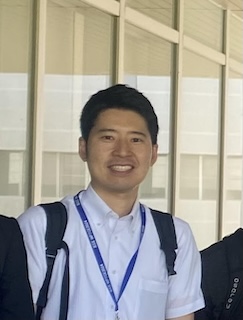}}]{Shuntaro Inoue} $(\text{Senior Member, IEEE})$ received a B.S. degree in engineering (robotics) from Osaka University, Suita, Japan, in 2011, an M.S. degree in engineering (sensor and actuator) from the University of Tokyo, Tokyo, Japan, in 2013, and a Ph.D. degree in electrical engineering from Utah State University, Logan, UT, USA, in 2023.
From 2013 to 2025, he was with Toyota Central R\&D Labs, Inc., Nagakute, Japan, where he developed high-power-density DC–DC and AC–DC converters and initiated research on dynamic wireless power transfer systems. He is currently an R\&D Staff Member in the Power Electronics Group at Oak Ridge National Laboratory, Oak Ridge, TN, USA, and a Fellow of the University of Tennessee–Oak Ridge Innovation Institute. He has authored more than 20 peer-reviewed publications and holds over 10 US patents.
His research focuses on advanced power electronics and wireless power transfer for transportation electrification, including high-power-density converters and neural network applications in power electronics design.
\end{IEEEbiography}

\vspace{-5mm}
\begin{IEEEbiography}[{\includegraphics[width=1in,height=1.25in,clip,keepaspectratio]{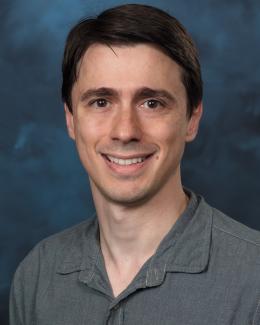}}]{Pedro Ribeiro} $(\text{Member, IEEE})$ received a B.Sc. degree in electrical engineering from the Federal University of Mato Grosso do Sul (UFMS), Brazil, in 2009 and an M.Sc. degree in electrical engineering from UFMS in 2012. He earned a Ph.D. degree in electrical engineering from the Federal University of Itajubá (UNIFEI), Brazil, in 2019. 

After completing his Ph.D., he gained experience in both industry and academia in power-electronics systems. In 2023, he joined Oak Ridge National Laboratory as a member of the Electric Drives Research Group. His research interests include high-performance AC drives, power module packaging, and applications of artificial intelligence.
\end{IEEEbiography}

\vspace{-4mm}
\begin{IEEEbiography}[{\includegraphics[width=1in,height=1.25in,clip,keepaspectratio]{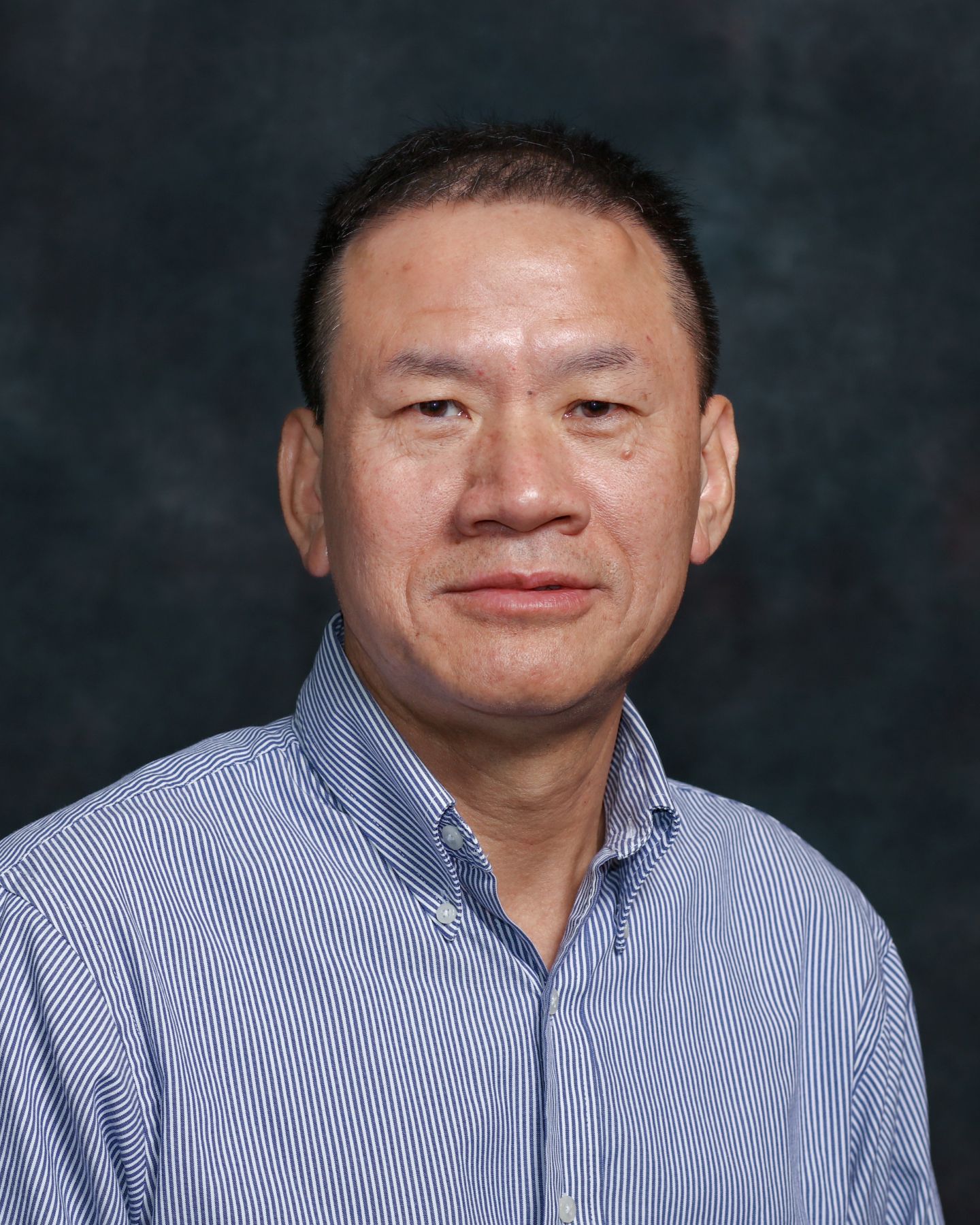}}]{Gui-Jia Su} $(\text{Senior Member, IEEE})$ (M'94--SM'01) received B.S., M.S., and Ph.D. degrees in 1985, 1989, and 1992, respectively, all in electrical engineering.

From 1992 to 1995, he was an assistant professor at Nagaoka University of Technology, Japan. From 1995 to 1998, he was with Sanken Electrical Co., Ltd., Japan, where he engaged in research and development of uninterruptible power supplies, sensorless PM motor drives, and power factor correction for single- and three-phase rectifiers. In 1998, he began working at the Power Electronics and Electric Machinery Research Center at Oak Ridge National Laboratory as a research scientist and is currently a distinguished member of the R\&D staff. His research interests include DC--DC converters, inverters, wired and wireless battery chargers, and traction motor drives for electric vehicle applications.

Dr. Su is a Battelle distinguished inventor and a recipient of the US Department of Energy Vehicle Technologies Office Distinguished Achievement Award in 2019.
\end{IEEEbiography}

\vspace{-5mm}
\begin{IEEEbiography}[{\includegraphics[width=1in,height=1.25in,clip,keepaspectratio]{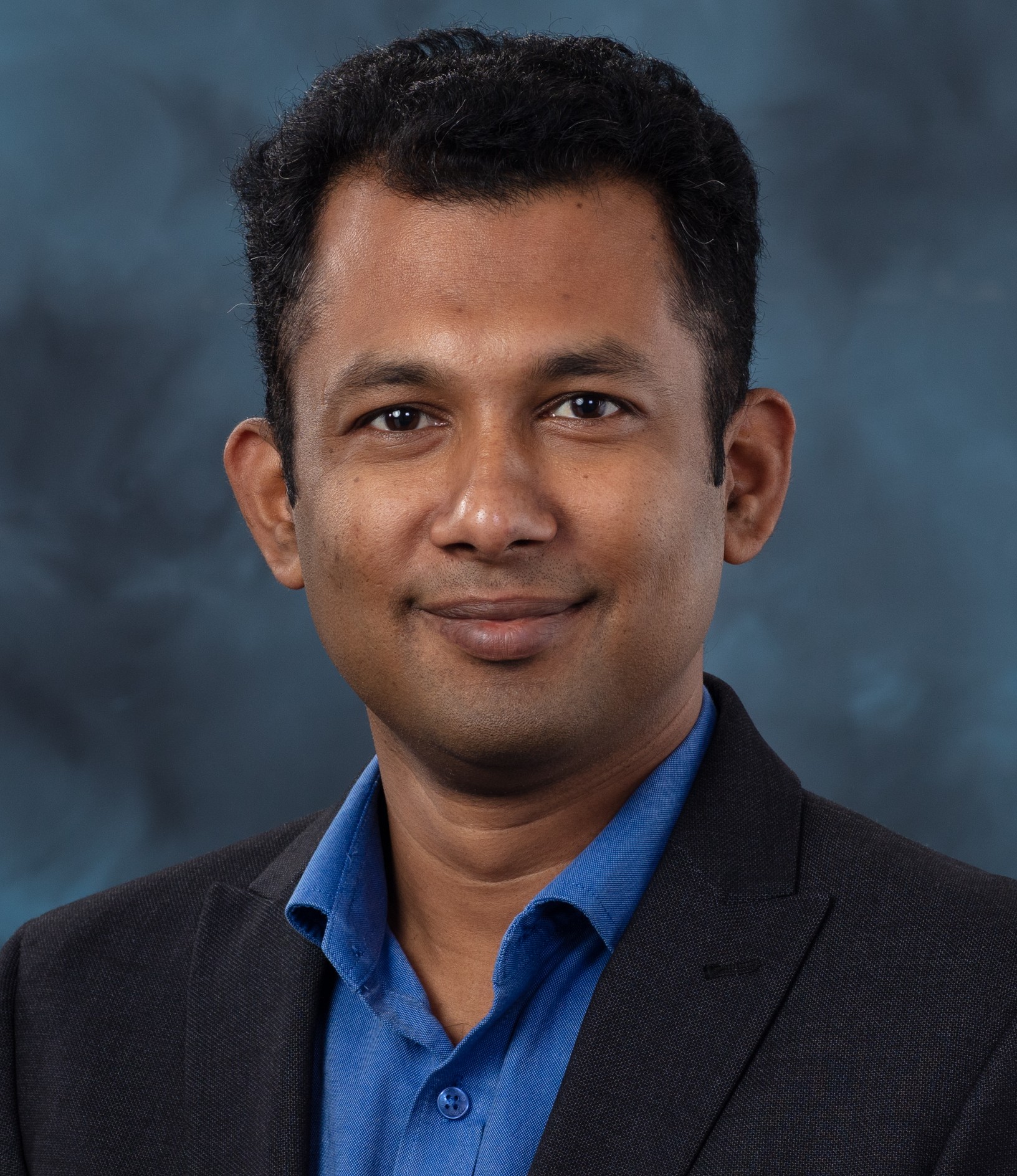}}]{Mostak Mohammad} $(\text{Senior Member, IEEE})$ received his B.Sc. degree in 2009 in electrical and electronics engineering from the Bangladesh University of Engineering and Technology (BUET), Bangladesh, and earned his Ph.D. in electrical engineering in 2019 from the University of Akron, Akron, OH, USA.

Dr. Mohammad is currently working as a Senior R\&D Staff Member in the National Transportation Research Center at Oak Ridge National Laboratory. He has been working on high-power (11 kW to 750 kW) poly-phase stationery and dynamic wireless charging systems. His research interests include high-fidelity multi-physics design and optimization, magnetic materials, electromagnetic shielding, and AI/ML in power electronics research.
\end{IEEEbiography}

\vspace{-5mm}
\begin{IEEEbiography}[{\includegraphics[width=1in,height=1.25in,clip,keepaspectratio]{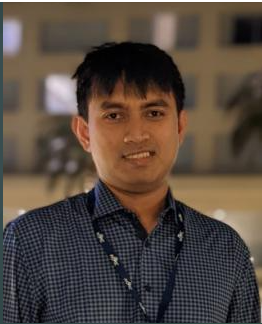}}]{Himel Barua} $(\text{Member, IEEE})$ received his B.S. degree in mechanical engineering from the Bangladesh University of Engineering and Technology (BUET), his M.S. degree from the University of Akron, and his Ph.D. in mechanical engineering from The University of Akron, Akron, OH, in 2021.

Himel Barua is an R\&D Associate in the Electric Drive Research Group, where his work focuses on thermal management system development, multiphysics simulation, and mechanical structural and vibration analysis of power electronics and electric machines. His research interests include the application of CFD and FEA techniques to power electronics, electric machines, chemical reactors, and manufacturing systems. He also has strong interests in mechatronic systems development and embedded system programming. 
\end{IEEEbiography}

\vspace{-5mm}
\begin{IEEEbiography}[{\includegraphics[width=1in,height=1.25in,clip,keepaspectratio]{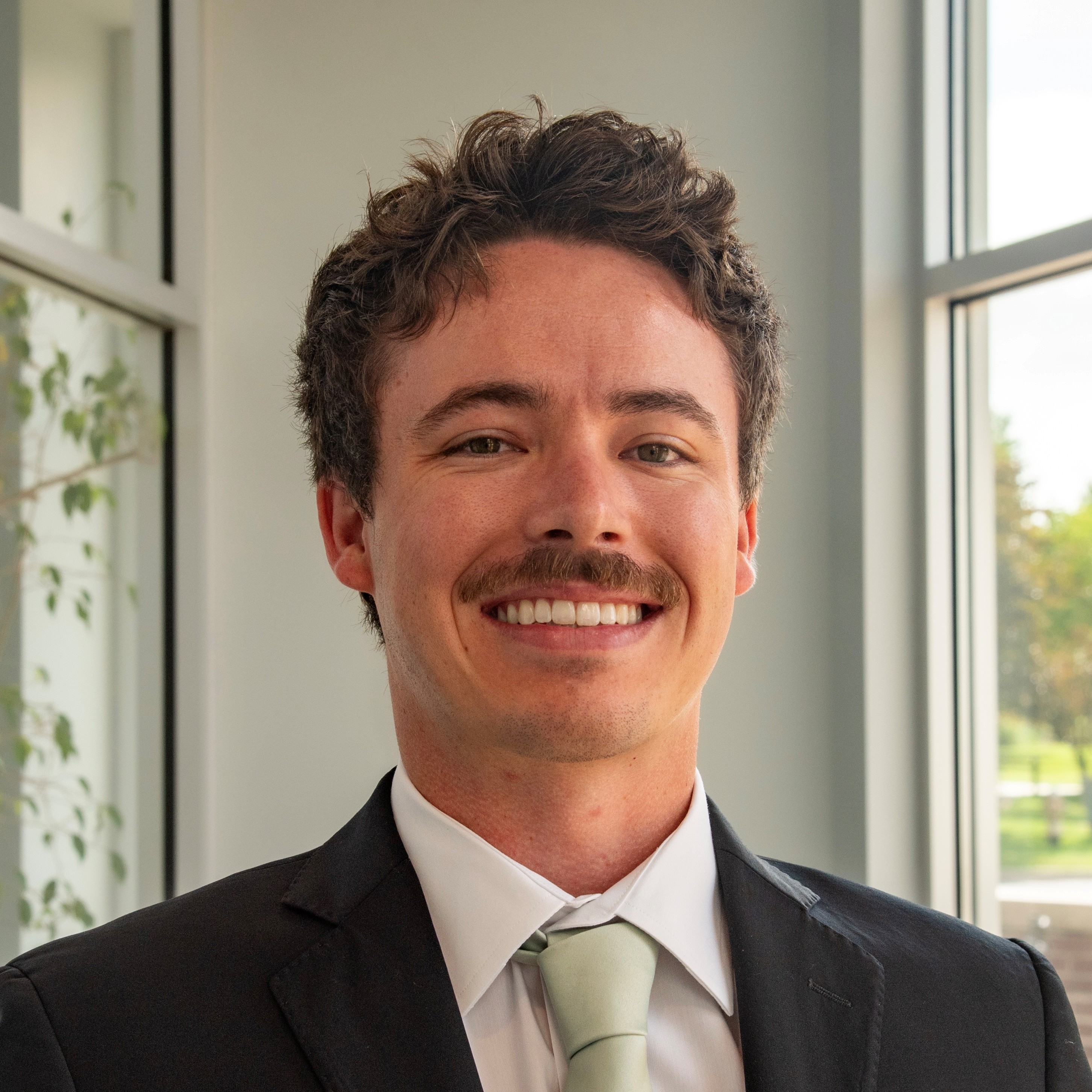}}]{Whit Vinson} $(\text{Member, IEEE})$ earned his B.S., M.S., and Ph.D. degrees in mechanical engineering from the University of Arkansas in Fayetteville in 2020, 2022, and 2025, respectively. His graduate studies focused on packaging and reliability evaluations in addition to thermal and mechanical analysis and design for micro- and power-electronics test beds and components. 

Since August 2025, Whit has been an R\&D Associate in Oak Ridge National Laboratory’s Vehicle Power Electronics Research Group and a Fellow in the University of Tennessee--Oak Ridge Innovation Institute.
\end{IEEEbiography}

\vspace{-5mm}
\begin{IEEEbiography}[{\includegraphics[width=1in,height=1.25in,clip,keepaspectratio]{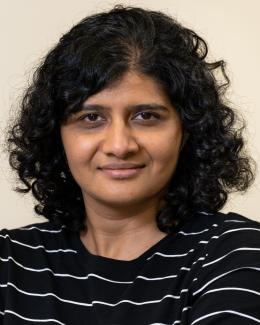}}]{Vandana P Rallabandi} $(\text{Senior Member, IEEE})$ has been a Senior R\&D Staff Member in Oak Ridge National Laboratory's Electric Drives Research Group since 2022. She has held previous positions at GE Research and was a postdoctoral researcher at the University of Kentucky, Lexington. Her research interests include electric machines, electric drives, inductive power transfer, and magnetics. 
\end{IEEEbiography}

\vspace{-5mm}
\begin{IEEEbiography}[{\includegraphics[width=1in,height=1.25in,clip,keepaspectratio]{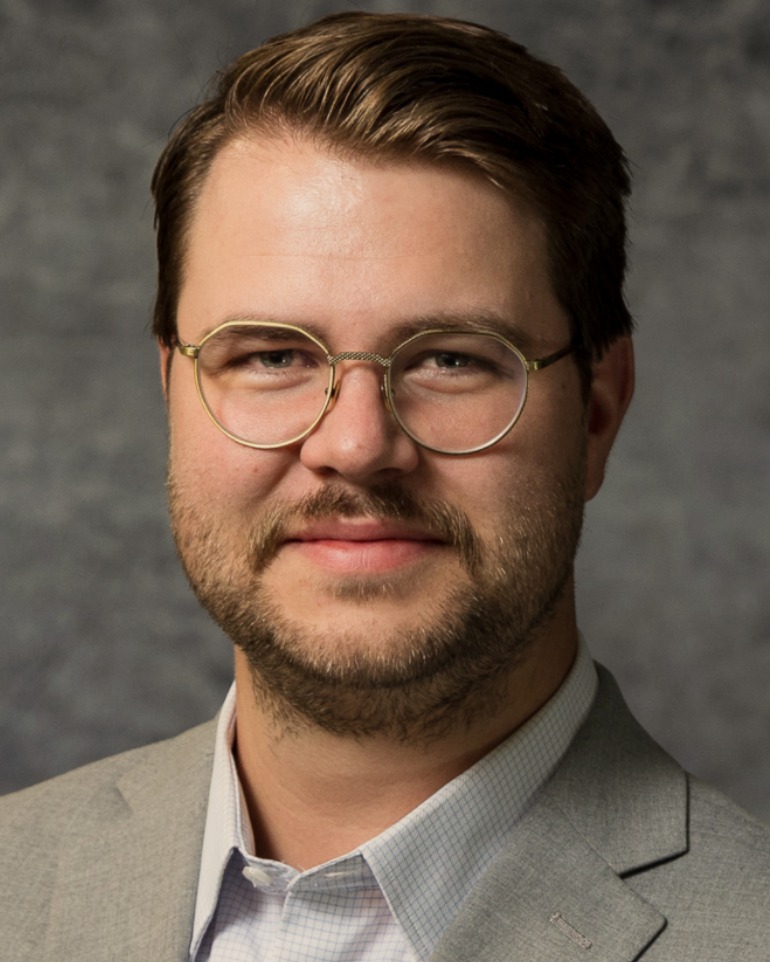}}]{Spencer Cochran} $(\text{Member, IEEE})$ received B.Sc., M.Sc., and Ph.D. degrees in electrical engineering from the University of Tennessee, Knoxville, TN, USA, in 2015, 2017, and 2021, respectively.
 
His graduate research focused on wireless power transfer at 150 kHz and 6.78 MHz, with emphasis on power-stage modeling and advanced control implementation. He was a Robert E. Bodenheimer Fellow during his master’s studies and a Tennessee Fellow for Graduate Excellence during his doctoral work. He spent 3.5 years at Astranis Space Technologies Corp. in San Francisco, CA, USA, contributing to the development of power-electronics systems for geostationary satellites. He is currently a UT–ORII Research Fellow at Oak Ridge National Laboratory, Knoxville, TN, USA. His research interests include high-switching-frequency converter design, wireless power transfer, fault tolerance, and multilevel power converter topologies.
\end{IEEEbiography}

\vspace{-5mm}
\begin{IEEEbiography}[{\includegraphics[width=1in,height=1.25in,clip,keepaspectratio]{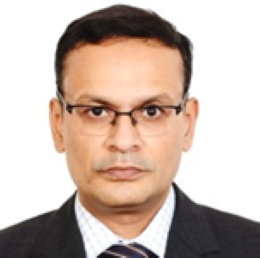}}]{Praveen Kumar} $(\text{Senior Member, IEEE})$ has an extensive career spanning nearly three decades in the field of electric motor design, with a specialized focus on high-power density motors, EV drivetrains, and the application of AI and ML to enhance the performance and reliability of these systems. He is currently a Senior Research Staff Member with Oak Ridge National Laboratory (ORNL), where he is leading initiatives to integrate AI and ML into the design and diagnostics of electric drivetrains, contributing to advancements in sustainable transportation technologies. His research has made significant strides in developing innovative motor architectures that push the boundaries of efficiency, power density, and sustainability. His work on rare earth--free magnet motors is particularly noteworthy, addressing critical supply chain challenges and environmental concerns associated with traditional high-performance motors. By leveraging advanced AI techniques, he has pioneered new motor designs that maintain high efficiency without relying on scarce and environmentally detrimental materials.
\end{IEEEbiography}

\vspace{-5mm}
\begin{IEEEbiography}[{\includegraphics[width=1in,height=1.25in,clip,keepaspectratio]{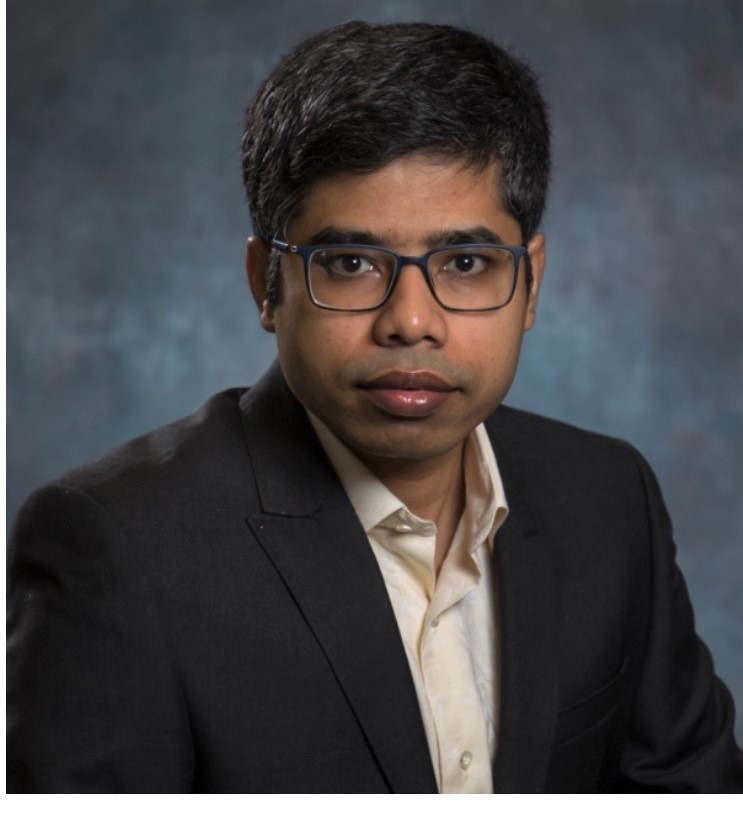}}]{Subhajyoti (“Subho”) Mukherjee} $(\text{Senior Member, IEEE})$ received a Ph.D. degree in electrical engineering from the Missouri University of Science and Technology, Rolla, MO, USA, in 2017. 

From 2018 to 2019, he was a Senior Engineer with Infineon Technologies, USA. He joined Oak Ridge National Laboratory (ORNL), Oak Ridge, TN, USA, in 2019 as a Postdoctoral Research Associate and returned in 2023 as an R\&D Staff Member in ORNL’s Vehicle Power Electronics Research Group. He previously served as an Assistant Professor at the Indian Institute of Technology (IIT) Bhubaneswar (2020–2021) and IIT Kharagpur (2021–2023). His research interests include wide-bandgap semiconductor–based power converters for electrified transportation, including integrated onboard and inductive chargers and power electronics for heavy-duty vehicle and fuel-cell power trains and grid-connected systems.
\end{IEEEbiography}

\vspace{-5mm}
\begin{IEEEbiography}[{\includegraphics[width=1in,height=1.25in,clip,keepaspectratio]{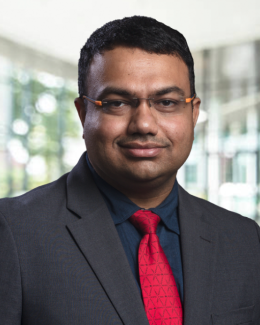}}]{Nishanth Gadiyar} $(\text{Senior Member, IEEE})$ is a native of Mangalore, India. He received his M.S. and Ph.D. degrees in electrical and computer engineering from the University of Wisconsin-Madison (UW-Madison), Madison, WI, USA, in 2020 and 2023, respectively. He is currently an Electric Machinery Design R\&D Staff Member with Oak Ridge National Laboratory (ORNL). 

Before joining ORNL, Dr. Gadiyar was a Research Engineer at the GE Research Center in Niskayuna, NY, USA. From 2018 to 2023, he was a Graduate Research Assistant and Teaching Assistant with the Wisconsin Electric Machines
and Power Electronics Consortium (WEMPEC), UW-Madison. His research interests are at the intersection of electric machine design, power electronics, and motor control, with a focus on sustainably advancing the state-of-the-art in power density, torque density, and efficiency of electromechanical drive systems across application domains. 

Dr. Gadiyar is an Associate Editor of \textit{IEEE Transactions on Industry Applications} and \textit{IEEE Transactions on Transportation Electrification}.
He has been recognized with the Wisconsin Power Engineering Award from UW-Madison in 2022, the Alvin M. Weinberg Distinguished Staff
Fellowship from Oak Ridge National Laboratory in 2024, and the Best Reviewer Award from the IEEE IAS Electric Machines Committee in 2025.
\end{IEEEbiography}

\vspace{-5mm}
\begin{IEEEbiography}[{\includegraphics[width=1in,height=1.25in,clip,keepaspectratio]{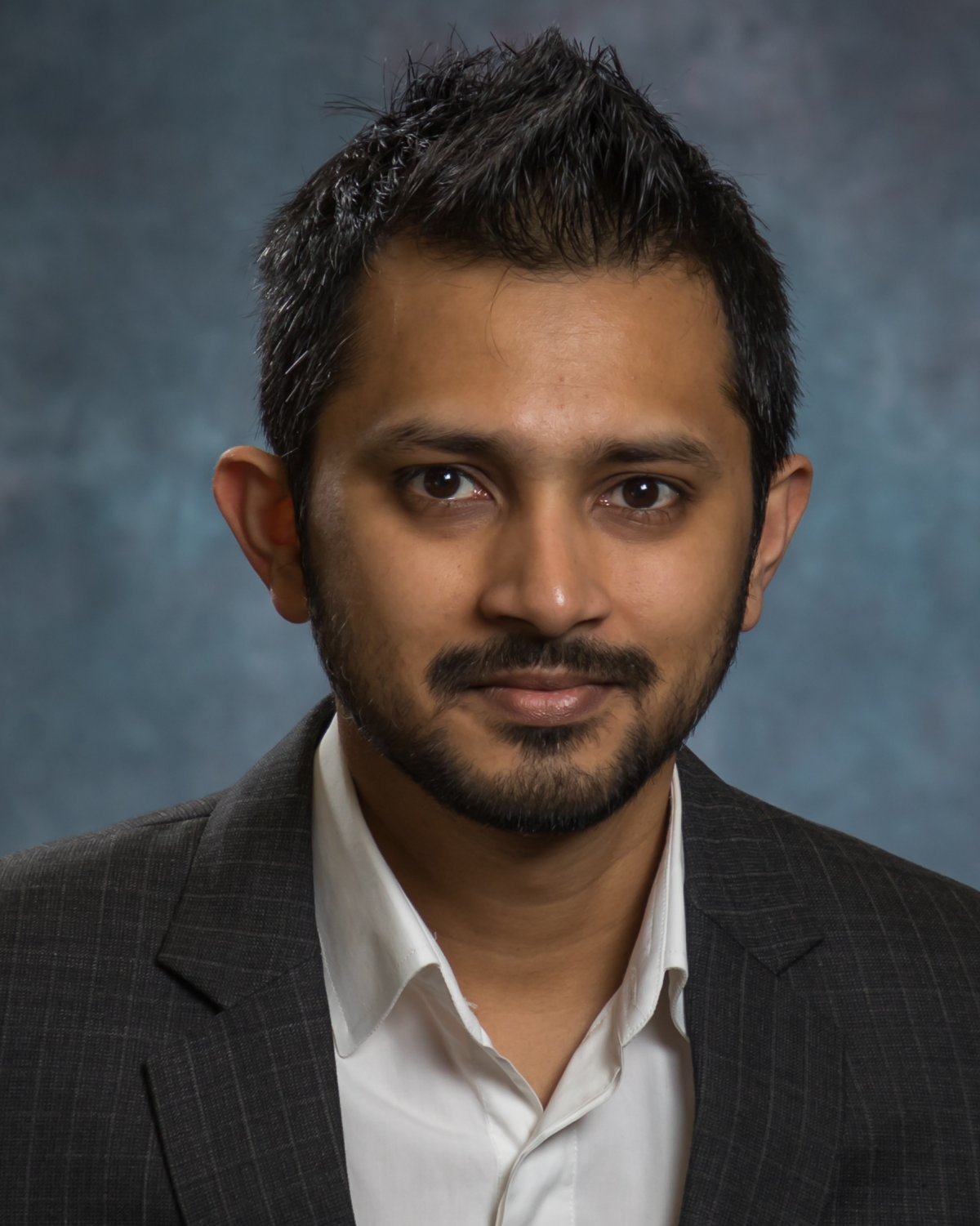}}]{Shajjad Chowdhury} $(\text{Senior Member, IEEE})$ (S’15–M’18) received a B.Sc. degree in electrical and electronics engineering from the American International University--Bangladesh, Dhaka, Bangladesh, in 2009, an M.Sc. degree in power and control engineering from Liverpool John Moores University, Liverpool, UK, in 2011, and a Ph.D. degree in electrical and electronics engineering from the University of Nottingham, Nottingham, UK, in 2016.

In January 2017, he joined the Power Electronics, Machines and Control Group, the University of Nottingham, as a Research Fellow. In 2018, he joined Oak Ridge National Laboratory, Oak Ridge, TN, USA, where he is currently leading the Electric Drive Research Group. His research interests include multilevel converters, modulation schemes, and high-performance ac drives.
\end{IEEEbiography}

\begin{IEEEbiography}[{\includegraphics[width=1in,height=1.25in,clip,keepaspectratio]{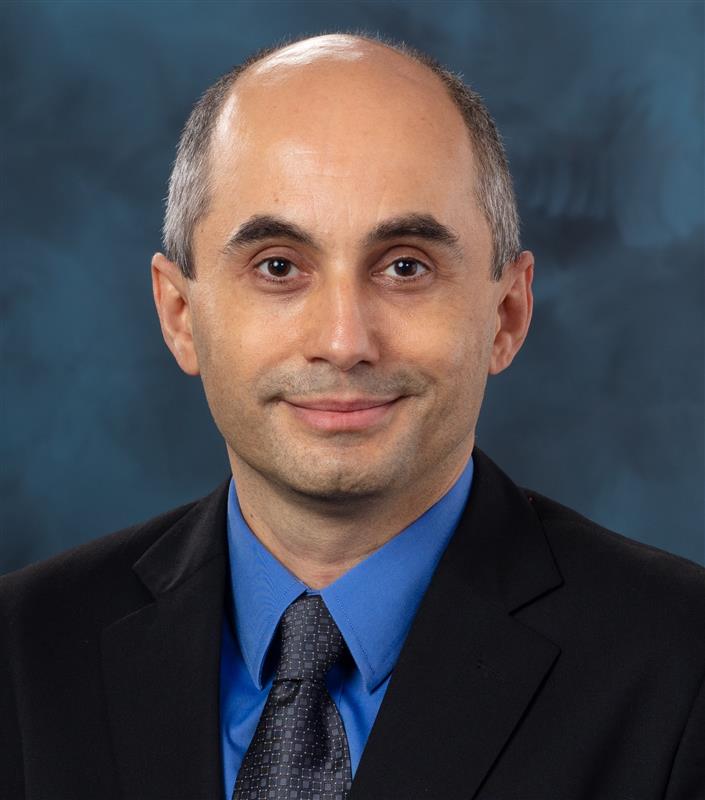}}]{Burak Ozpineci} $(\text{Fellow, IEEE})$ earned his B.S. degree in electrical engineering from Orta Dogu Technical University, Ankara, Turkey, in 1994. He then completed his M.S. and Ph.D. degrees in electrical engineering at the University of Tennessee, Knoxville, in 1998 and 2002, respectively. 

Since 2001, he has been with Oak Ridge National Laboratory, where he began as a student and has held positions as a researcher, founding group leader of the Power and Energy Systems Group, and group leader of the Power Electronics and Electric Machinery Group. He currently serves as a Corporate Fellow and the Section Head of the Vehicle and Mobility Systems Research Section. Additionally, he has a joint faculty appointment with The University of Tennessee. Dr. Ozpineci is a Fellow of IEEE.
\end{IEEEbiography}
\end{document}